\documentclass{aa} 
\usepackage{graphicx}
\usepackage{txfonts}
\def\Q{\ifmmode\mathcal{Q}\else$\mathcal{Q}$\fi}
\newcommand{\lt}{\ensuremath <}

\begin{document} 

   \title{The structure and characteristic scales of molecular clouds}

   \author{Sami Dib \inst{1}\and Sylvain Bontemps\inst{1}\and Nicola Schneider\inst{2}\and Davide Elia\inst{3}\and Volker Ossenkopf-Okada\inst{2}\and Mohsen Shadmehri\inst{4}\and Doris Arzoumanian\inst{5}\and Fr\'{e}d\'{e}rique Motte\inst{6}\and Mark Heyer\inst{7}\and {\AA}ke Nordlund\inst{8}\and Bilal Ladjelate\inst{9}}
 
   \institute{Laboratoire d'Astrophysique de Bordeaux, Universit\'{e} de Bordeaux, CNRS, B18N,  all\'{e}e Geoffroy Saint-Hilaire, 33615, Pessac, France
      \email{sami.dib@gmail.com}
     \and
              I. Physikalisches Institut, Universit\"{a}t zu K\"{o}ln, Z\"{u}lpicher Stra{\ss}e 77, 50937 K\"{o}ln, Germany 
     \and 
              Istituto di Astrofisica e Planetologia Spazialli, INAF, via Fosso del Cavaliere 100, Roma, 00133, Italy   
      \and 
              Department of Physics, Faculty of Sciences, Golestan University, Gorgan 49138-15739, Iran   
      \and
              Instituto de Astrof\'{i}sica e Ci\^{e}ncias do Espa\c{c}o, Universidade do Porto, CAUP, Rua das Estrelas, PT4150-762, Porto, Portugal                        
      \and
              Institut de Plan\'{e}tologie et d'Astrophysique de Grenoble, Universit\'{e} Grenoble Alpes, CNRS, Grenoble, France
     \and 
              Department of Astronomy, University of Massachusetts, Amherst, MA 01003, USA         
      \and  
              Centre for Star and Planet Formation, the Niels Bohr Institute and the Natural History Museum of Denmark, University of Copenhagen, {\O}ster Voldgade 5-7, DK-1350, Denmark        
      \and
              Instituto de Radioastronom\'{i}a Milim\'{e}trica, IRAM Avenida Divina Pastora 7, Local 20, 18012, Granada, Spain                                                             
          }
          
\authorrunning{Dib et al.}
\titlerunning{Characteristic scales in molecular clouds}
         
 
\abstract{The structure of molecular clouds holds important clues regarding the physical processes that lead to their formation and subsequent dynamical evolution. While it is well established that turbulence imprints a self-similar structure onto the clouds, other processes, such as gravity and stellar feedback, can break their scale-free nature. The break of self-similarity can manifest itself in the existence of characteristic scales that stand out from the underlying structure generated by turbulent motions. In this work, we investigate the structure of the Cygnus-X North and Polaris Flare molecular clouds, which represent two extremes in terms of their star formation activity. We characterize the structure of the clouds using the delta-variance ($\Delta$-variance) spectrum. In the Polaris Flare, the structure of the cloud is self-similar over more than one order of magnitude in spatial scales. In contrast, the $\Delta$-variance spectrum of Cygnus-X North exhibits an excess and a plateau on physical scales of $\approx 0.5-1.2$ pc. In order to explain the observations for Cygnus-X North, we use synthetic maps where we overlay populations of discrete structures on top of a fractal Brownian motion (fBm) image. The properties of these structures, such as their major axis sizes, aspect ratios, and column density contrasts with the fBm image, are randomly drawn from parameterized distribution functions. We are able to show that, under plausible assumptions, it is possible to reproduce a $\Delta$-variance spectrum that resembles that of the Cygnus-X North region. We also use a "reverse engineering" approach in which we extract the compact structures in the Cygnus-X North cloud and reinject them onto an fBm map. Using this approach, the calculated $\Delta$-variance spectrum deviates from the observations and is an indication that the range of characteristic scales ($\approx 0.5-1.2$ pc) observed in Cygnus-X North is not only due to the existence of compact sources, but is a signature of the whole population of structures that exist in the cloud, including more extended and elongated structures.}

   \keywords{stars: formation - ISM: clouds, general, structure - galaxies: ISM, star formation}

 \maketitle
%

\section{Introduction}\label{introduction}

The interstellar medium (ISM), both in the Milky Way and in external galaxies, exhibits a scale-free nature that extends over many physical scales. This is observed both in the diffuse \ion{H}{i} gas (e.g., Elmegreen 2001; Dickey et al. 2001; Dib \& Burkert 2005; Begum et al. 2006; Dutta et al. 2009, Zhang et al. 2012; Dutta et al. 2013, Miville-Desch\^{e}nes et al. 2016) and in the molecular phase (e.g., Stutzki et al. 1998; Heyer \& Brunt 2004; Heyer et al. 2009; Schneider et al. 2011; Roman-Duval et al. 2011; Rebolledo et al. 2015; Panopoulou et al. 2017; Traficante et al. 2018; Hirota et al. 2018; Kong et al. 2018; Dib \& Henning 2019; Henshaw et al. 2020). This self-similarity is also observed in the spatial distribution of young clusters in galactic disks (e.g., Elmegreen et al. 2006; Gouliermis et al. 2017; Grasha et al. 2019).

\begin{figure*}
\centering
\vspace{0cm}
\includegraphics[width=0.497\textwidth]{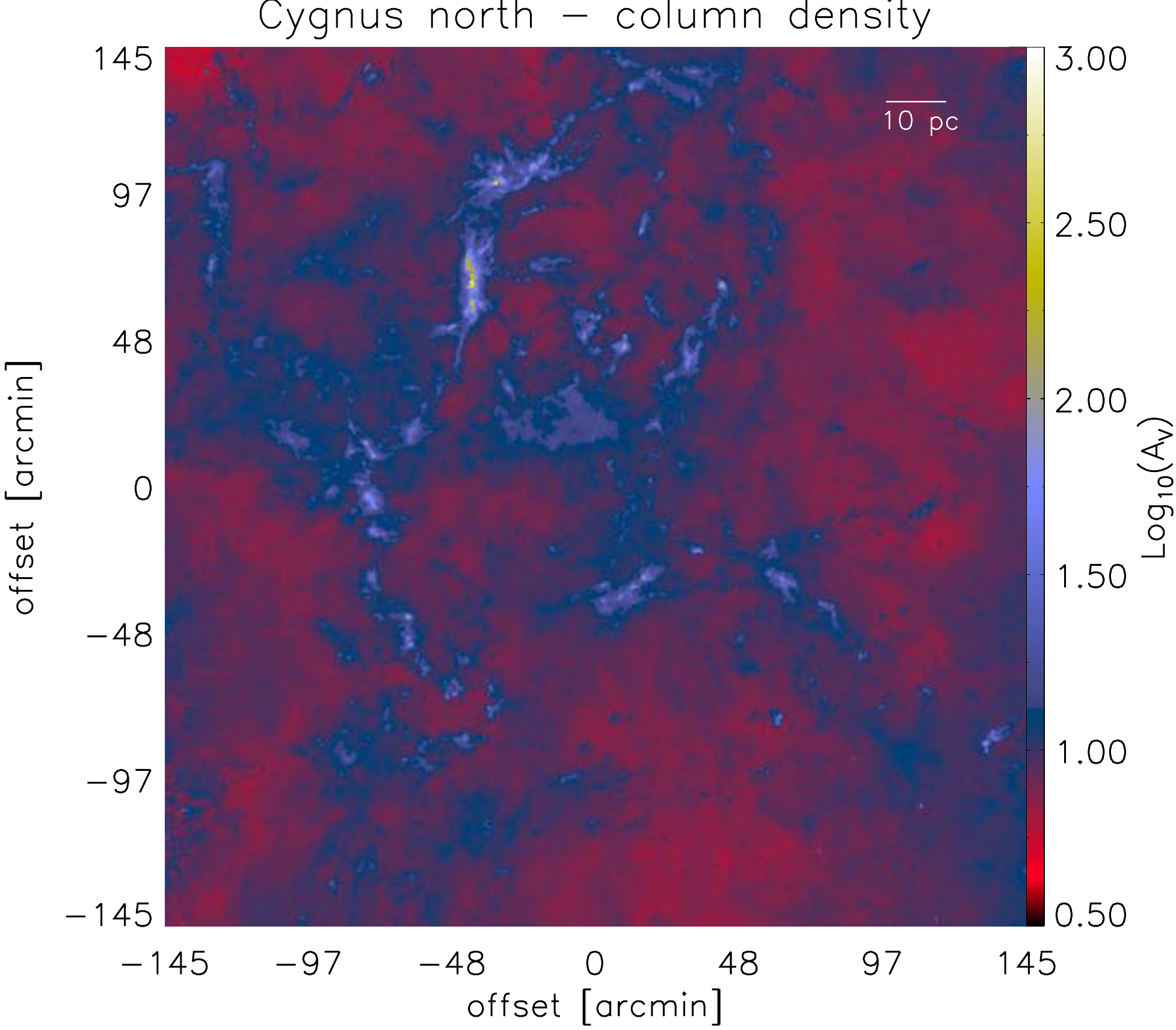}
\includegraphics[width=0.497\textwidth]{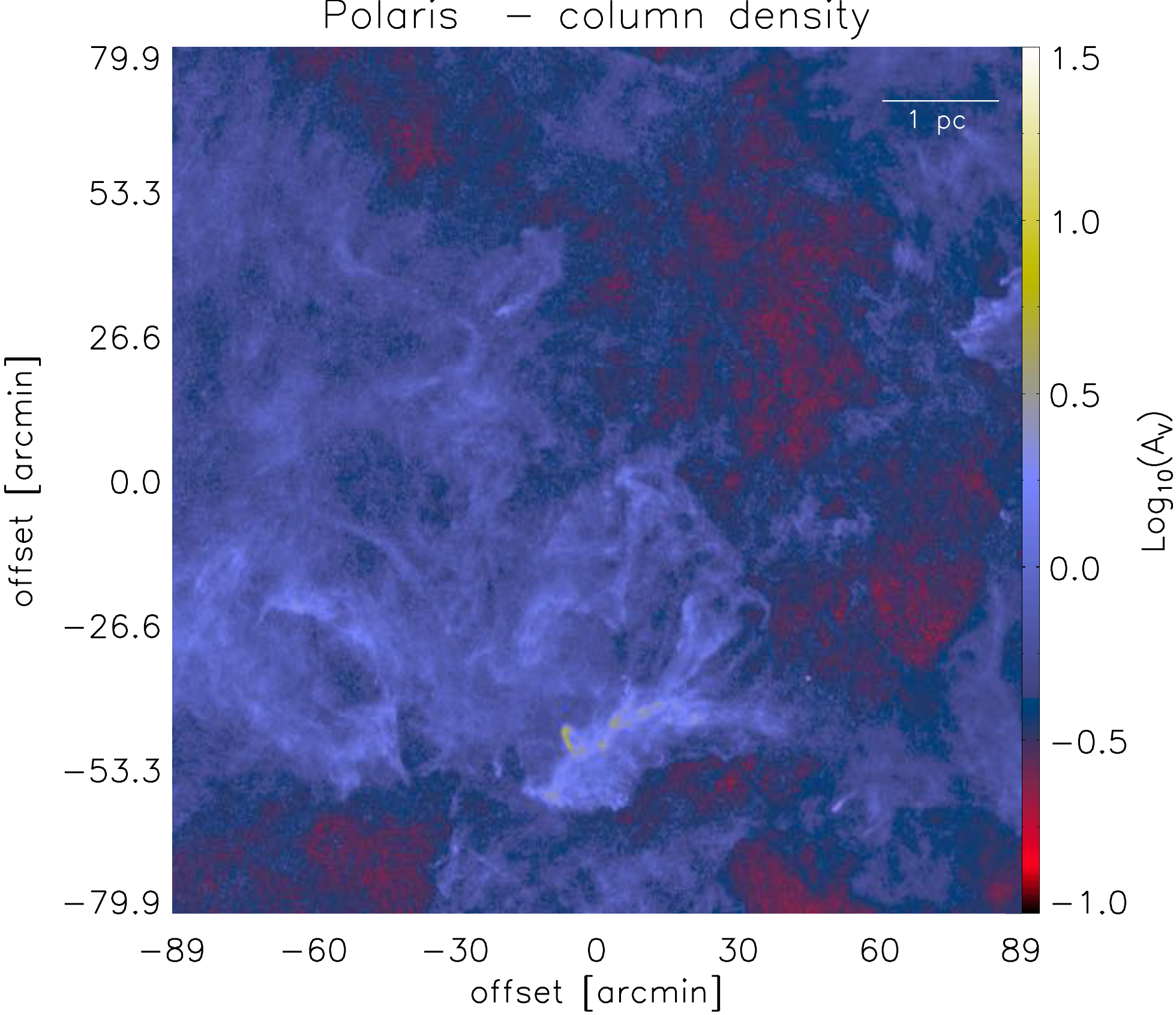}\\
\vspace{0cm}
\caption{Column density maps of the Cygnus-X North cloud (left) and the Polaris Flare cloud (right). Column densities are displayed in units of the visual extinction using the conversion $N_{\rm H_{2}}/{\rm A_{V}}=0.94\times 10^{21}$ cm$^{-2}$.}
\label{fig1}
\end{figure*}

Turbulence is ubiquitously observed in all phases of the interstellar gas. It is thought to be the main regulator of the ISM structure and dynamics in cold, neutral gas and, hence, is responsible for setting a self-similar behavior in this regime (e.g., Elmegreen \& Scalo 2004; Dib et al. 2008; Burkhart et al. 2013). This self-similarity can be broken on various scales. This can happen when specific physical processes dominate the injection of energy and momentum in the ISM. In galactic disks, various forms of feedback from massive stars (i.e., ionizing radiation, radiation pressure, stellar winds, and supernova explosions) impart significant amounts of energy and momentum onto the ISM on intermediate scales, that is, $\approx 50-500$ pc (e.g., Dib et al. 2006; Ostriker et al. 2010; Dib 2011; Dib et al. 2011,2013; Hennebelle \& Iffrig, 2014; Hony et al. 2015; Padoan et al. 2016; Ntormousi et al. 2017; Dib et al. 2017; Seifried et al. 2020). Some of these scales could be detected as characteristic scales in the ISM. Dib et al. (2009) found that the orientations of molecular clouds in the outer Galaxy are correlated on spatial scales that are on the order of the expected sizes of supernova remnants, which are prevalent in those regions of the Galactic disk. On small scales, particularly within molecular clouds, the self-similarity can be broken on physical scales where the self-gravity of the gas becomes important and dictates the motions of the gas (e.g., Dib et al. 2008). When dynamically important, and due to their anisotropic nature, magnetic fields can also play a role in breaking the self-similar nature of the gas (e.g., Soler 2019). Scales at which there might be a departure from self-similarity are the ones associated with the sizes of filaments and fragments within filaments (e.g., Andr\'{e} et al. 2010; Arzoumanian et al. 2011; Hacar et al. 2013; K\"{o}nyves et al. 2015) as well as scales on which filaments interact, with hubs and ridges forming at the intersection of two or more filaments (e.g., Schneider et al. 2012; Samal et al. 2015; Dewangan et al. 2017; Trevi\~{n}o-Morales et al. 2019; Clarke et al. 2019). Stellar feedback can also perturb the self-similarity of the gas on small scales. Russeil et al. (2013) find, in a study of the massive star-forming region NGC6334, that characteristic scales around 1-10 pc can be caused by the injection of energy due to expanding \ion{H}{ii} regions. 
 
\begin{figure*}
\centering
\includegraphics[width=0.497\textwidth]{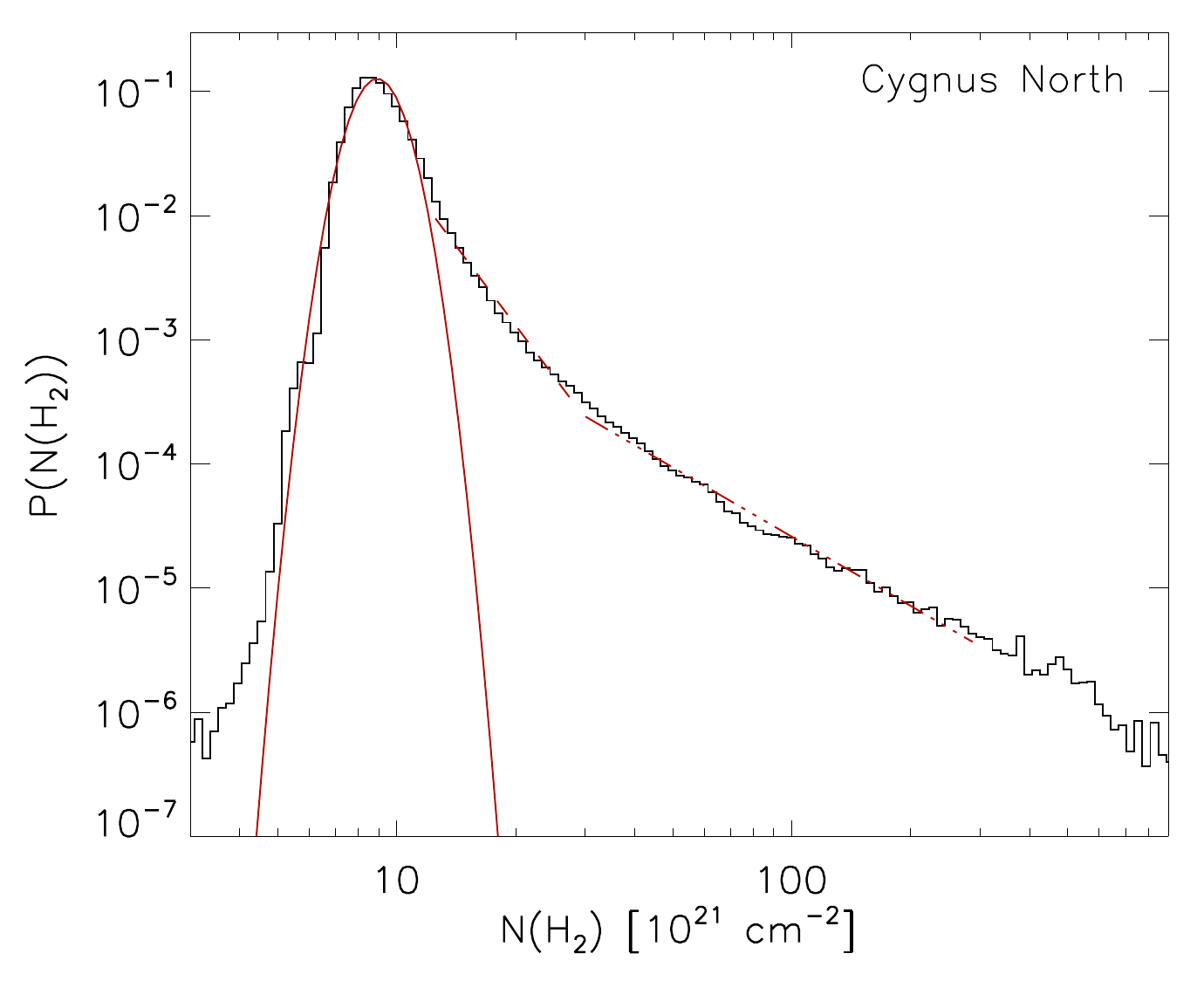}
\includegraphics[width=0.497\textwidth]{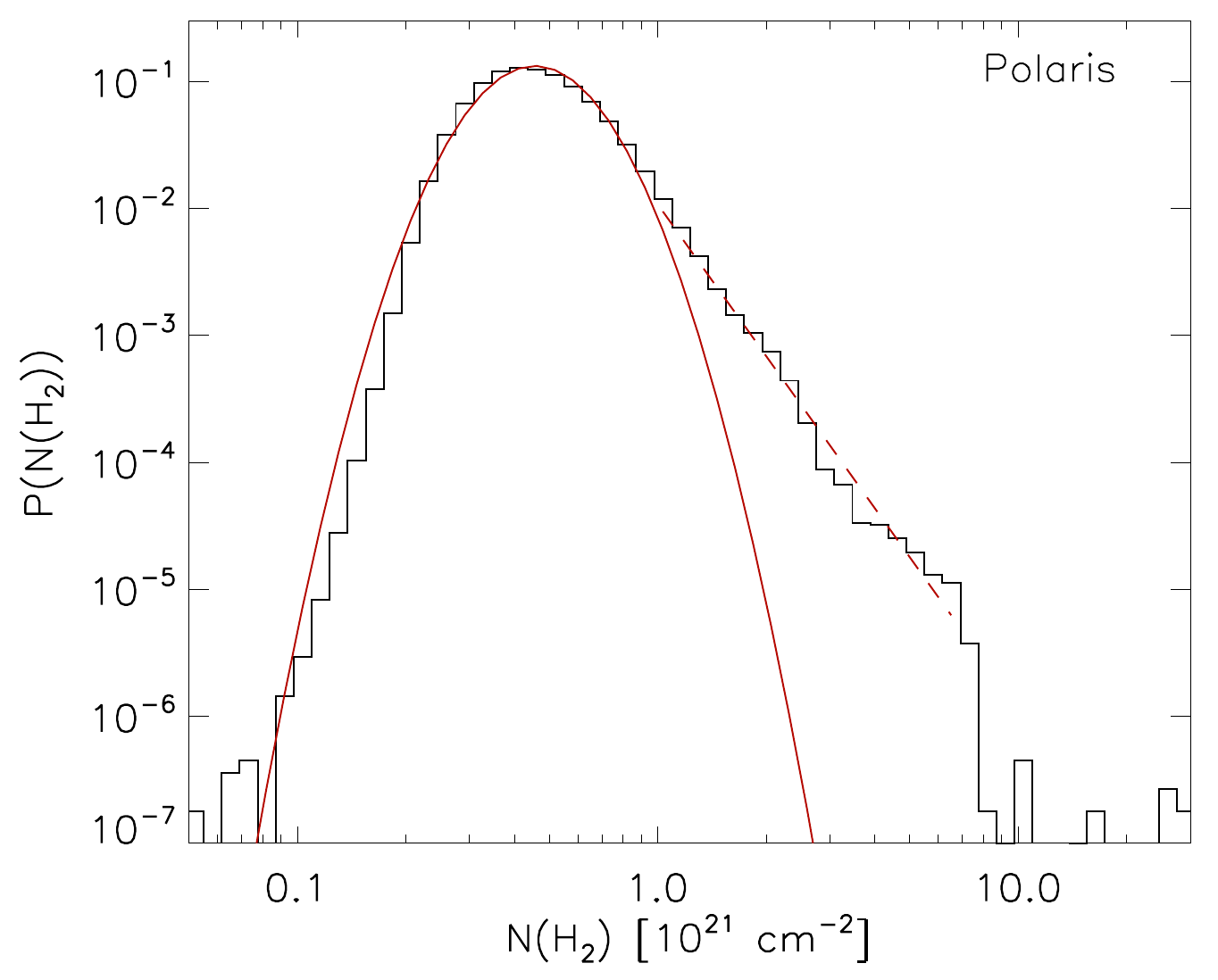}\\
\caption{Column density distribution function in the Cygnus-X North cloud (left panel) and the Polaris Flare cloud (right panel). The full red lines in both panels show a fit by a lognormal function in the low column density regime ($\lesssim 5\times10^{21}$ cm$^{-2}$ in Cygnus-X North and $\lesssim 1\times 10^{21}$ cm$^{2}$ in Polaris). The dashed line is a fit to the power-law regime that is observed in both regions in the intermediate column density regime, while the triple-dot dashed red line in the case of the Cygnus-X North region is a fit to the shallower power law in the high column density regime.}
\label{fig2}
\end{figure*}

\begin{figure*}
\centering
\includegraphics[width=0.497\textwidth]{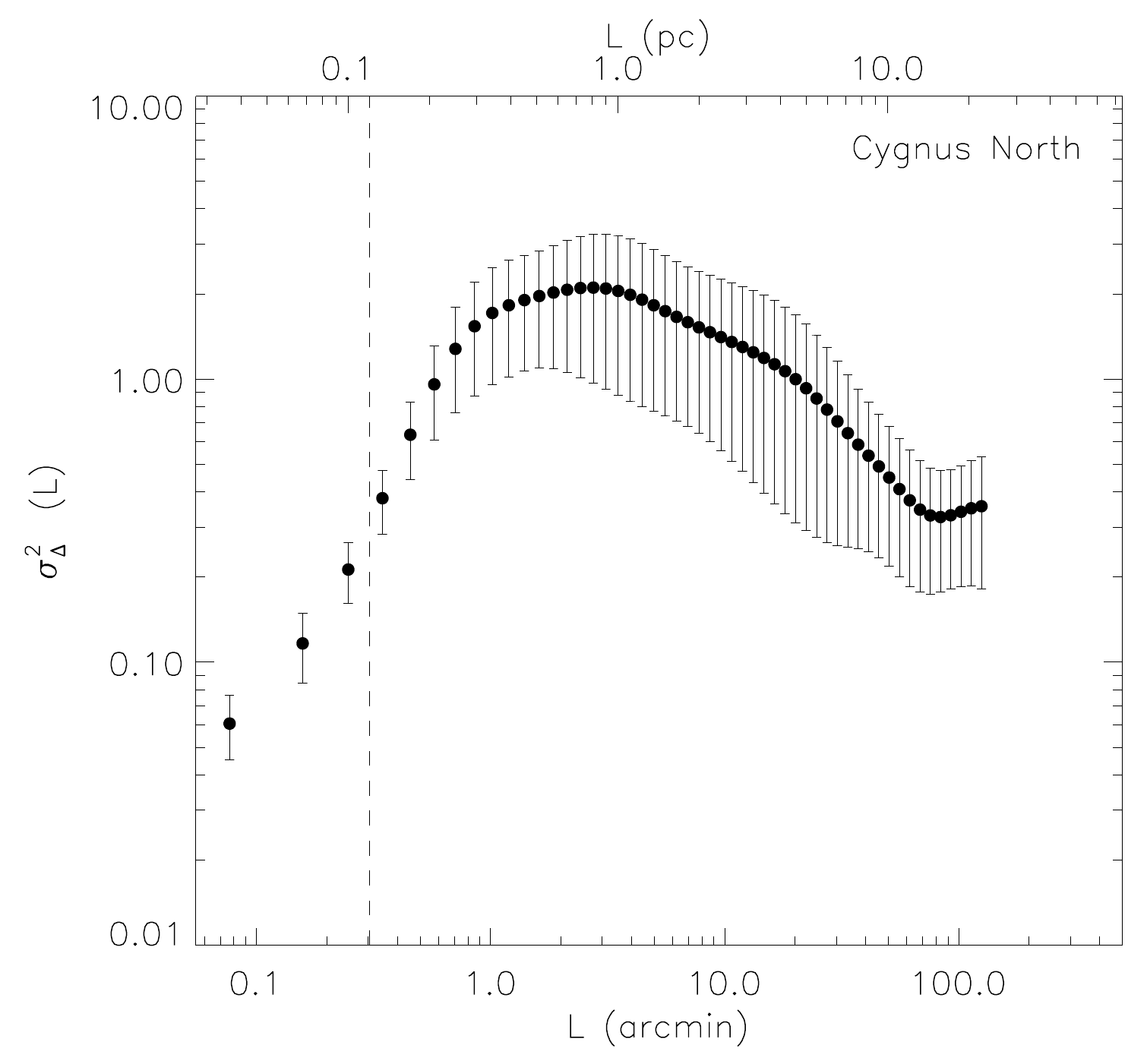}
\includegraphics[width=0.497\textwidth]{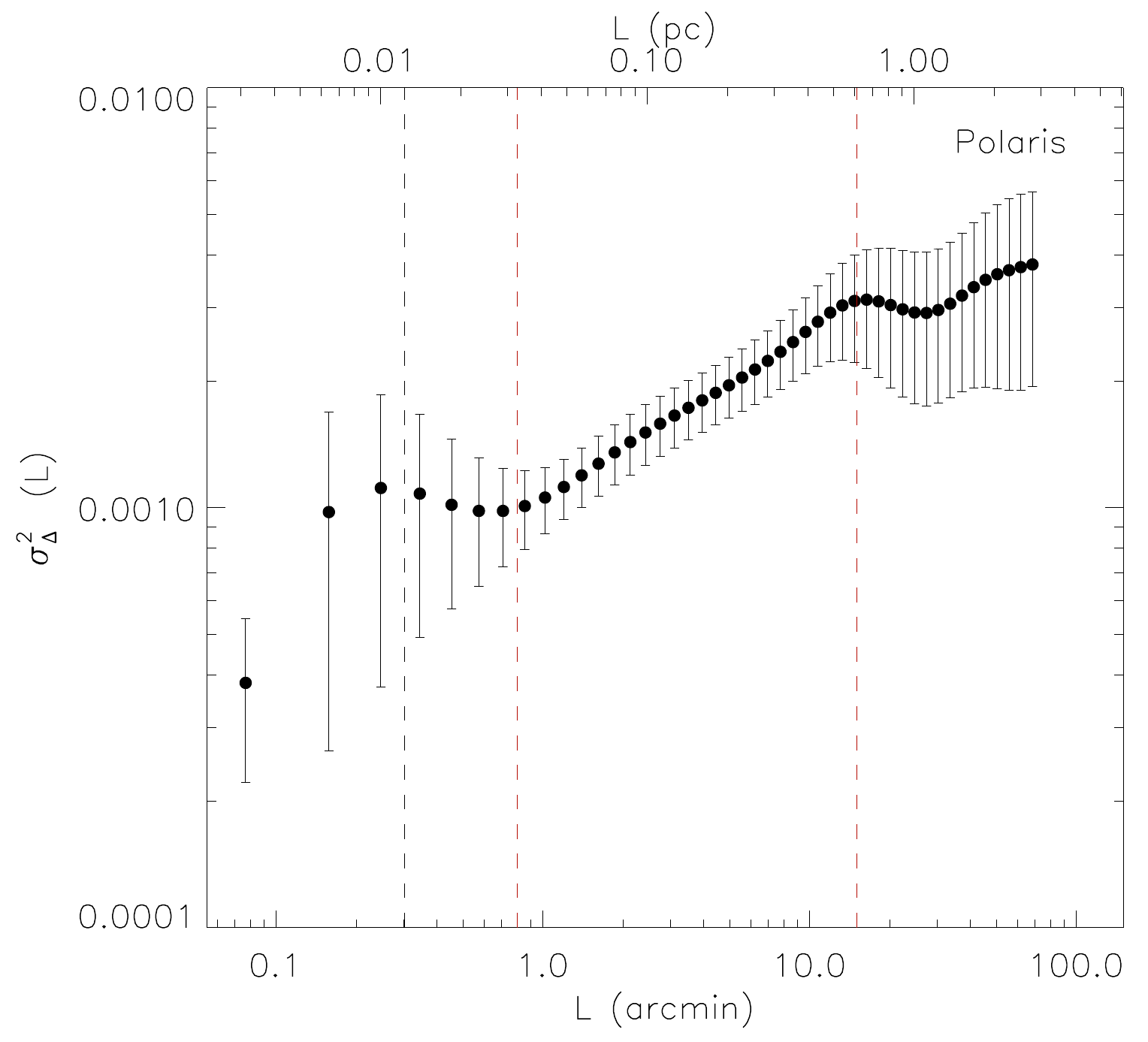}\\
\caption{Delta-variance functions calculated for the Cygnus-X North cloud (left) and the Polaris Flare cloud (right). The vertical dashed black lines in both panels mark the position of the spatial resolution for each of these two regions. The vertical dashed red lines in the case of Polaris mark the spatial range over which the power-law fit is performed. We do not attempt any fit in Cygnus-X North because the underlying self-similar regime is heavily perturbed by the presence of structures (see Sect.~\ref{realistic})}
\label{fig3}
\end{figure*}

Using the delta-variance ($\Delta$-variance) spectrum (Stutzki et al. 1998), we analyzed the spatial structure of two Galactic molecular clouds that lie at the extreme ends of what can be found in the Galaxy in terms of their star formation activity, namely the Cygnus-X North region and the Polaris Flare. The Cygnus-X North molecular cloud complex is an active region of star formation where many sub-regions of high-mass star formation  can be found (e.g., Schneider et al. 2006; Motte et al. 2007; Reipurth \& Schneider 2008; Schneider et al. 2010; Bontemps et al. 2010; Csengeri et al. 2011; Hennemann et al. 2012; Kryukova et al. 2014; Maia et al. 2016). In contrast, the Polaris Flare is essentially a translucent, non-star forming cloud (e.g., Ward-Thompson et al. 2010; Miville-Desch\^{e}nes et al. 2010). In Sect.~\ref{obsdata}, we briefly summarize the {\it Herschel} satellite data that are analyzed in this work, and in Sect.~\ref{npdf} we present and discuss the column density distribution functions of both regions. The $\Delta$-variance method is discussed in Sect.~\ref{deltavar}, and its application to the {\it Herschel} satellite maps of Cygnus-X North and Polaris is presented in Sect.~\ref{obsresults}. In Sect.~\ref{interpret}, we interpret our findings with the help of simulated synthetic observations and discuss the shape of the $\Delta$-variance in models of increasing complexity. We start with models of pure fractal Brownian motion (fBm) images (Sect.~\ref{fbm}) and continue to models where individual structures are superimposed on an fBm (Section.~\ref{fbmplus}). We finish with models in which an entire population of structures is superimposed on an fBm and which have sets of properties (such as the size of major axis, elongation, and column density contrast) that are described by parameterized probability distribution functions (Sect.~\ref{realistic}). In Sect.~\ref{discussion}, we discuss different caveats and limitations pertaining to the observations and the models, and in Sect.~\ref{conclusions} we summarize our results and conclude. 

\section{The data: {\it\bf Herschel} maps of star-forming regions}\label{obsdata}

The observations that are analyzed in this work were performed using the {\it Herschel} space observatory (Pilbratt et al. 2010). In particular, we made use of data products from the {\it Herschel} Gould Belt Survey (HGBS\footnote{http://gouldbelt-herschel.cea.fr/archives}, Andr\'{e} et al. 2010) for the Polaris Flare region and the {\it Herschel} imaging survey of OB Young Stellar objects (HOBYS, Motte et al. 2010) program for the Cygnus-X North region. The column density maps were determined from a pixel-to-pixel greybody fit to the red wavelength of PACS (Poglitch et al. 2010) observations at 160 $\mu$m ($11.7\arcsec$ angular resolution), and the three SPIRE (Griffin et al. 2010) wavelengths are $250$, $350$, and $500$ $\mu$m at the resolutions of $18.2\arcsec$, $24.9\arcsec$, and $36.3\arcsec$, respectively. For the SPIRE data reduction, we used the HIPE pipeline (versions 10 to 13), including the destriper task for SPIRE as well as HIPE and scanamorphos (Roussel 2013) for PACS. The SPIRE maps were calibrated for extended emission. All maps have an absolute flux calibration using offset values determined in Bernard et al. (2010). For the spectral energy distribution (SED) fit, the specific dust opacity per unit mass (dust plus gas) is approximated by a power law $\kappa_{\nu} = 0.1 (\nu/1000 {\rm GHz})^{\beta_{dust}}$ cm$^{2}$ g$^{-1}$ with $\beta_{dust}=2$ and the dust temperature and column density left as free parameters. The description of how high angular resolution maps were derived is detailed in Palmeirim et al. (2013). The concept is to employ a multi-scale decomposition of the flux maps and assume a constant line-of-sight temperature. The final map at $18.2\arcsec$ angular resolution is constructed from the difference maps of the convolved surface density SPIRE maps (at 500, 350, and 250 $\mu$m) and the temperature information  from the color temperature derived from the $160~\mu{\rm m}/250~\mu{\rm m}$ ratio.

The molecular hydrogen (H$_{2}$) column densities were transformed into visual extinction ($\rm {A_{V}}$) using the conversion formula $N(\rm{H_{2}})/{\rm A_{V}}= 0.94\times10^{21}$ cm$^{-2}$ mag$^{-1}$ (Bohlin 1978). The column density maps for the Cygnus-X North and Polaris Flare clouds are displayed in Fig.~\ref{fig1} (left- and right-hand panels, respectively). For Cygnus-X, column density maps were already presented in  Hennemann et al. (2012) and Schneider et al. (2016a,b), and for Polaris in Robitaille et al. (2019). These maps have a lower angular resolution of 36.3$''$ and cover different areas than those presented in the current study. The Polaris Flare cloud is located at a distance of $140$ pc (Falgarone et al. 1998), and hence each pixel on the map corresponds to a spatial size of $\approx 0.002$ pc. The Cygnus-X cloud is located at a distance of $1.7$ kpc (Schneider et al. 2006), and, in this case, each pixel corresponds to a spatial size of $\approx 0.025$ pc. The maps of Cygnus-X North and Polaris contain $5740\times5740$ pixels and $3538\times3164$ pixels, respectively. The total physical size covered by the maps of Cygnus-X North and Polaris is $\approx 143.5~{\rm pc} \times 143.5~{\rm pc}$ and $7.07~{\rm pc} \times 6.32~{\rm pc}$, respectively. The full width at half maximum (FHWM) of the beam is sampled with six pixels and thus the spatial resolution for the Cygnus-X North and Polaris maps are $\approx 0.15$ pc and $\approx 0.012$ pc, respectively.

\section{Column density distribution functions}\label{npdf}

Here, we only wish to highlight the differences between the Cygnus-X North and Polaris regions in terms of their column density distributions before analyzing the spatial structure of the clouds. Figure~\ref{fig2} displays the column density probability distribution function (N-PDF) for Cygnus-X North (left-hand panel) and Polaris (right-hand panel). Both N-PDFs resemble those shown in Schneider et al. (2016b) for Cygnus-X and Schneider et al. (2013) for Polaris. The N-PDFs for both regions exhibit a lognormal behavior at low ${\rm A_{V}}$ ($2.5 \le {\rm A_{V}} \lt 12$ for Cygnus-X North and $\le 1$ for Polaris) with a significant difference in the position of the peak between the two regions. A fit to the lognormal part of the N-PDF yields $\rm {A_{V,peak}}=8.92^{+1.40}_{-1.21}$ in Cygnus-X North and $\rm {A_{V,peak}}=0.45^{+0.19}_{-0.13}$ in Polaris. The width of the lognormal is $\sigma_{\rm {A_{V}}}=1.14\pm1.10$ and $\sigma_{\rm {A_{V}}}=1.39\pm1.25$ in Cygnus-X North and Polaris, respectively. At larger column densities (${\rm A_{V}} \ge 12$ in Cygnus-X and ${\rm A_{V}} \ge 1$ in Polaris), the N-PDF turns into a power-law distribution. In Cygnus-X, there are two distinct power laws: a steep power law with an exponent of $-4.24\pm0.15$ in the ${\rm A_{V}}$ range of [$12,30$] and a shallower power law with an exponent of $-1.85\pm0.02$ in the ${\rm A_{V}}$ range [$30,300$]. In contrast, the N-PDF for Polaris exhibits a single power-law tail (PLT) starting from ${\rm A_{V}} \gtrsim 1$ with an exponent of $-3.97\pm 0.13$. The exponent of the PLT we find for Polaris matches the one found in other studies for the same cloud (e.g., Alves et al. 2007; Schneider et al. 2013). The N-PDF parameters we find for the two regions are only slightly different from what was obtained in earlier studies, and this difference is due to the fact that the considered areas of the clouds are different. 
 
A PLT in the N-PDF is connected to the existence of a power-law distribution in volume density and is commonly attributed to the effects of the self-gravity of the gas in generating dense structures in the cloud (e.g., Klessen 2000; Dib 2005; Dib \& Burkert 2005; Kainulainen et al. 2009; Kritsuk et al. 2011; Ward et al. 2014; Girichidis et al. 2014; Schneider et al. 2015; Donkov et al. 2018; Corbelli et al. 2018; Veltchev et al. 2019). Another interpretation for the origin of the first, steep PLT has been proposed by Auddy et al. (2018;2019). These authors showed, using numerical simulations of molecular clouds with non-ideal magnetohydrodynamics (MHD), that in the case of a magnetically subcritical cloud, a steep PLT (slope $\approx -4$) can emerge as a result of gravitational contraction driven by ambipolar diffusion. The second, shallower PLT is only associated with regions of the highest column densities in Cygnus-X. This was reported for the first time in high-mass star-forming regions (Tremblin et al. 2014, Schneider et al. 2016b) and interpreted as arising from gravitational collapse of cores with either internal sources (protostars, ultra-compact \ion{H}{ii} regions) that lead to internal ionization compressions, or external compression from the associated \ion{H}{ii} region. However, the picture is probably more complicated since a second shallower PLT was also detected in low-mass star-forming regions (Schneider et al. 2020, in prep.). It is not within the scope of this paper to discuss the N-PDFs in extensive detail. In summary, and despite the fact that gravity is suspected to be the primary culprit of the formation of both PLTs, we conclude that it currently is not straightfoward to explain the different parts of the PLT as the consequence of a hierarchical gravitational collapse, whereby the first steep PLT can be attributed to the formation of compact structures (filaments or clumps) and the second, shallower PLT to the collapse of dense cores.

\section{Analysis: The $\Delta$-variance method}\label{deltavar}

We quantified the structure of molecular clouds using the $\Delta$-variance method. The method was originally introduced in Stutzki et al. (1998) and Zielinsky et al. (1999) and is a generalization of the Allan variance (Allan 1966). In this work, we used an improved version of the method presented in Ossenkopf et al. (2008a)\footnote{The IDL package for calculating the $\Delta$-variance can be found at \url{https://hera.ph1.uni-koeln.de/~ossk/Myself/deltavariance.html}}.
 
Here, we briefly present a summary of the main steps and characteristics of the method. For a 2D field  $A(x,y)$, the $\Delta$-variance on a scale $L$ is defined as being the variance of the convolution of $A$ with a filter function $\sun_{L}$ such that

\begin{equation}
\sigma_{\Delta}^{2}(L)=\frac{1}{2\pi} \langle (A * \sun_{L})^{2}  \rangle_{x,y}.
\label{eq1}
\end{equation}

For the filter function, Ossenkopf et al. (2008a) recommend the use of a "Mexican hat" that is defined as

\begin{equation}
\sun_{L} \left(r\right)= \frac{4}{\pi L^{2}} e^{\frac{r^{2}} {(L/2)^{2}}} - \frac{4}{\pi L^{2} (v^{2} -1)} \left[ e^{\frac{r^{2}}{(vL/2)^{2}}} -e^{\frac{r^{2}}{(L/2)^{2}}}\right],
\label{eq2}
\end{equation}

\noindent where the two terms on the right side of Eq.~\ref{eq2} represent the core and the annulus of the Mexican hat function, respectively, and $v$ is the ratio of their diameters (we used $v=1.5$). For a faster and more efficient computation of Eq.~\ref{eq1}, Ossenkopf et al. (2008a) performed the calculation as a multiplication in Fourier space, and thus, the $\Delta$-variance is given by

\begin{equation}
\sigma_{\Delta}^{2}(L)=\frac{1}{2 \pi} \int \int P \left| \bar{\sun}_{L} \right |^{2} dk_{x} dk_{y},
\label{eq3}
\end{equation}  

\noindent where $P$ is the power spectrum of $A$ and $\bar{\sun}_{L}$ is the Fourier transform of the filter function. If $P$ can be described by a power law, and if $\beta$ is the exponent of the power spectrum, then a relation exists between the exponent of the power law that describes the $\Delta$-variance ($\alpha$) and $\beta$ (Stutzki et al. 1998), and this is given by

\begin{equation}
\sigma_{\Delta}^{2}(L) \propto L^{\alpha} \propto L^{\beta-2}
\label{eq4}
.\end{equation}

The value of $\alpha$ can be inferred from the range of spatial scales over which the $\Delta$-variance displays a self-similar behavior and can be tied to the value of $\beta$. The error bars of the $\Delta$-variance on a given scale are computed from the counting error determined by the finite number of statistically independent measurements in the filtered map and the variance of the variances (i.e., the fourth moment of the filtered map). Characteristic scales are scales at which there is a break of the self-similarity and which show up in the $\Delta$-variance spectra as break points, peaks, or inflection points. Any underlying self-similar behavior of the cloud can be entirely perturbed on many or all physical scales if there is a variety of structures that coexist in the cloud. The $\Delta$-variance has been employed to analyze the structure of observed molecular clouds (e.g., Bensch et al. 2001; Campeggio et al. 2004; Sun et al. 2006; Ossenkopf et al. 2008b; Rowles \& Froebrich 2011; Schneider et al. 2011; Russeil et al. 2013; Elia et al. 2014) as well as simulated molecular clouds (e.g., Ossenkopf et al. 2001; Mac Low \& Ossenkopf 2000; Ossenkopf \& Mac Low 2002; Federrath et al. 2009; Bertram et al. 2015). In most cases, the $\Delta$-variance has been used to investigate the self-similar nature of the clouds and examine whether the slope of the $\Delta$-variance in the self-similar regime varies from cloud to cloud and, in the case of simulations, whether it depends on the properties of the turbulent motions that are generated in the clouds. However, it has already been demonstrated that the method is capable of detecting break points. Ossenkopf \& Mac Low (2002) found, when applying the method to numerical models of molecular clouds where turbulence is driven on various physical scales, that the $\Delta$-variance departs from the self-similar regime on physical scales where turbulence is injected into the clouds. Using extinction maps, Schneider et al. (2011) found that  low-mass star-forming clouds have a double-peak structure in the $\Delta$-variance with characteristic size scales around $\approx 1$ pc and $\approx 4$ pc. They propose that the physical process governing structure formation could be the scale at which either a large-scale supernova shock or an expanding \ion{H}{ii} region sweeping through the diffuse medium are broken at dense clouds, which turns the well-ordered velocity into turbulence. 

\section{Spatial structure of Cygnus-X North and Polaris}\label{obsresults}

We applied the $\Delta$-variance method to the column density maps of Cygnus-X North and Polaris. As stated above, these two regions were selected because they are significantly different, both in terms of their column density distribution (i.e., Fig.~\ref{fig2}) and their star formation activity. While Polaris harbors a population of starless cores, it is still a region with no ongoing star formation and a modest contrast in column density. On the other hand, the Cygnus-X North cloud is a region with a much higher star formation rate and a much larger contrast in column densities (see Fig.~\ref{fig2}, also Hennemann et al. 2012; Schneider et al. 2016). The $\Delta$-variance spectra for both clouds are displayed in Fig.~\ref{fig3}. The $\Delta$-variance spectrum of Polaris displays a self-similar behavior above the resolution limit, and this self-similarity extends for more than one order of magnitude in spatial scales (from $\approx 0.03$ pc to $\approx 0.6$ pc). A power-law fit to the $\Delta$-variance of Polaris in the range [0.035-0.6] pc yields a value of the power-law exponent of $\alpha=0.4\pm0.003$, and this implies a value of $\beta=2.4$. On scales larger than $0.6$ pc, the self-similarity is perturbed, possibly due to the existence of a large filamentary structure (i.e., the MCLD 123.5+24.9 structure), though substructured, in the region. The scale-free nature of the $\Delta$-variance spectrum of Polaris is consistent with earlier findings using the $\Delta$-variance technique for the same cloud (Bensch et al. 2001; Ossenkopf-Okada \& Stepanov 2019). However, Bensch et al. (2001) found larger values of $\beta$  ($\approx 3$ from observations in the $^{12}$CO ($J=2-1$) line and $\beta \approx 3.2$ from observations in the $^{13}$CO $(J=1-0)$ line) when the $\Delta$-variance spectrum is fitted over a spatial range that is roughly similar to the one used in this study. The spatial resolution of the observations they used are $2.2\arcmin$ and $0.78\arcmin$, respectively, and are lower than the resolution of the observations presented in this work ($\approx 0.3\arcmin$). Ossenkopf (2002) showed that the use of low-$J$ CO isotopologues leads to somewhat steeper $\Delta$-variance spectra than the one corresponding to the underlying column density structure. The exact relative effect of the lower spatial resolution, which effectively smoothes the map and possibly increases the values of $\beta$, compared to the role of the optical depths of these molecular tracers in steepening the $\Delta$-variance spectrum, is not yet entirely clear.

 In contrast to Polaris, the $\Delta$-variance of Cygnus-X North displays a more complex shape with a steep slope above the spatial resolution limit (dashed black lines in Fig.~\ref{fig3}) and a broad peak at around $\approx 0.5-0.12$ pc. A reasonable assumption to make is that the existence of many small-scale dense structures (e.g., cores, clumps, and filaments) in Cygnus-X North alters the underlying (i.e., primordial) self-similar structure of the gas that had existed before these structures formed. However, it remains an open question whether a massive star-forming region such as Cygnus-X North had, at an earlier stage, a spatial distribution of (column) density similar to that of Polaris. This is a plausible assumption given that, prior to the formation of massive stars in the region, turbulence in the Cygnus-X North cloud, like in Polaris and elsewhere in the ISM, must have been dominated by shearing motions. Large-scale converging flows may be responsible for aggregating gas in specific regions that would be the parental structures of ridges and hubs. Compressive motions due to feedback from massive stars in specific regions of Cygnus-X North can also modify the spatial distribution of the (column) density field. However, massive star formation is localized in Cygnus-X North and not distributed across the entire cloud (Beerer et al. 2010). We speculate here that the underlying, "primordial," structure in the Cygnus-X North cloud resembled that of Polaris and use this as a working hypothesis. In what follows, we focus our attention on the Cygnus-X North cloud and adopt the Polaris value of $\beta=2.4$ as the exponent of the underlying self-similar fBm structure in Cygnus-X. We explore, using synthetic data, if and how the addition of dense structures with specific properties on top of a cloud with a self-similar structure modifies the $\Delta$-variance spectrum.  

\section{Interpretation}\label{interpret}

While one of our aims is to understand the structure of the Cygnus-X North region as revealed by its $\Delta$-variance spectrum, a broader goal is to investigate how the existence of compact and dense structures (cores, clumps, and filaments) with diverse characteristics can alter the self-similar nature of a molecular cloud and modify the $\Delta$-variance spectrum. We first summarize some of the basic properties of fBm images that are known to possess a self-similar structure. In a second step, we include discrete structures with specific characteristics on top of an existing fBm and investigate how the inclusion of these structures impacts the shape of the $\Delta$-variance spectrum. Lastly, we investigate how the shape of the $\Delta$-variance is modified in the presence of an entire population of structures that are characterized by distribution functions of their sizes, elongations, and column density contrasts.

\subsection{Fractal Brownian motion maps}\label{fbm}

\begin{figure}
\centering
\includegraphics[width=0.47\columnwidth] {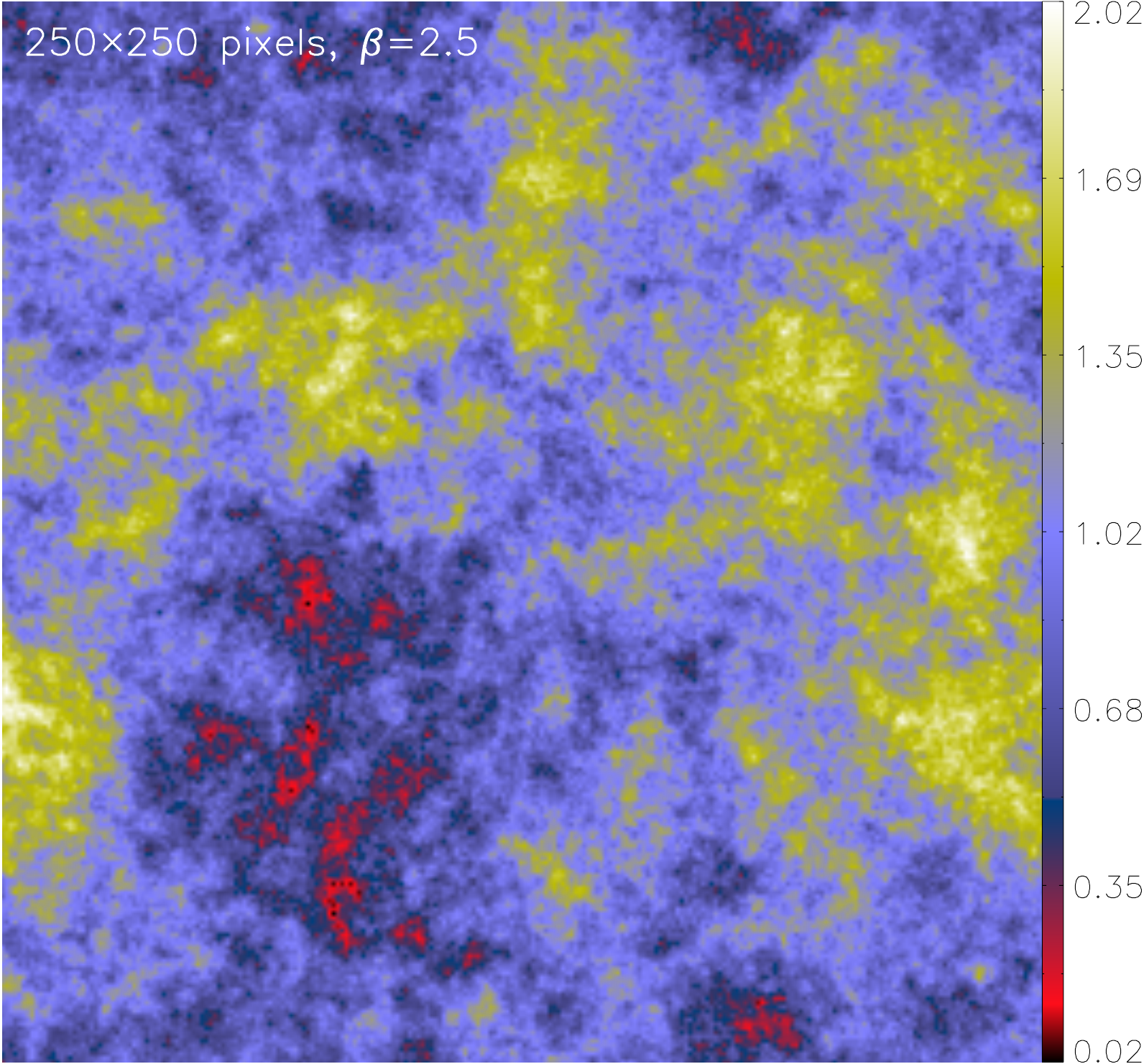}
\hspace{0.2cm}
\includegraphics[width=0.47\columnwidth] {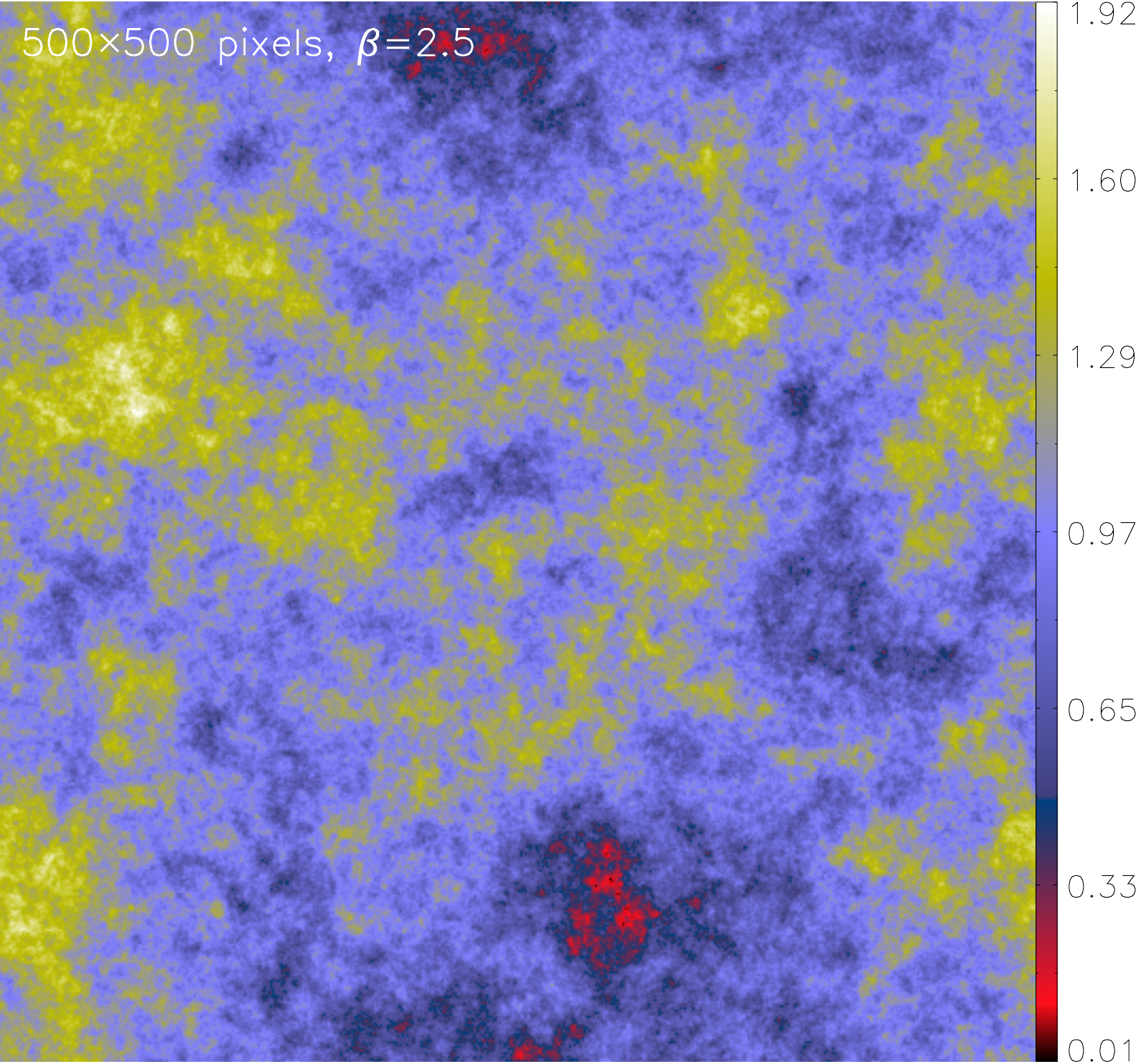}\\
\vspace{0.5cm}
\includegraphics[width=0.47\columnwidth] {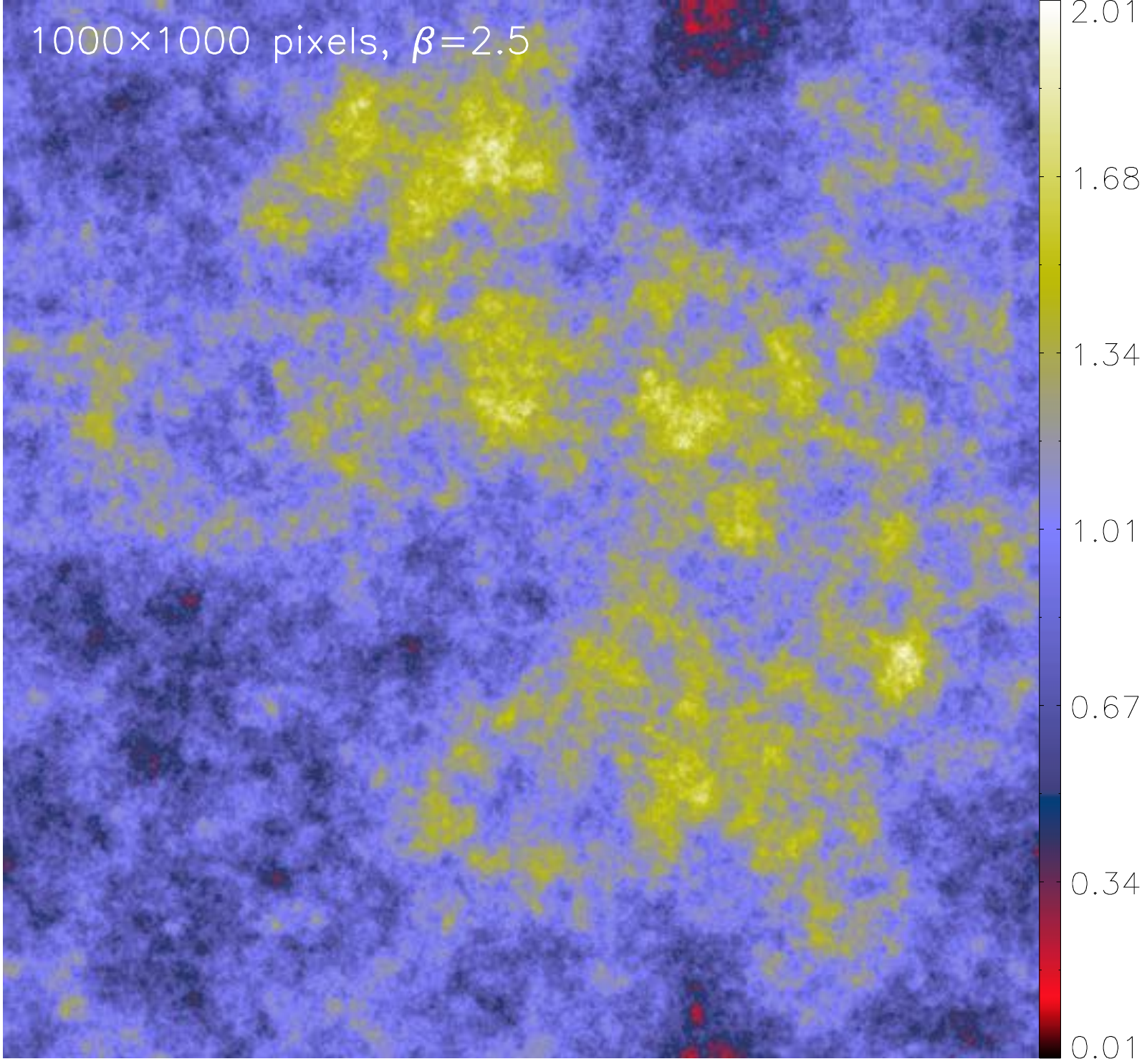}
\hspace{0.2cm}
\includegraphics[width=0.47\columnwidth] {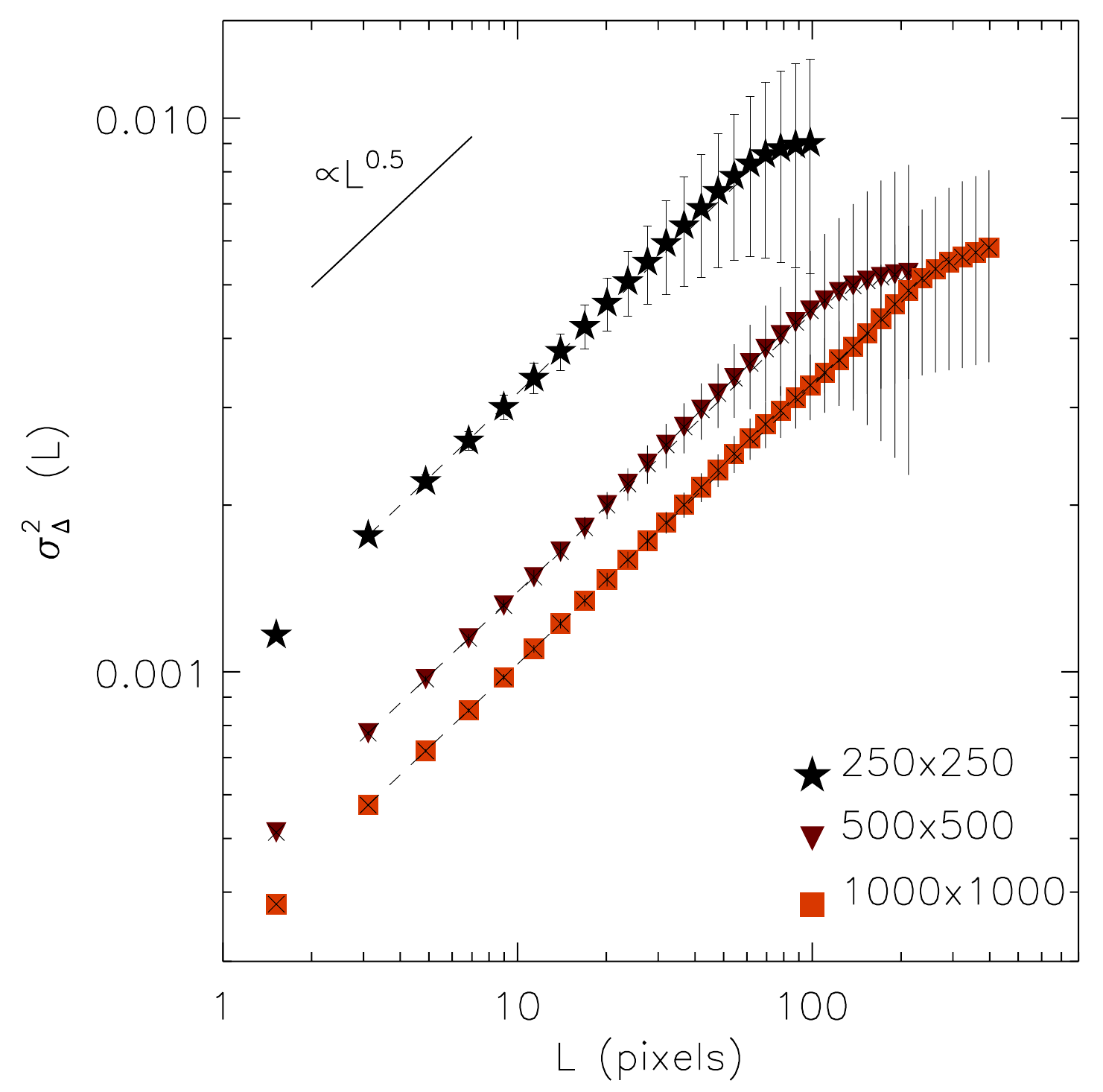}
\caption{fBm images with $\beta=2.5$ and with resolutions of $250\times250$ pixels (top left), $500\times500$ pixels (top right), and $1000\times1000$ pixels (bottom left). The $\Delta$-variance spectra for all three cases are compared in the bottom-right subpanel. All display a self-similar regime with an exponent of the power law of $\alpha=\beta-2$ (i.e., Eq.~\ref{eq4}). All maps are normalized by their own mean value, and the vertical offset between the three $\Delta$-variance functions simply reflects the effect of this different normalization.}
\label{fig4}
\end{figure}

Fractal Brownian motion images (Peitgen \& Saupe 1988) are often used as a surrogate of ISM maps thanks to their visual similarity with cloud features (e.g., Stutzki et al. 1998; Bensch et al. 2001; Miville-Desch\^{e}nes et al. 2003; Elia et al. 2014; Elia et al. 2018). A full description of their analytic properties is presented in Stutzki et al. (1998). Here, we simply review their basic properties. Firstly, their radially averaged power spectrum exhibits a power-law behavior with an exponent $\beta=E+2 H$, where $E$ is the Euclidian dimension (E=2 for 2D images) and $H$ is the Hurst exponent whose value ranges from $0$ to $1$. For 2D maps, $\beta$ can take values between $2$ and $4$. Secondly, the distribution of the phases of their Fourier transform is completely random. Thus, it is possible to generate fBm maps by defining the value of $\beta$ and a random phase distribution. If expressed in terms of the fractal dimension, the fractal dimension of an fBm image has been shown to be given by $D=E+1-H$, and this leads to a direct relation between $D$ and $\beta$ that is given by:

\begin{equation} 
D=\frac{3E+2-\beta}{2}.
\label{eq5}
\end{equation}

Stutzki et al. (1998) showed that the power spectrum of the $(E-1)$ projection of an E-dimensional fBm is also a power law with the same spectral index (i.e., the same $\beta$). Using this property, it is possible to establish the link between the 2D ($E=2$) and 3D ($E=3$) fractal dimensions. This will be given by:
 
 \begin{equation}
 D_{2}=D_{3}-\frac{3}{2}.
 \label{eq6}
 \end{equation} 

\begin{figure}
\centering
\includegraphics[width=0.47\columnwidth] {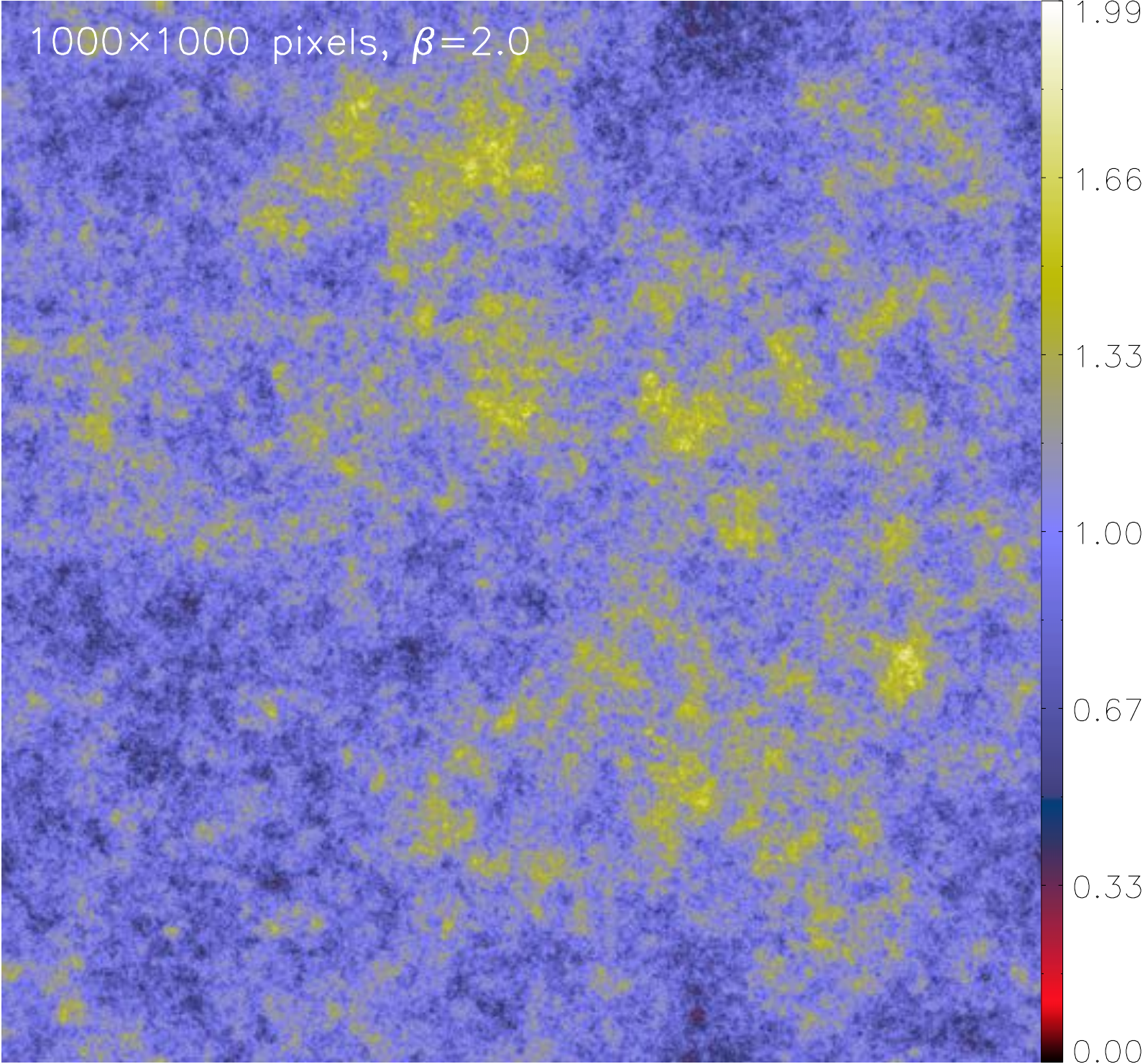}
\hspace{0.2cm}
\includegraphics[width=0.47\columnwidth] {f9.pdf}\\
\vspace{0.5cm}
\includegraphics[width=0.47\columnwidth] {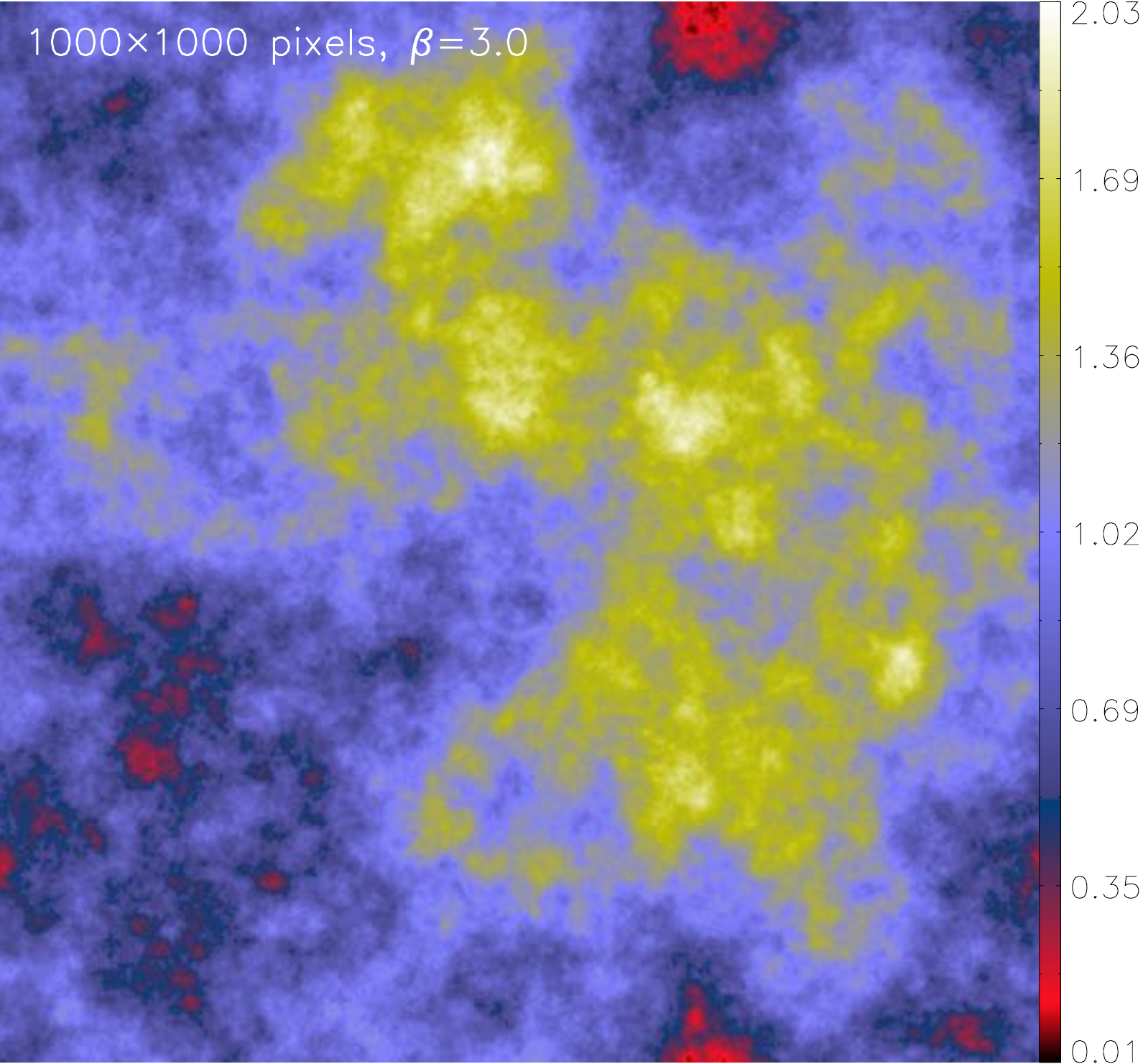} 
\hspace{0.2cm}
\includegraphics[width=0.47\columnwidth] {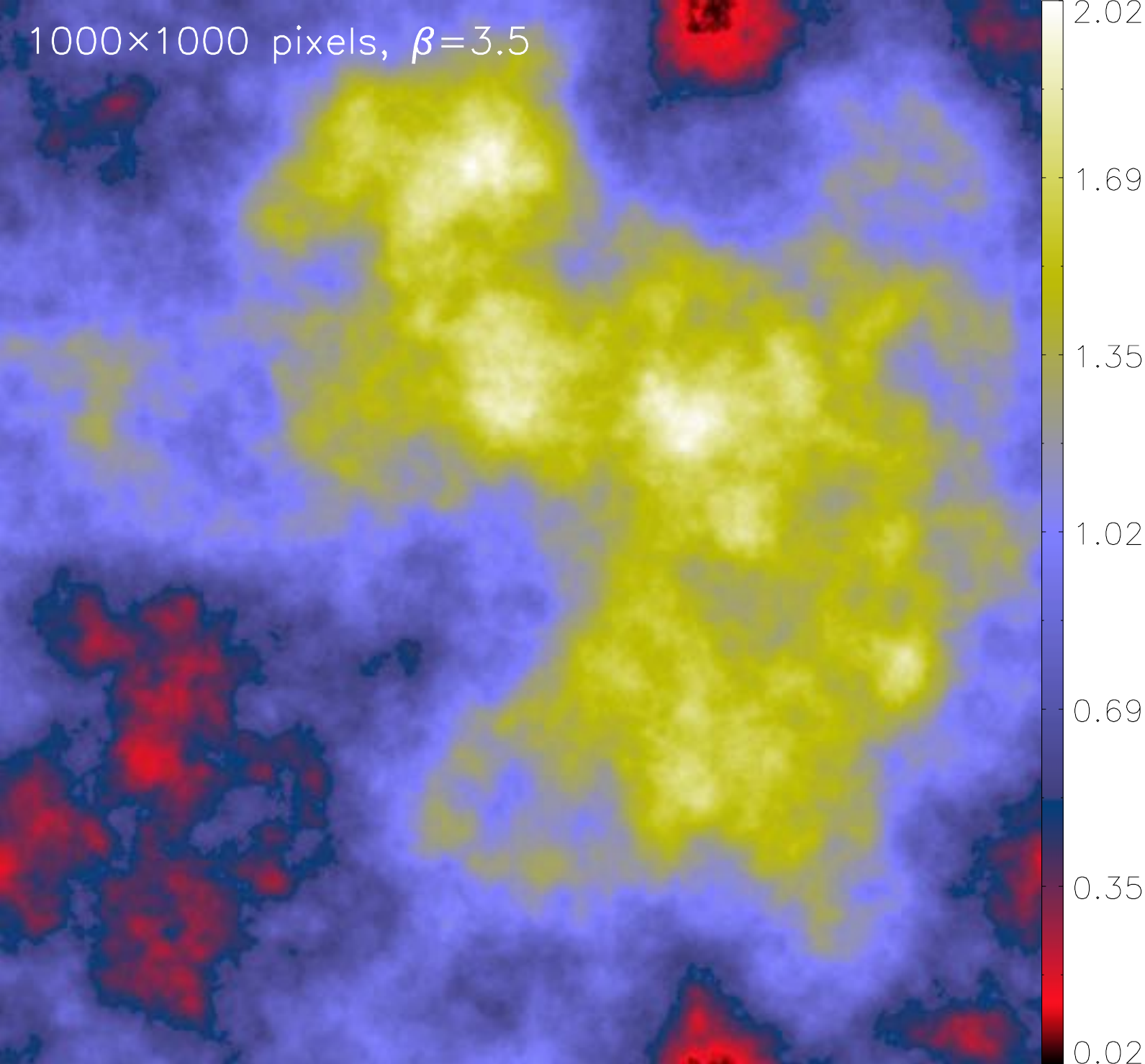}\\
\vspace{0.5cm}
\includegraphics[width=0.47\columnwidth] {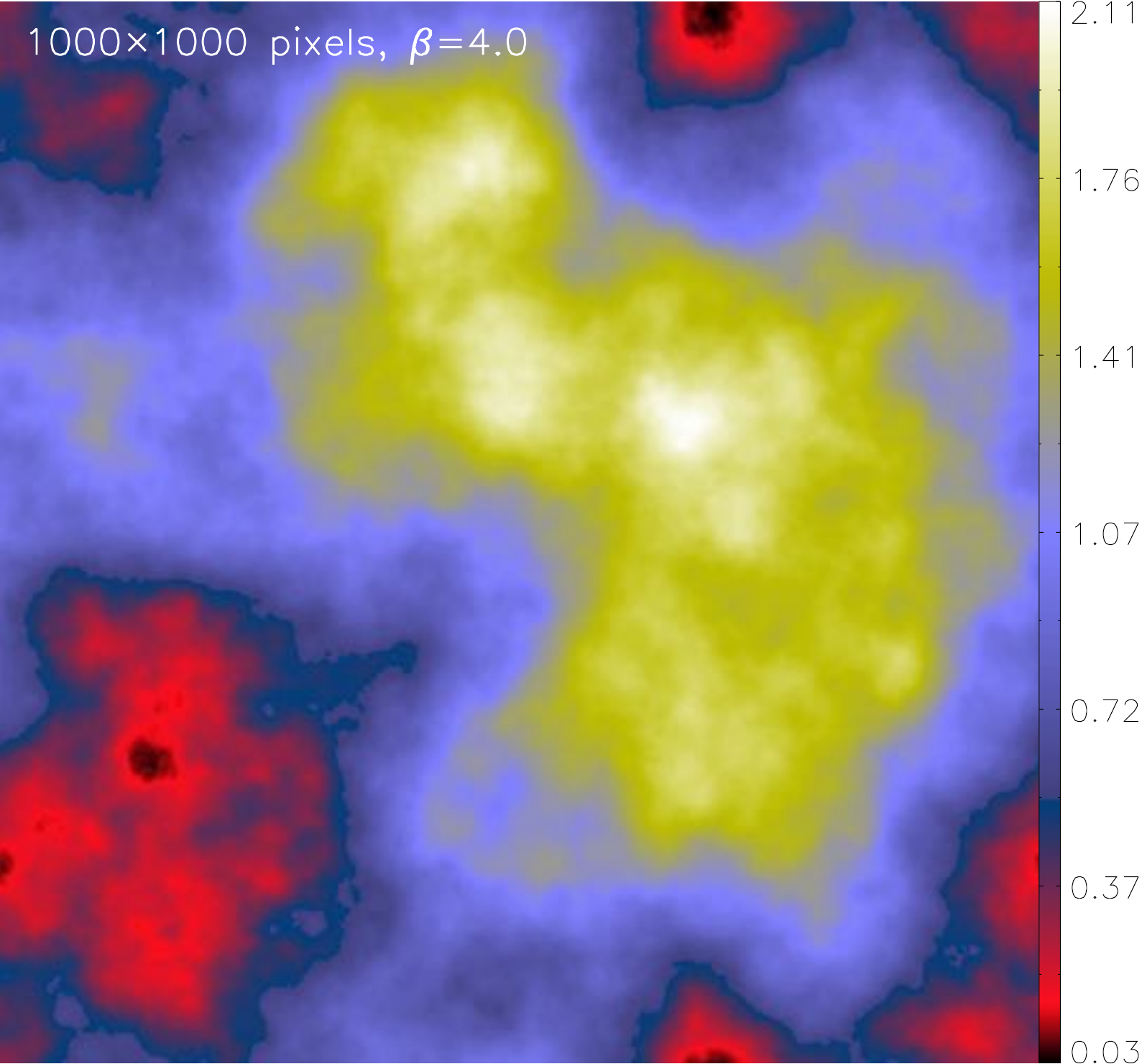}
\hspace{0.2cm} 
\includegraphics[width=0.47\columnwidth] {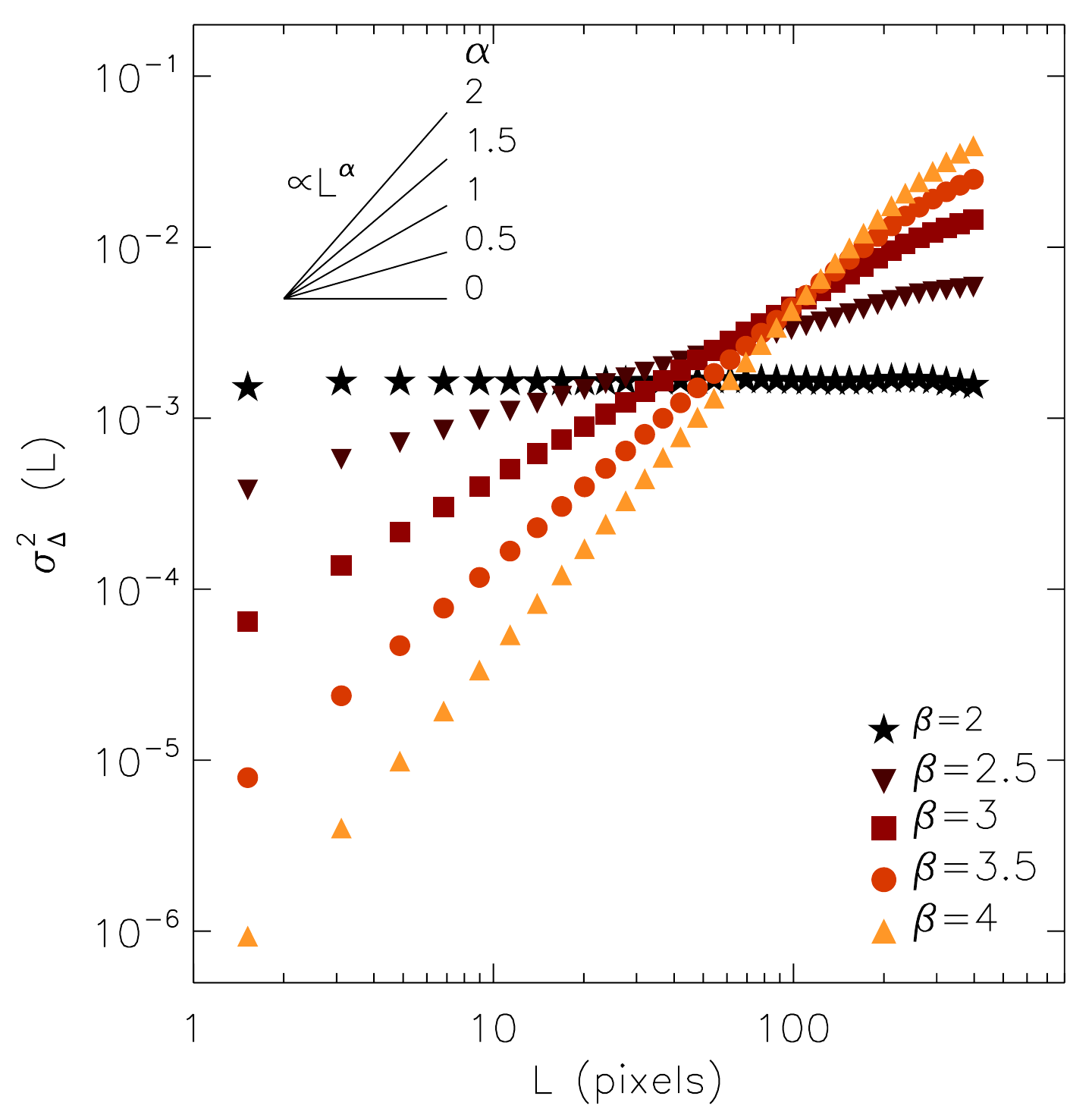}
\caption{fBm images with values of $\beta$ ranging from $2$ to $4$ in steps of $0.5$. All maps have a resolution of $1000\times1000$ pixels. The $\Delta$-variance figures for all cases are compared in the bottom-right subpanel. All display a self-similar regime with an exponent of the power law of $\alpha=\beta-2$ (i.e., Eq.~\ref{eq4}). All maps are normalized by their own mean value. The random number series generating the phases has been kept the same in all maps.}
\label{fig5}
\end{figure}

In this paper, we used fBm images as a reference for self-similar structures since they can be obtained with preconditioned statistical properties (e.g., Stutzki et al. 1998; Shadmehri \& Elmegreen 2011; Elia et al. 2018). Figure~\ref{fig4} displays three fBm images generated with a value of $\beta=2.5$ and for three resolutions: $250\times250$ pixels (top left), $500\times500$ pixels (top right), and $1000\times1000$ pixels (bottom left)\footnote{By construction, the mean value of the fBm is zero. We applied an arbitrary offset to the maps in order to insure that all values were positive. The maps were then normalized by their mean value. The addition of a constant offset for the whole map does not alter the shape of the $\Delta$-variance spectrum since the relative differences between pixels remain the same.}. The bottom-right panel in Fig.~\ref{fig4} displays the $\Delta$-variance functions calculated for these three fBm images. The self-similar regime is observed in all cases and extends to larger spatial scales for cases with a higher spatial resolution. The fBm images are periodic, and, if we were using a periodic analysis, the $\Delta$-variance spectra would be perfect power laws. However, in order to compare them to the observational data, we performed the calculations of the $\Delta$-variance with a cut at the map boundaries. This cut has two effects. First, there is a natural limit to the size of any structure so that the $\Delta$-variance spectrum flattens at the largest scales. Second, the statistical significance of the structures close to the map boundaries is reduced. This changes the denominator in the normalization of the $\Delta$-variance when computing mean properties of the map according to the area-to-boundary ratio of maps of different sizes so that the absolute scale of the $\Delta$-variance is only comparable in the limit of very large maps. Figure~\ref{fig5} displays the same type of fBm images, but in this case all images have a fixed resolution of $1000\times1000$ pixels and the value of $\beta$ is varied between $2$ and $4$ (in steps of 0.5). While in Fig.~\ref{fig4} the phase distribution varies from image to image, in Fig~\ref{fig5} the same distribution of phases is kept, so that the basic "shape" of the image remains the same. Increasing the value of $\beta$ produces a gradual smoothing of the image due to the transfer of power from high to low spatial frequencies. The bottom-right panel in Fig.~\ref{fig5} displays the corresponding $\Delta$-variance functions for each of these cases. As expected, the $\Delta$-variance functions are scale-free power-law functions whose exponent is given by $\alpha=\beta-2$. 

\subsection{Fractal Brownian motion maps with additional structure}\label{fbmplus}

We now explore the effect of discrete structures on the $\Delta$-variance spectrum. The structures we superimpose on top of fBm images are generalized 2D Gaussian functions that are given by:

\begin{equation}
N_{G}(x,y)=N_{peak}{\rm exp}[-a(x-x_{0})^{2}+2b(x-x_{0})(y-y_{0})+c(y-y_{0})^{2})],
\label{eq7}
\end{equation}

\noindent where $N_{G}$ is the added column density of the 2D Gaussian structure, $N_{peak}$ is its peak value, $x_{0}$ and $y_{0}$ are the coordinates of the peak, and the terms $a$, $b$, and $c$ are given by: 

\begin{align}
a=& \frac{\cos^{2}\left(\theta\right)} {2\sigma_{1}^{2}}+\frac{\sin^{2}\left(\theta\right)}{2\sigma_{2}^{2}} \nonumber\\
b=& \frac{\sin(2\theta)}{4\sigma_{1}^{2}} -\frac{\sin(2\theta)}{4\sigma_{2}^{2}} \nonumber\\
c= &\frac{\sin^{2}\left(\theta\right)}{2\sigma_{1}^{2}}+\frac{\cos^{2}\left(\theta\right)}{2\sigma_{2}^{2}}.
\label{eq8}
\end{align}

In Eq.~\ref{eq8}, $\sigma_{1}$ and $\sigma_{2}$ are the standard deviations of the 2D Gaussian along the major and minor axes, respectively, and $\theta$ is the angle between the Gaussian function major axis and the {\it x}-axis, defined in the counterclockwise direction. We generated a very large number of synthetic maps on which we superimposed one or several 2D Gaussian structures over fBm images in a controlled manner.  All fBm images have a value of $\beta=2.4$, similar to the value found in the Polaris cloud, and a resolution of $1000\times1000$ pixels. For individual structures, we varied the aspect ratio of the 2D Gaussian, $f=\left(\sigma_{1}/\sigma_{2}\right)$, over a range of 1 to 10. The peak value of the 2D Gaussians is expressed in terms of the mean value of the fBm image, $N_{peak}=\delta_{c} \left<N_{\rm fBm} \right>,$ and the column density contrast between the peak of the 2D Gaussian and the mean value of the fBm, $\delta_{c}$, is varied between 1 and 10. We also explored the effect of varying the absolute size of the Gaussian function with respect to the image size, as well as the effect of including multiple Gaussian functions in the fBm images. 

\begin{figure}
\centering
\includegraphics[width=0.47\columnwidth] {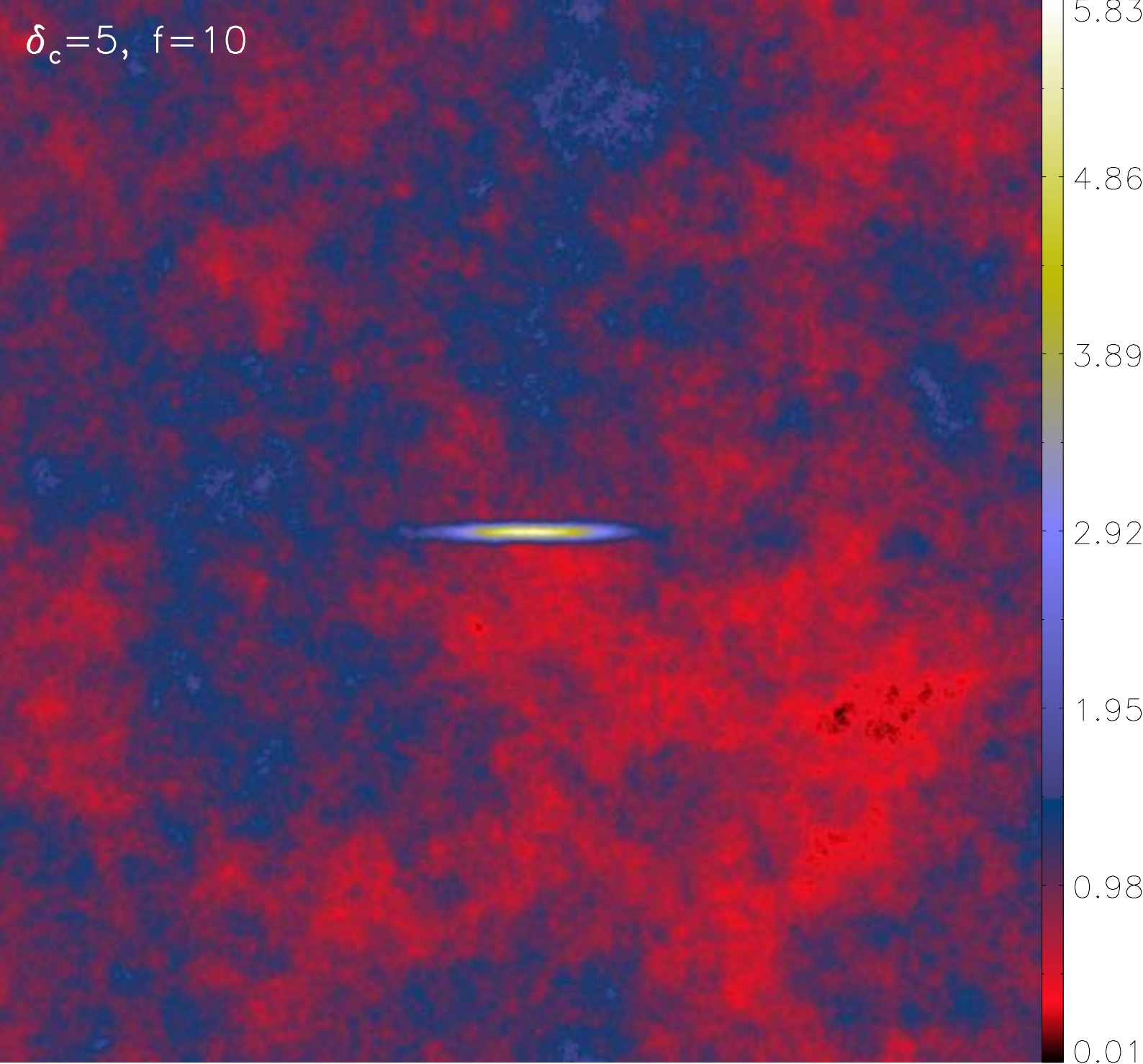}
\hspace{0.2cm}
\includegraphics[width=0.47\columnwidth] {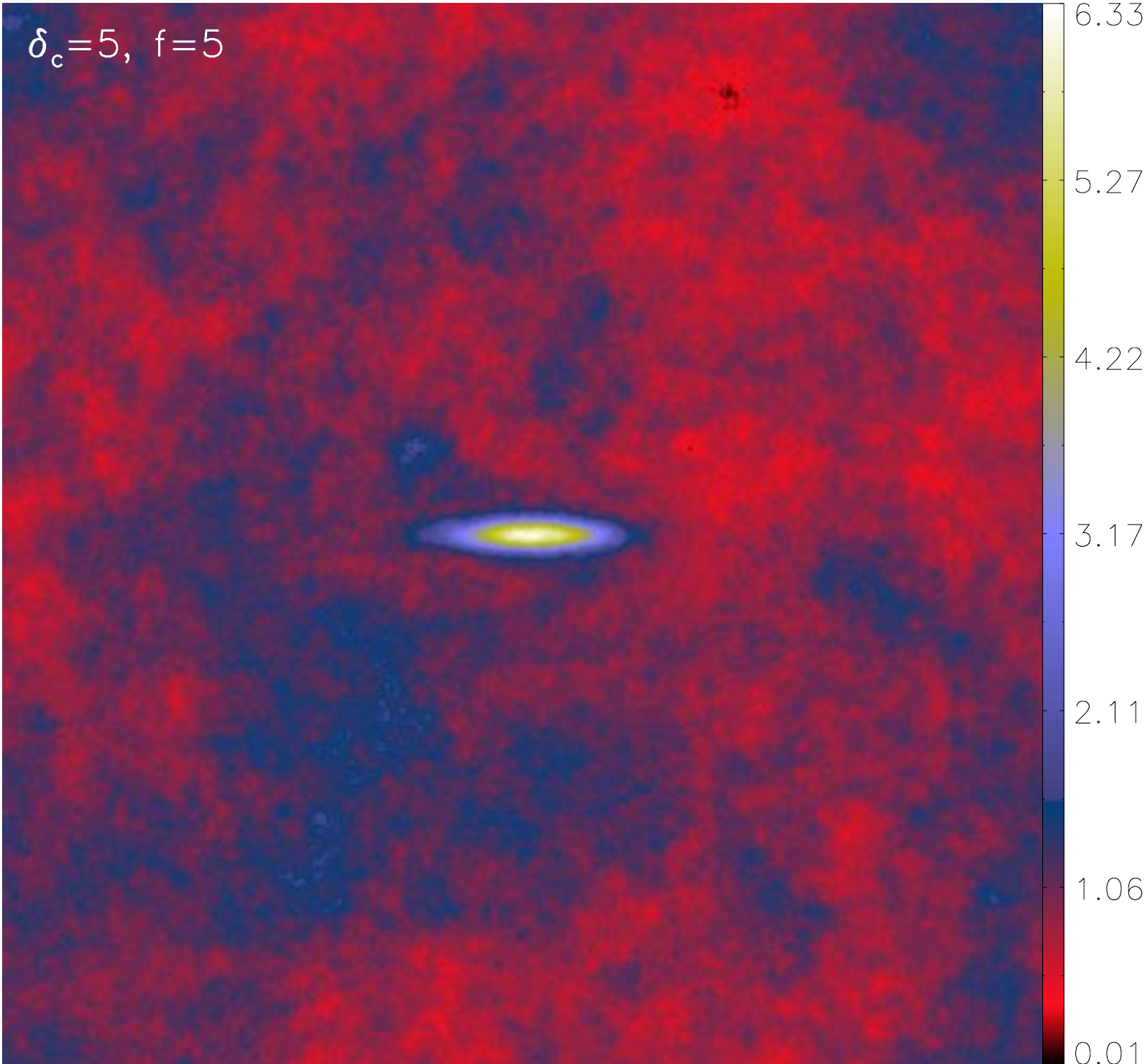}\\
\vspace{0.5cm}
\includegraphics[width=0.47\columnwidth] {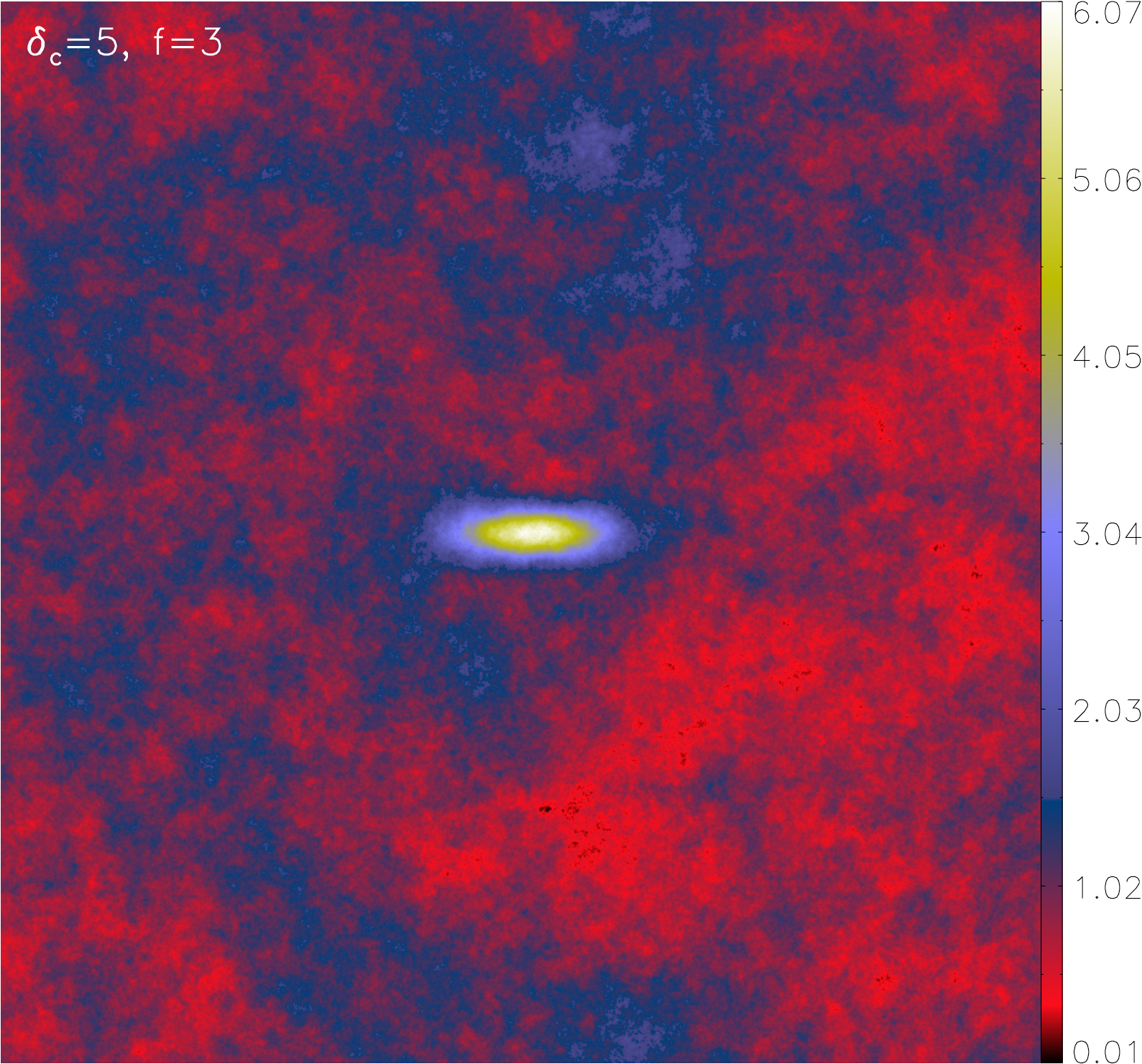}
\hspace{0.2cm}
\includegraphics[width=0.47\columnwidth] {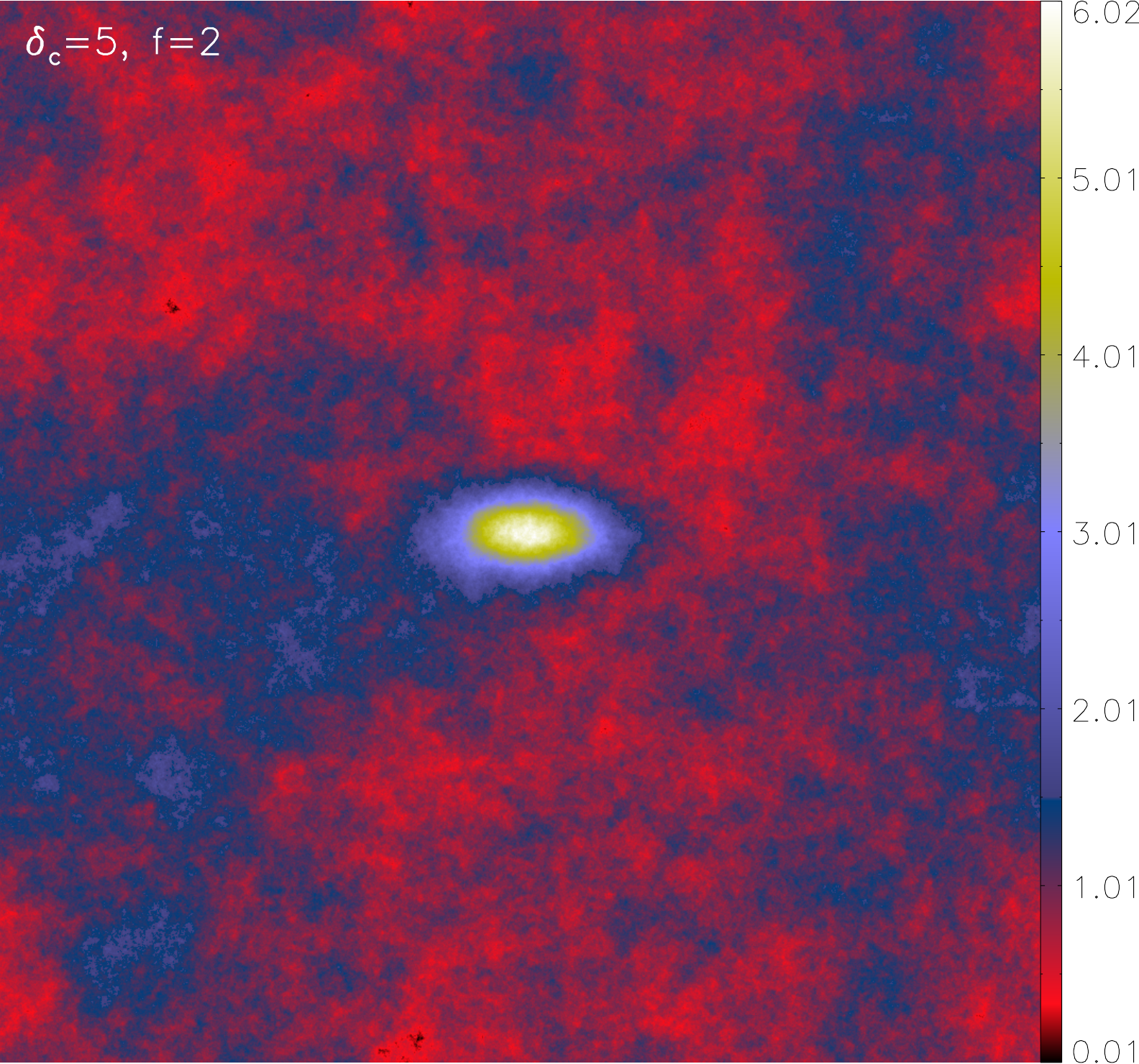}\\
\vspace{0.5cm}
\includegraphics[width=0.47\columnwidth] {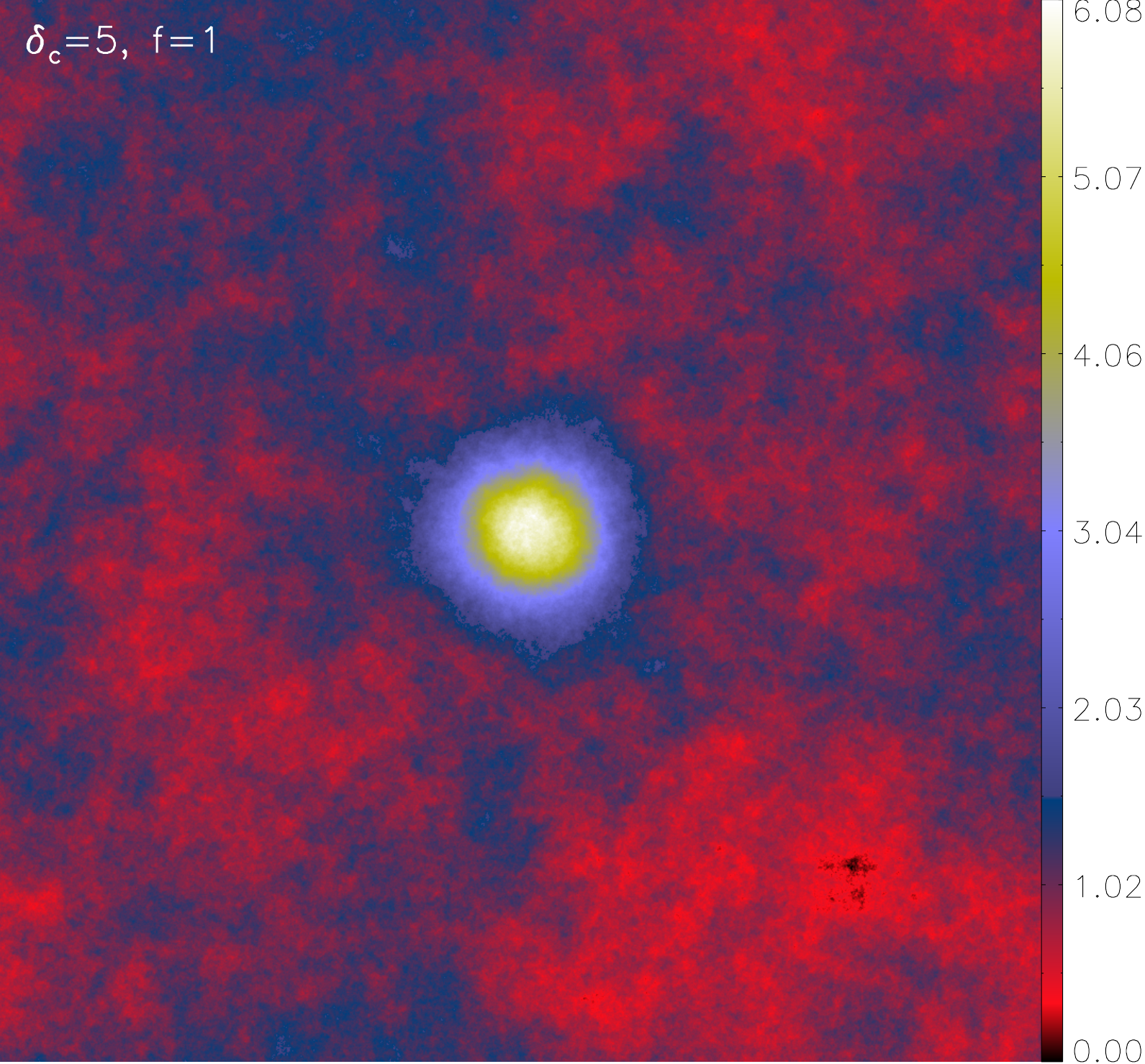}
\hspace{0.2cm} 
\includegraphics[width=0.47\columnwidth] {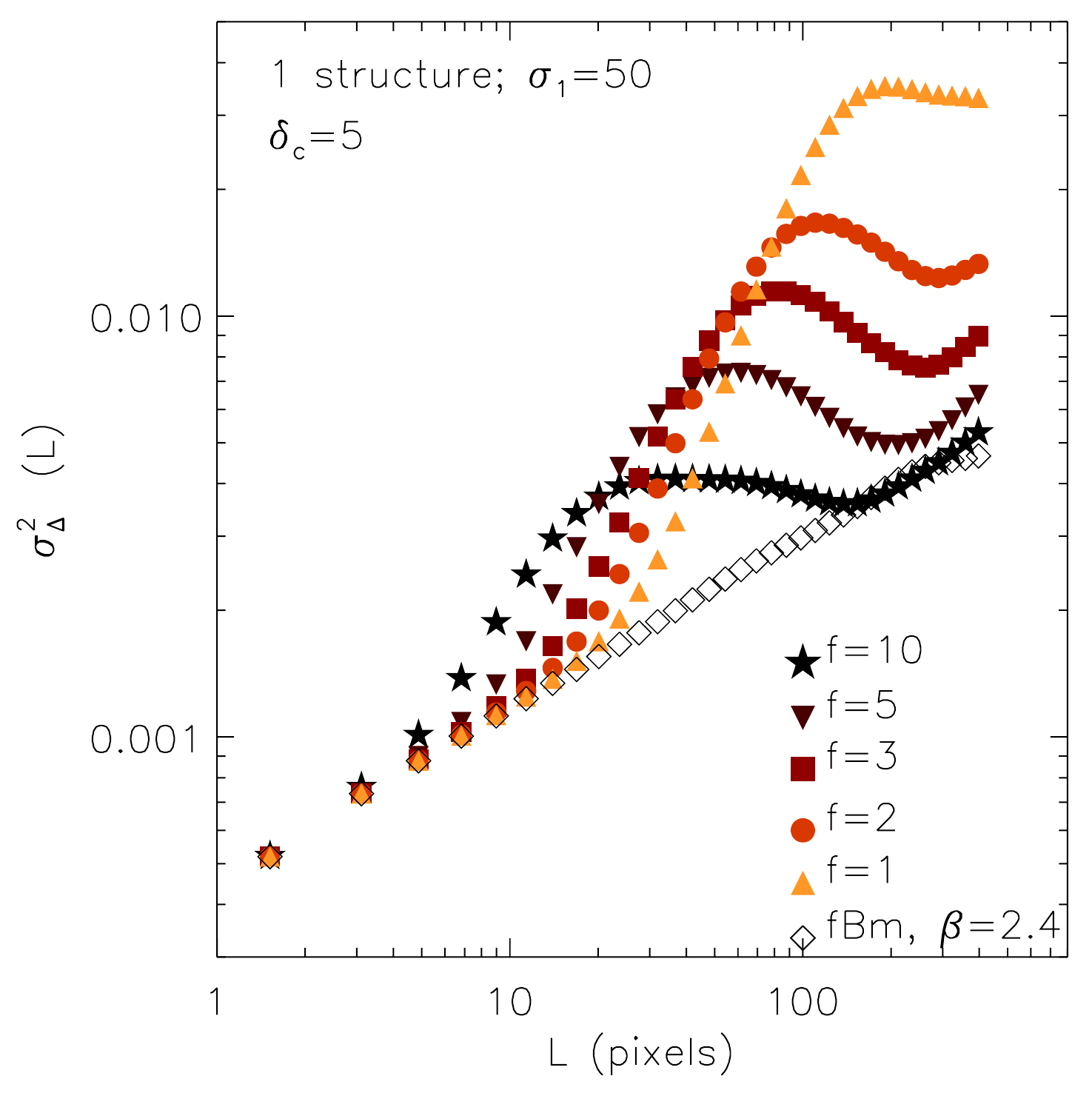}
\caption{2D Gaussian structures injected on top of an fBm image with $\beta=2.4$. The 2D Gaussian functions have an aspect ratio ($f=\sigma_{1}/\sigma_{2}$) that is varied in the range [1-10], and all have a value of $\delta_{c}$=5 and a fixed size of $\sigma_{1}=50$ pixels. All maps are normalized to their mean value. The bottom-right figure displays the corresponding $\Delta$-variance functions calculated for each case, and these are compared to the $\Delta$-variance function of the underlying fBm image.}
\label{fig6}
\end{figure}

Figure~\ref{fig6} displays five realizations of an fBm with superimposed Gaussian structures. The 2D Gaussians all have $\delta_{c}=5$, $\sigma_{1}=50$ pixels, and an aspect ratio $f=\left(\sigma_{1}/\sigma_{2}\right)$ that is varied between 1 and 10. The $\Delta$-variance functions of these maps are displayed in the bottom-right panel of Fig.~\ref{fig6} and are compared to the $\Delta$-variance function of a pure fBm image with $\beta=2.4$. The inclusion of an additional structure in the fBm image increases the value of $\sigma_{\Delta}^{2}$ on all spatial scales. The increment of the $\Delta$-variance function with respect to the $\Delta$-variance of the pure fBm reaches a maximum on a scale that is on the order of the equivalent diameter of the injected structure. Figure~\ref{fig6} shows that as the aspect ratio is reduced, the position of the point where the deviation from the fBm is maximized in the $\Delta$-variance function moves to larger spatial scales. If we approximate the surface of the 2D Gaussian by the area that lies within $[2\sigma_{1}, 2\sigma_{2}]$ (i.e., where most of the signal lies), the equivalent diameter is then given by $D_{eq} \approx 4 \sqrt{\sigma_{1}\sigma_{2}}$. The measured positions of the points of maximum deviation in the $\Delta$-variance functions in Fig.~\ref{fig6} do indeed confirm that the position of maximum deviation is well approximated by $D_{eq}$. A deviation from this value (by up to $\approx 30\%$) can be observed for smaller structures (i.e., in this case, the most elongated) as they are less well resolved on the grid. 

\begin{figure}
\centering
\includegraphics[width=0.47\columnwidth] {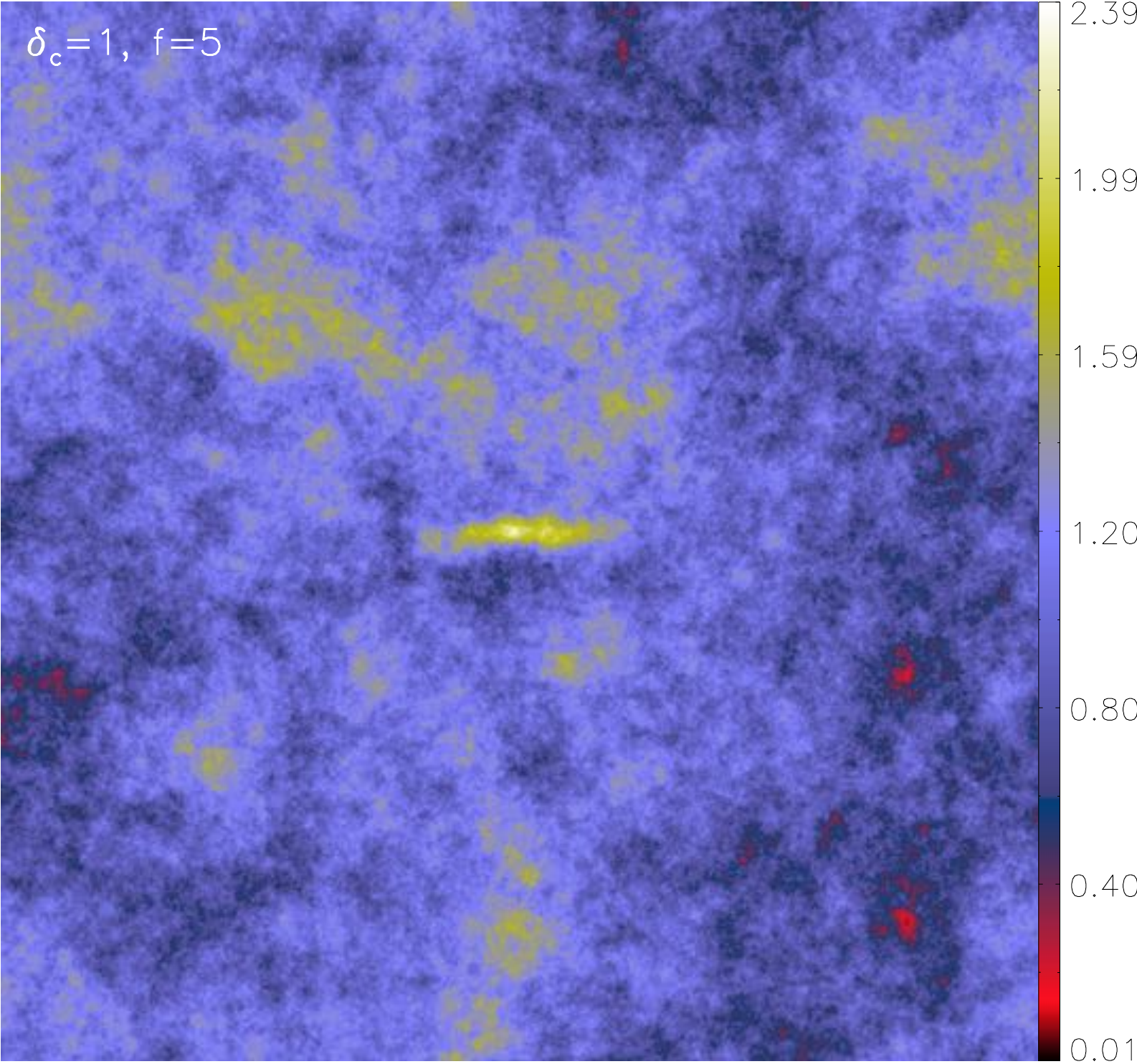}
\hspace{0.2cm}
\includegraphics[width=0.47\columnwidth] {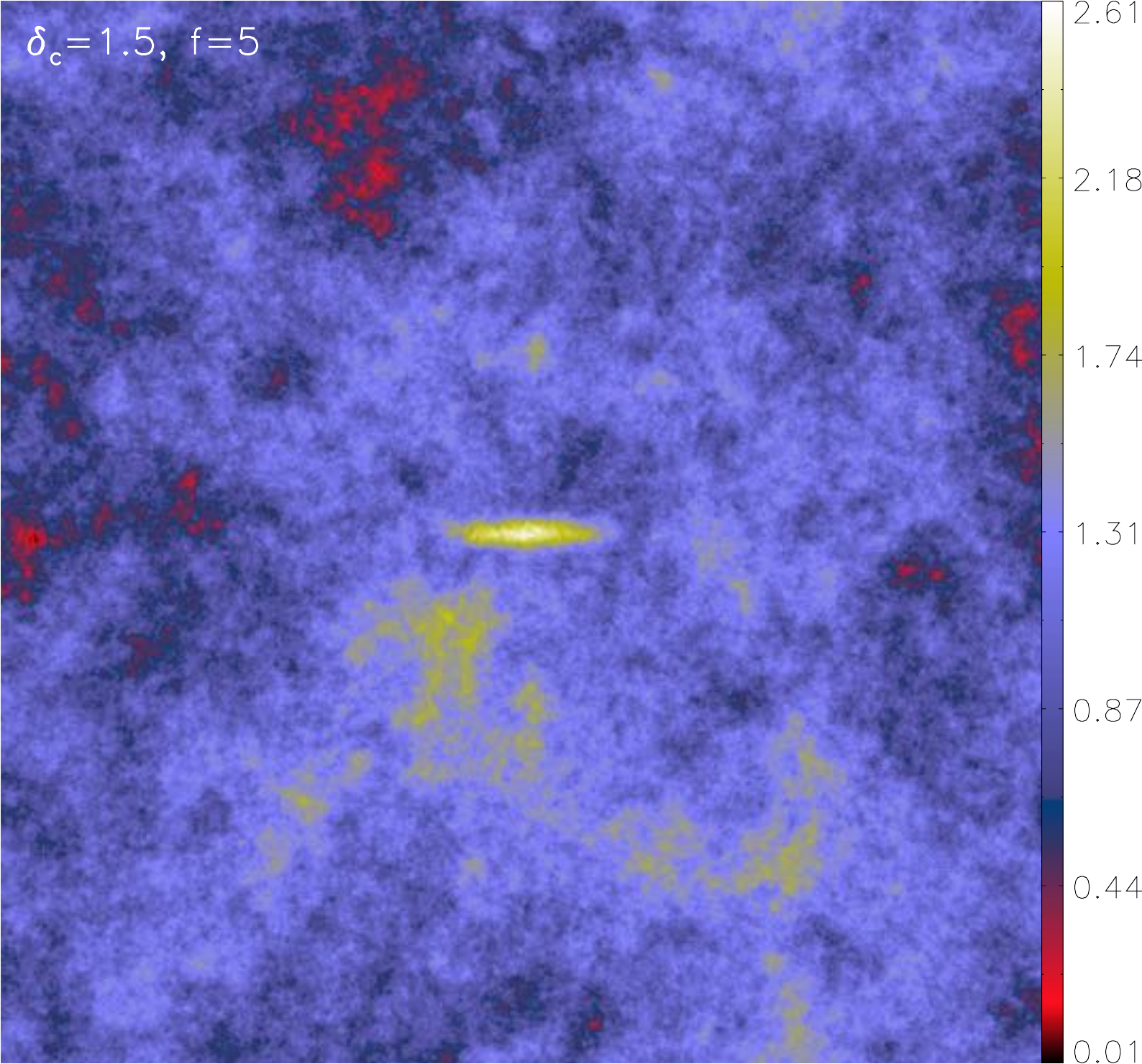}\\
\vspace{0.5cm}
\includegraphics[width=0.47\columnwidth] {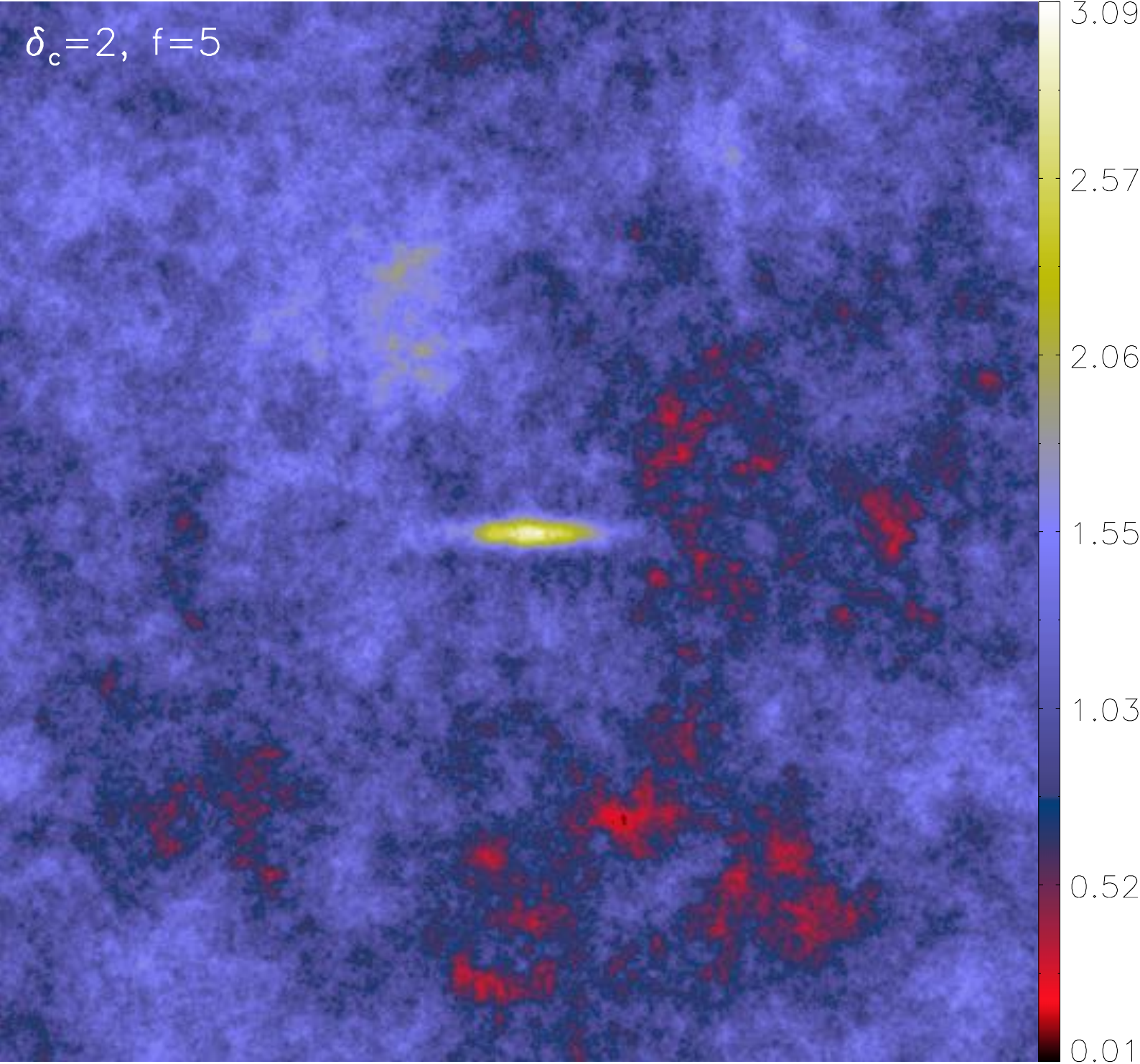} 
\hspace{0.2cm}
\includegraphics[width=0.47\columnwidth] {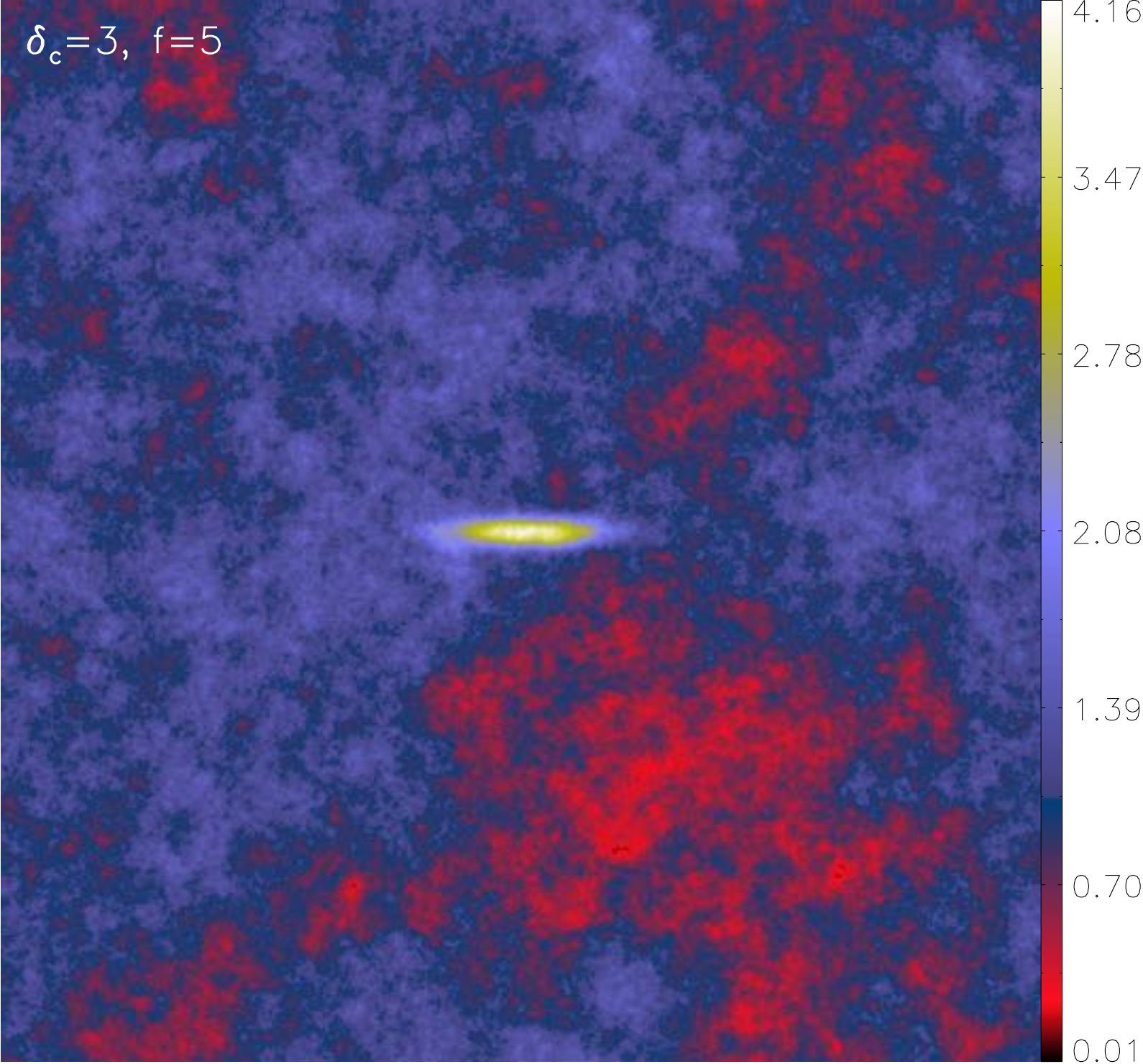}\\
\vspace{0.5cm}
\includegraphics[width=0.47\columnwidth] {f17.pdf}
\hspace{0.2cm} 
\includegraphics[width=0.47\columnwidth] {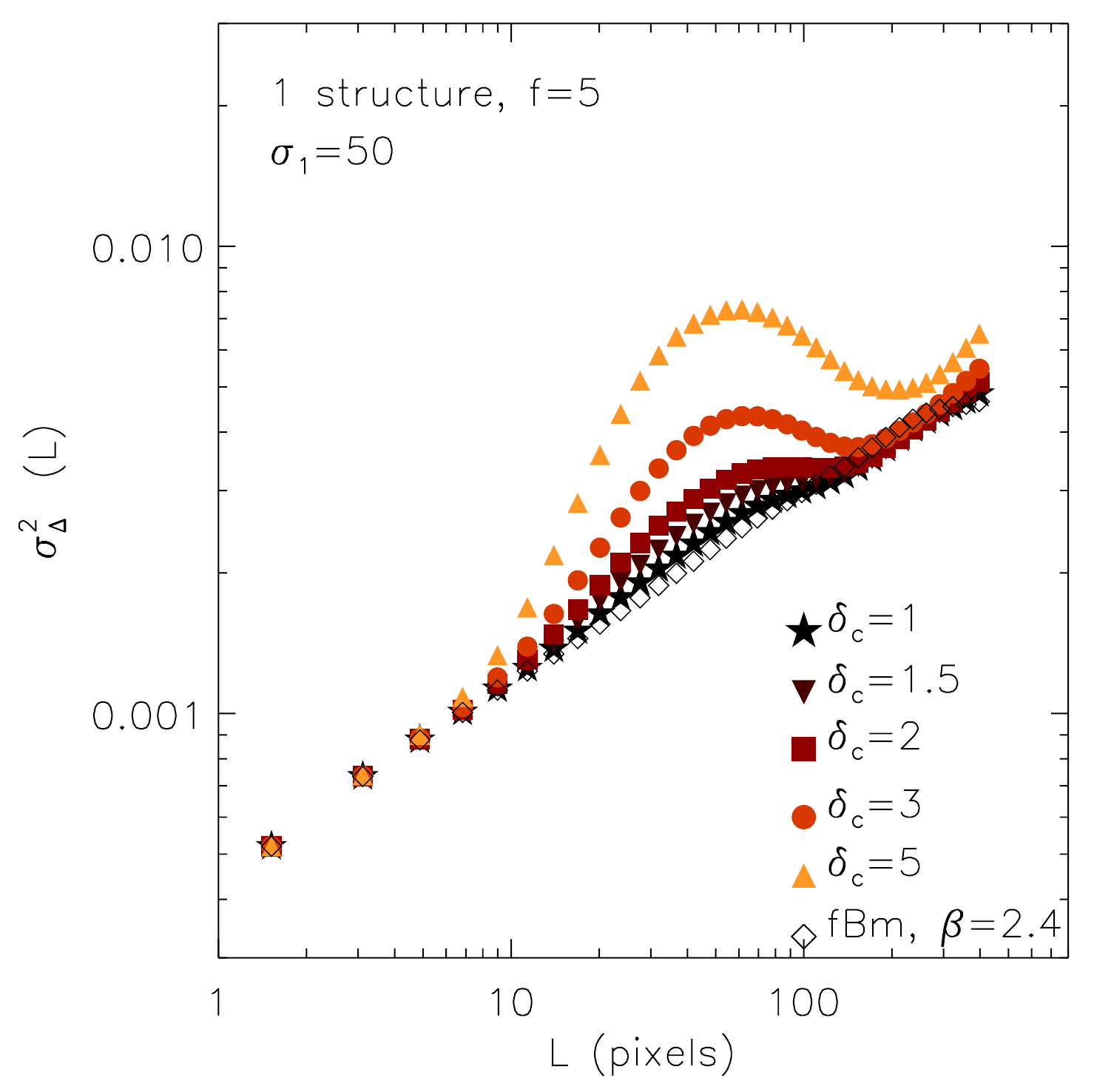}
\caption{2D Gaussian structures injected on top of an fBm image with $\beta=2.4$. The Gaussian functions have an aspect ratio $f=5$, a value of $\sigma_{1}=50$ pixels, and a column density contrast between the peak of the 2D Gaussian and the mean value of the fBm, $\delta_{c}$, that is varied between 1 and 5. All maps are normalized to their mean value. The bottom-right figure displays the corresponding $\Delta$-variance functions calculated for each case, and these are compared to the $\Delta$-variance function calculated for the underlying fBm image.}
\label{fig7}
\end{figure}

We explored the effect of varying the contrast between the injected 2D Gaussian structure and the underlying fBm image. Figure~\ref{fig7} displays five realizations where the value of $\delta_{c}=N_{peak}/\left<N_{\rm fBm}\right>$ is varied between 1 and 5. For all cases, the other parameters are fixed to $f=5$ and $\sigma_{1}=50$ pixels. All images have a resolution of $1000\times1000$ pixels, and the underlying fBm has an exponent of $\beta=2.4$. The lower-right panel in Fig.~\ref{fig7} displays the corresponding $\Delta$-variance functions, which, here once again, are compared to the $\Delta$-variance of the fBm image. Figure~\ref{fig7} shows that higher contrasts ($\delta_{c}$) between the self-similar fBm and the injected structure lead to higher values of the $\sigma_{\Delta}^{2}$ on spatial scales equal to $D_{eq}$.

\begin{figure}
\centering
\includegraphics[width=0.47\columnwidth] {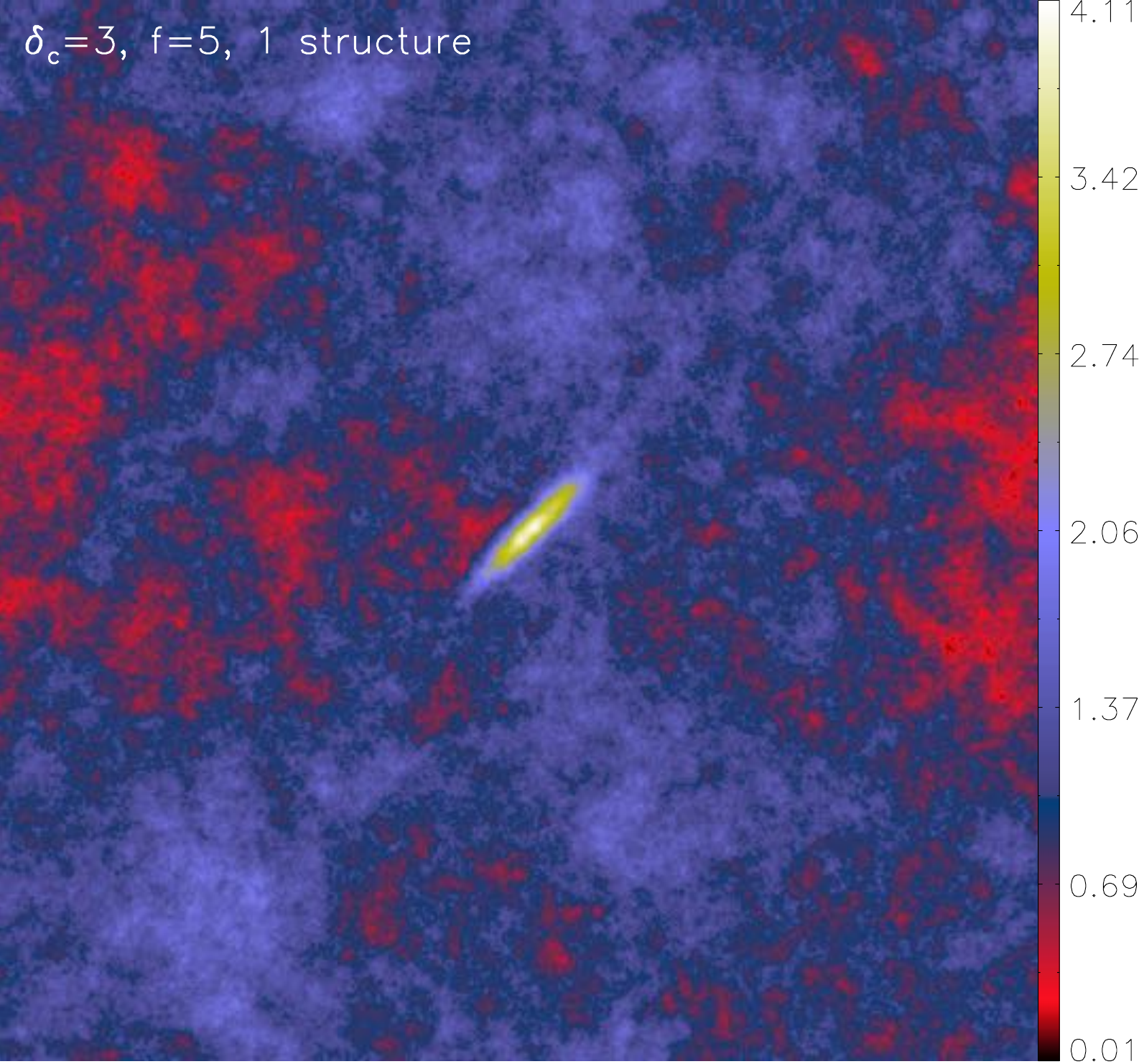}
\hspace{0.2cm}
\includegraphics[width=0.47\columnwidth] {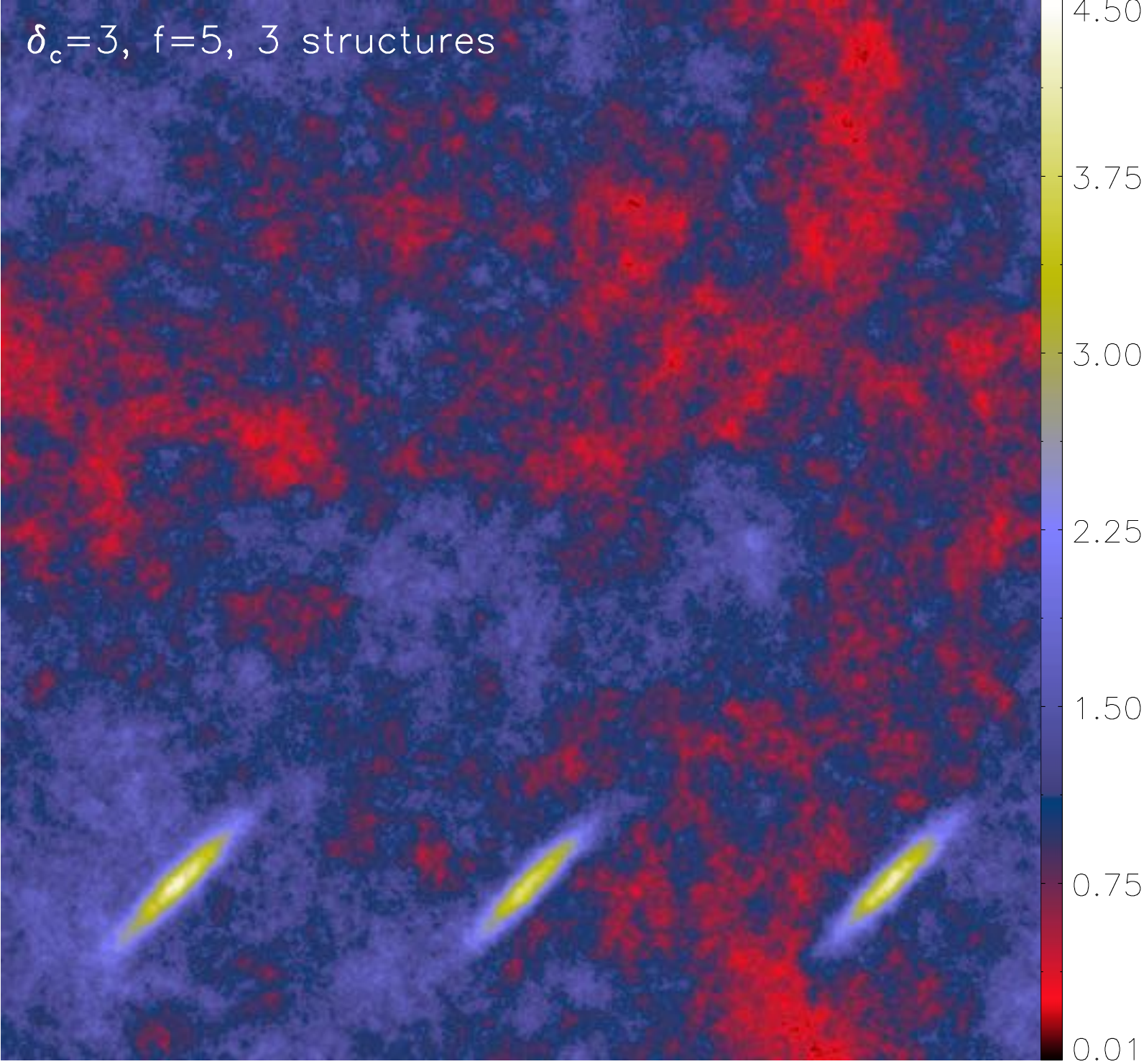}\\
\vspace{0.5cm}
\includegraphics[width=0.47\columnwidth] {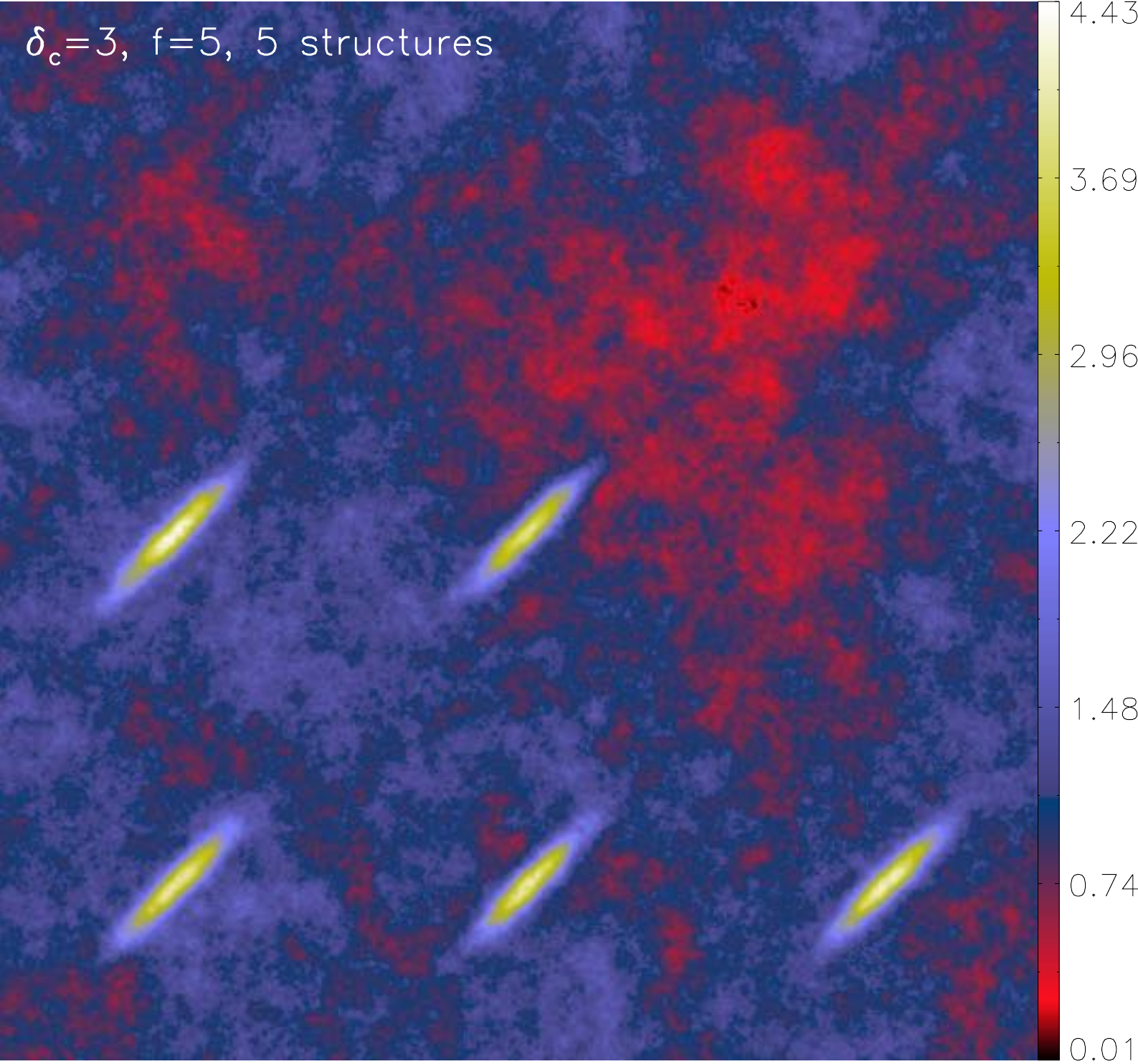}
\hspace{0.2cm}
\includegraphics[width=0.47\columnwidth] {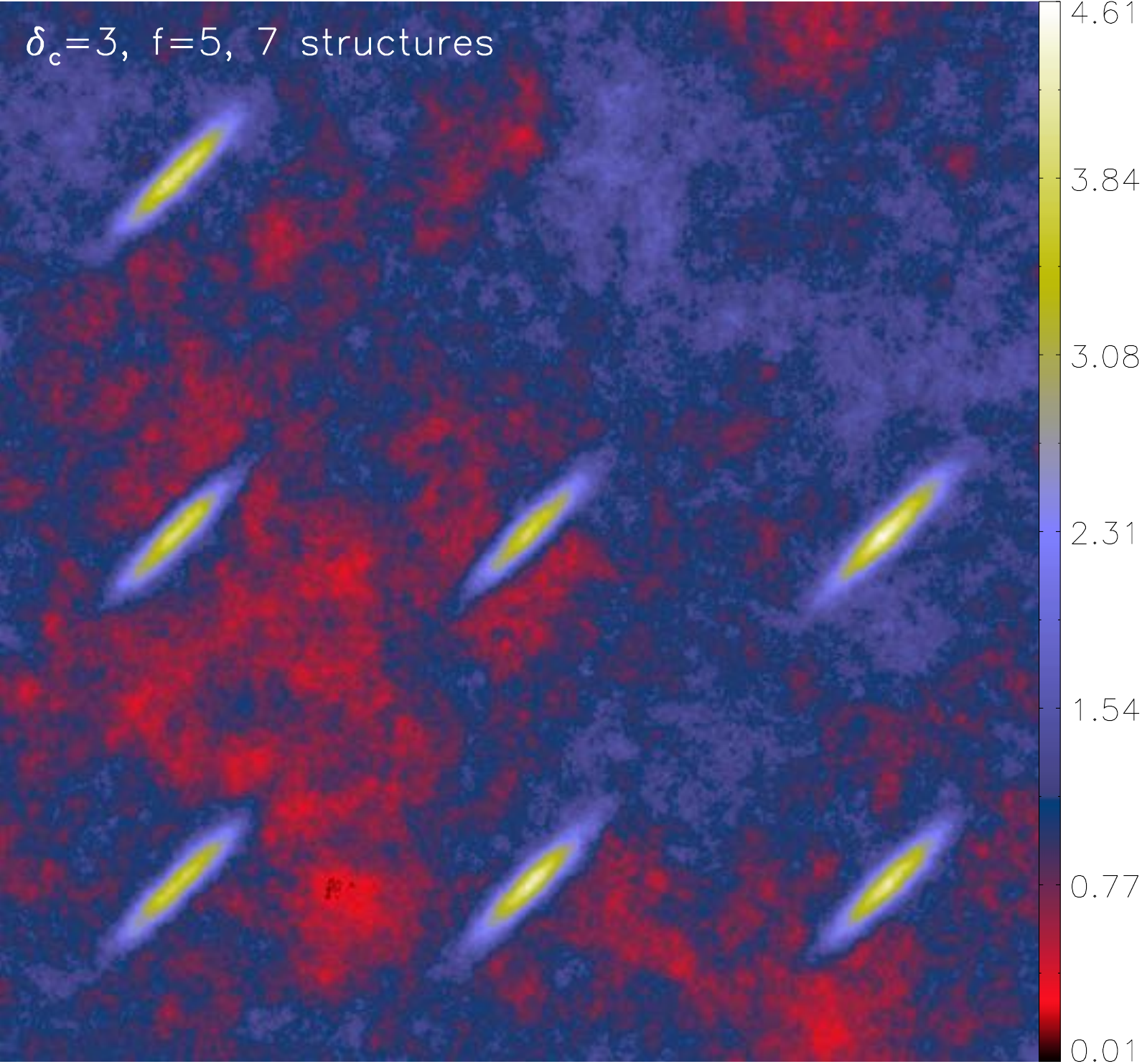}\\
\vspace{0.5cm}
\includegraphics[width=0.47\columnwidth] {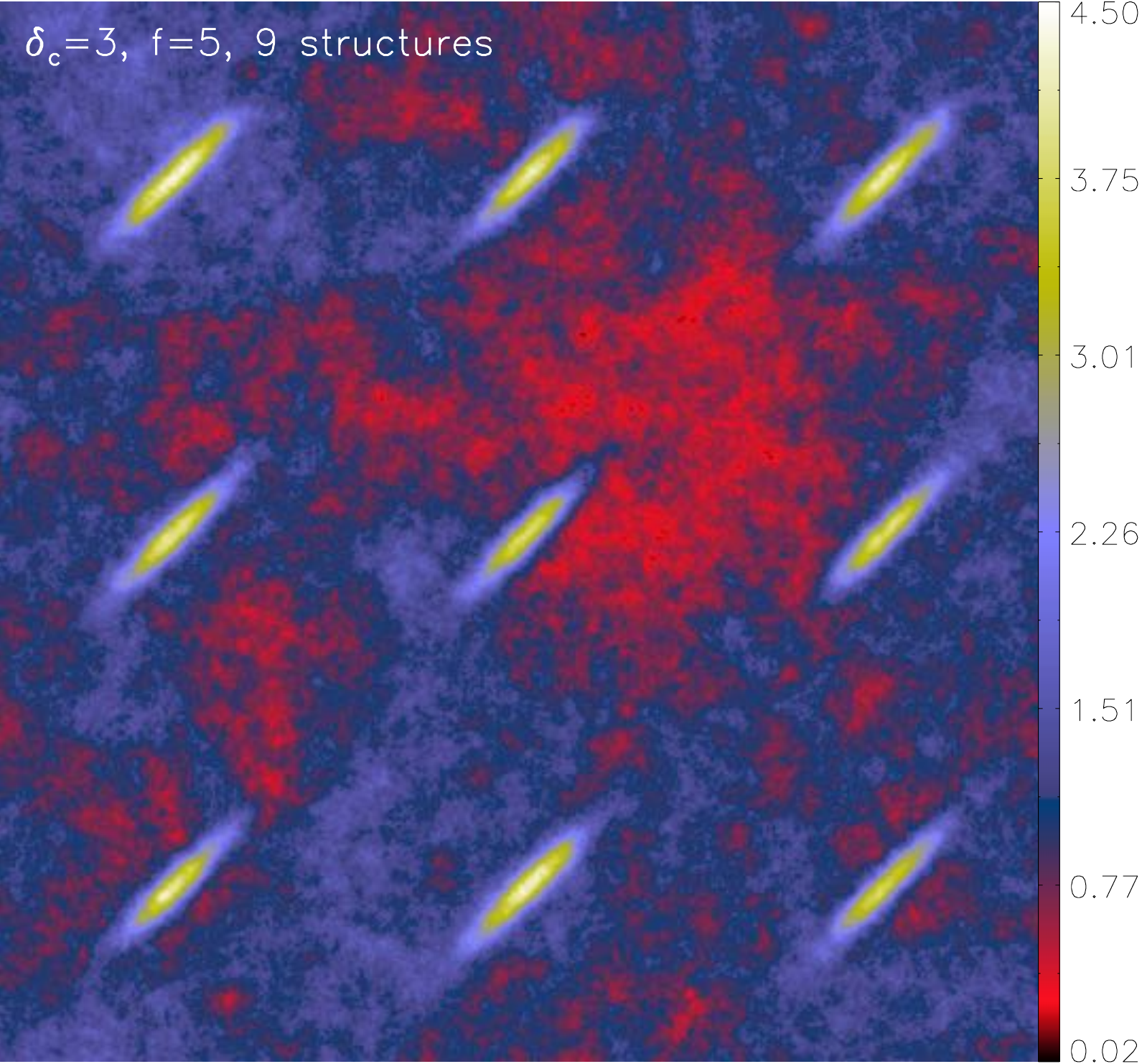}
\hspace{0.2cm} 
\includegraphics[width=0.47\columnwidth] {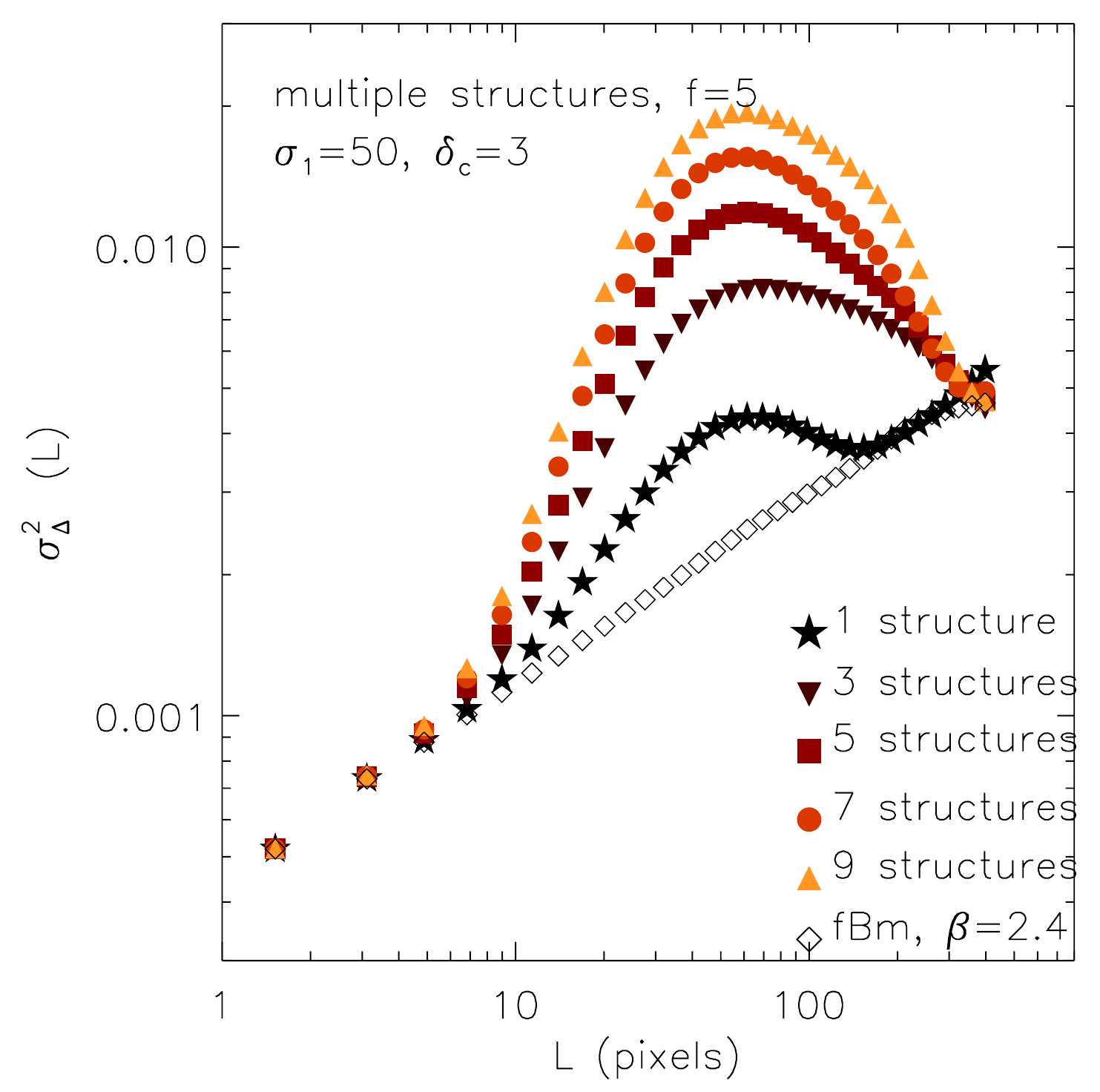}
\caption{One or several similar 2D Gaussian structures injected on top of an fBm image with $\beta=2.4$. The Gaussian functions have an aspect ratio $f=5$, a value  $\sigma_{1}=50$ pixels, and $\delta_{c}=3$. All maps are normalized to their mean value. The bottom-right figure displays the corresponding $\Delta$-variance functions calculated for each case, and these are compared to the $\Delta$-variance function calculated for the underlying fBm image.}
\label{fig8}
\end{figure}

\begin{figure}
\centering
\includegraphics[width=0.47\columnwidth] {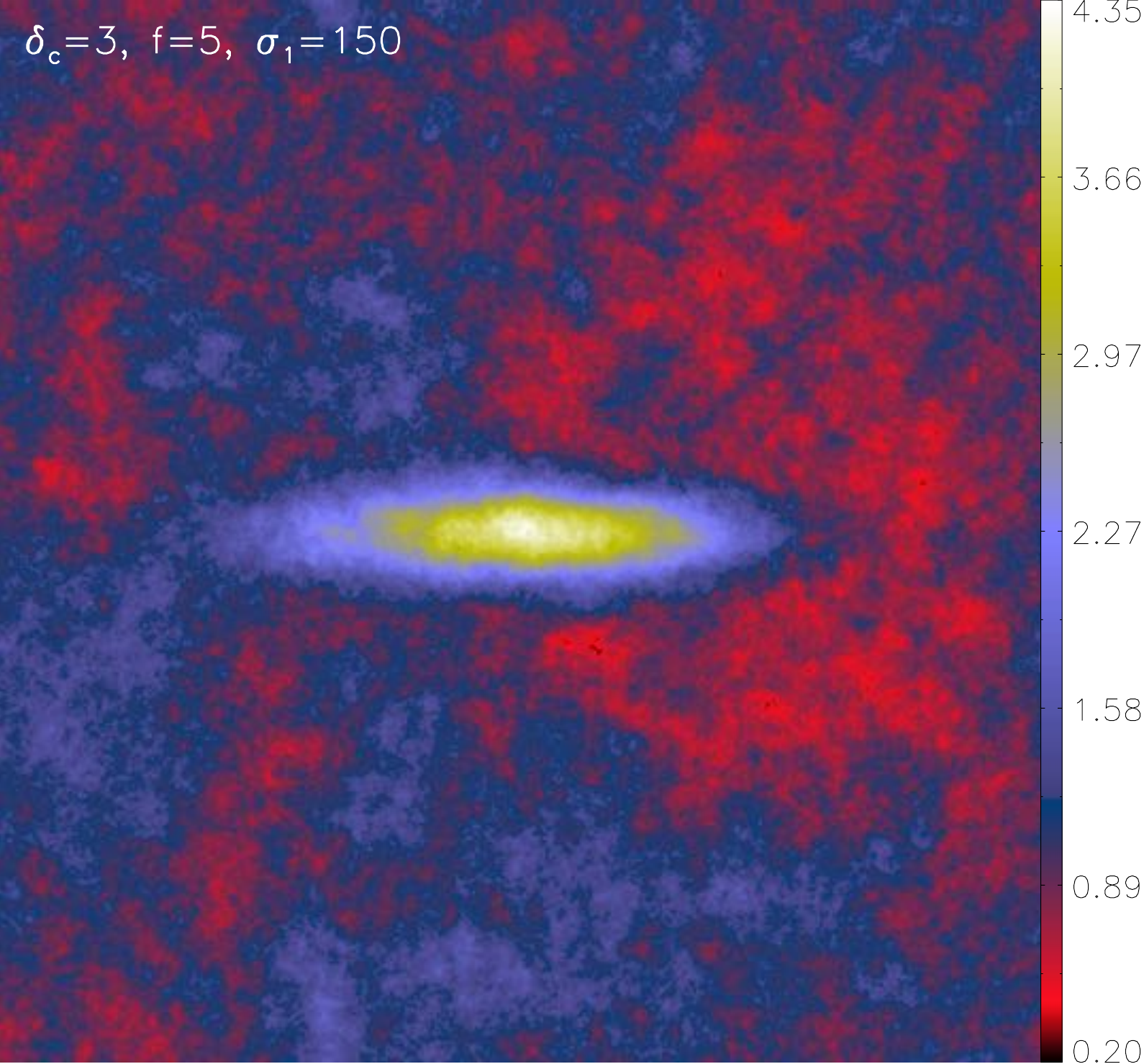}
\hspace{0.2cm}
\includegraphics[width=0.47\columnwidth] {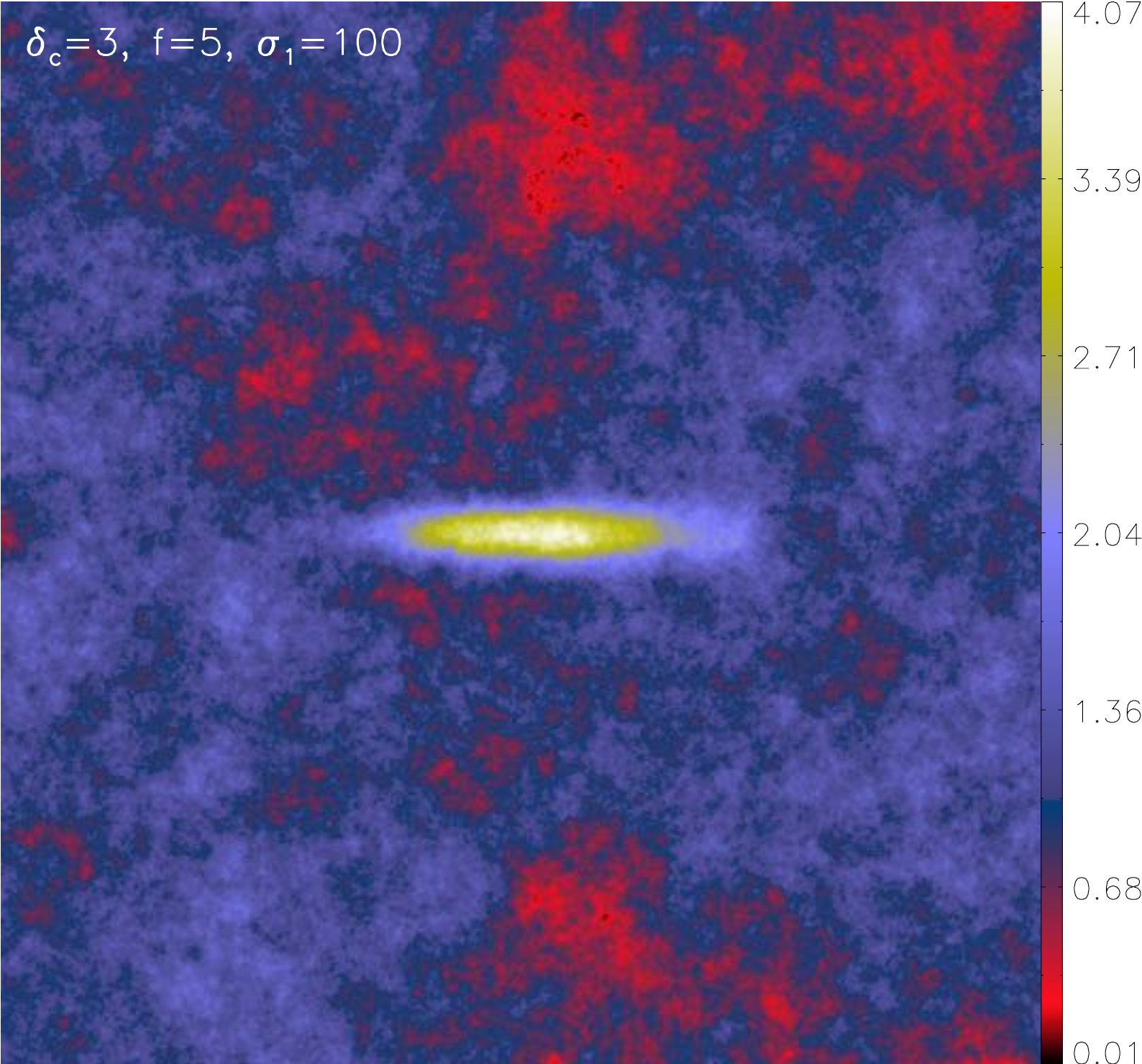}\\
\vspace{0.5cm}
\includegraphics[width=0.47\columnwidth] {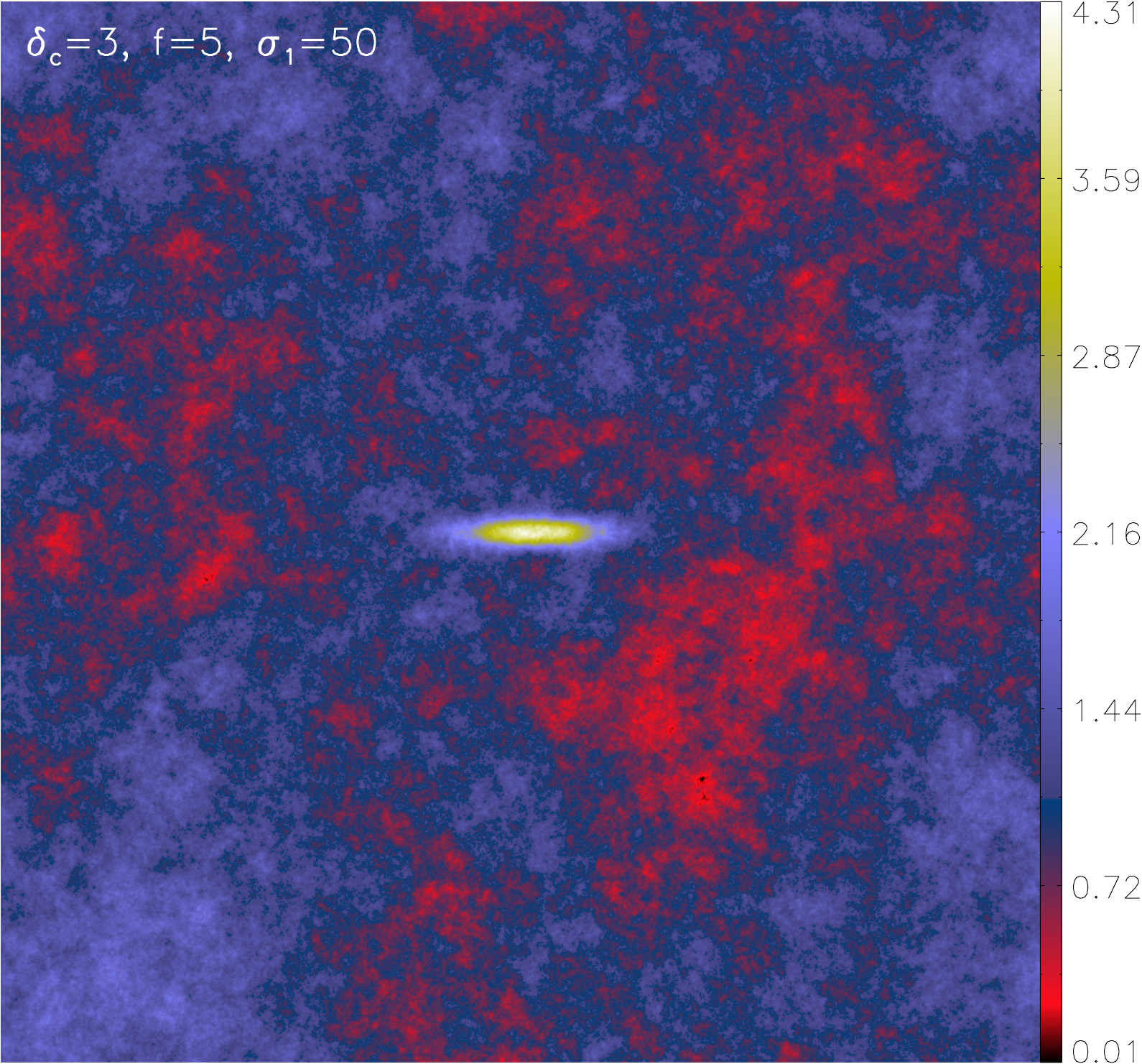}
\hspace{0.2cm}
\includegraphics[width=0.47\columnwidth] {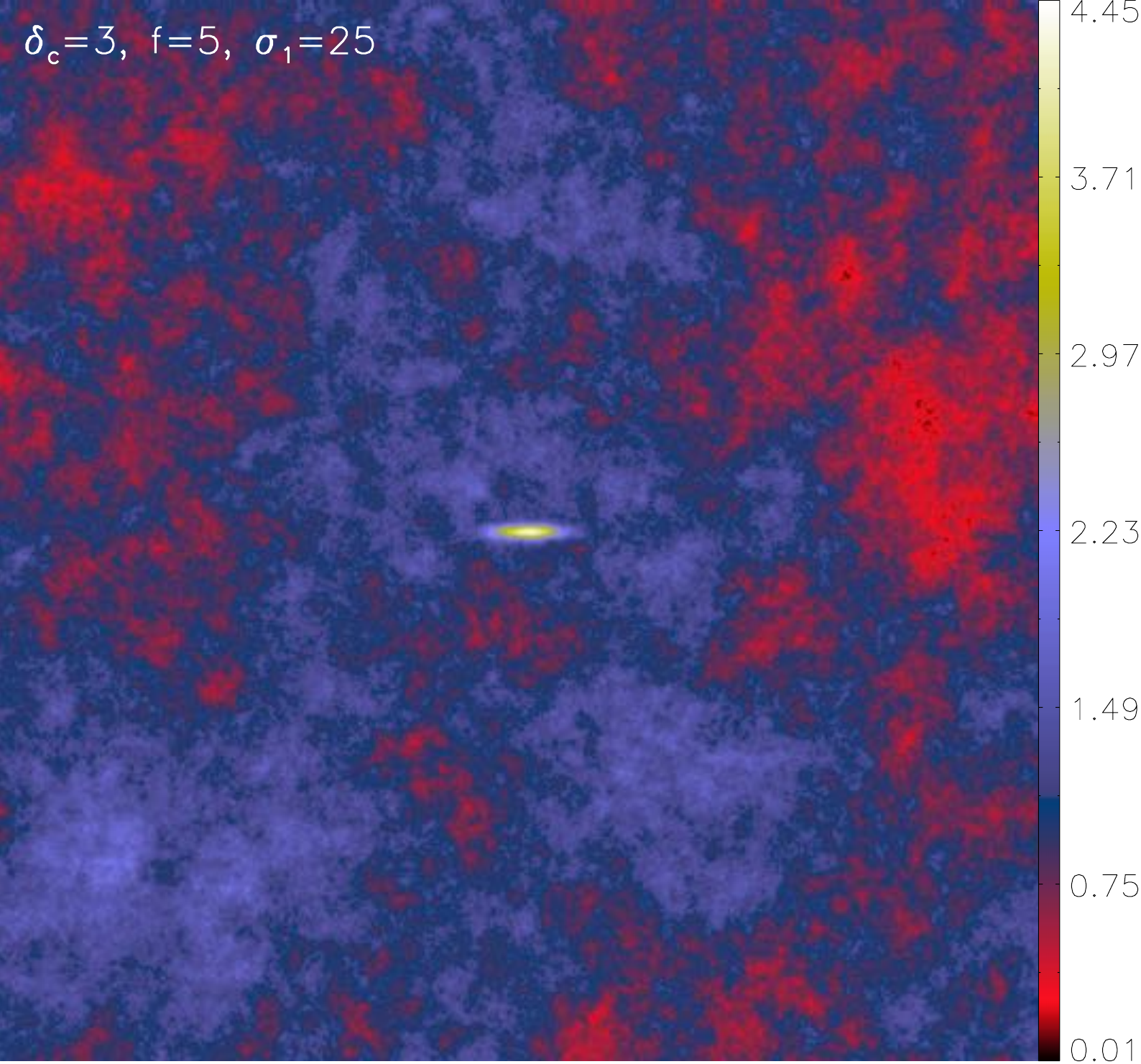}\\
\vspace{0.5cm}
\includegraphics[width=0.47\columnwidth] {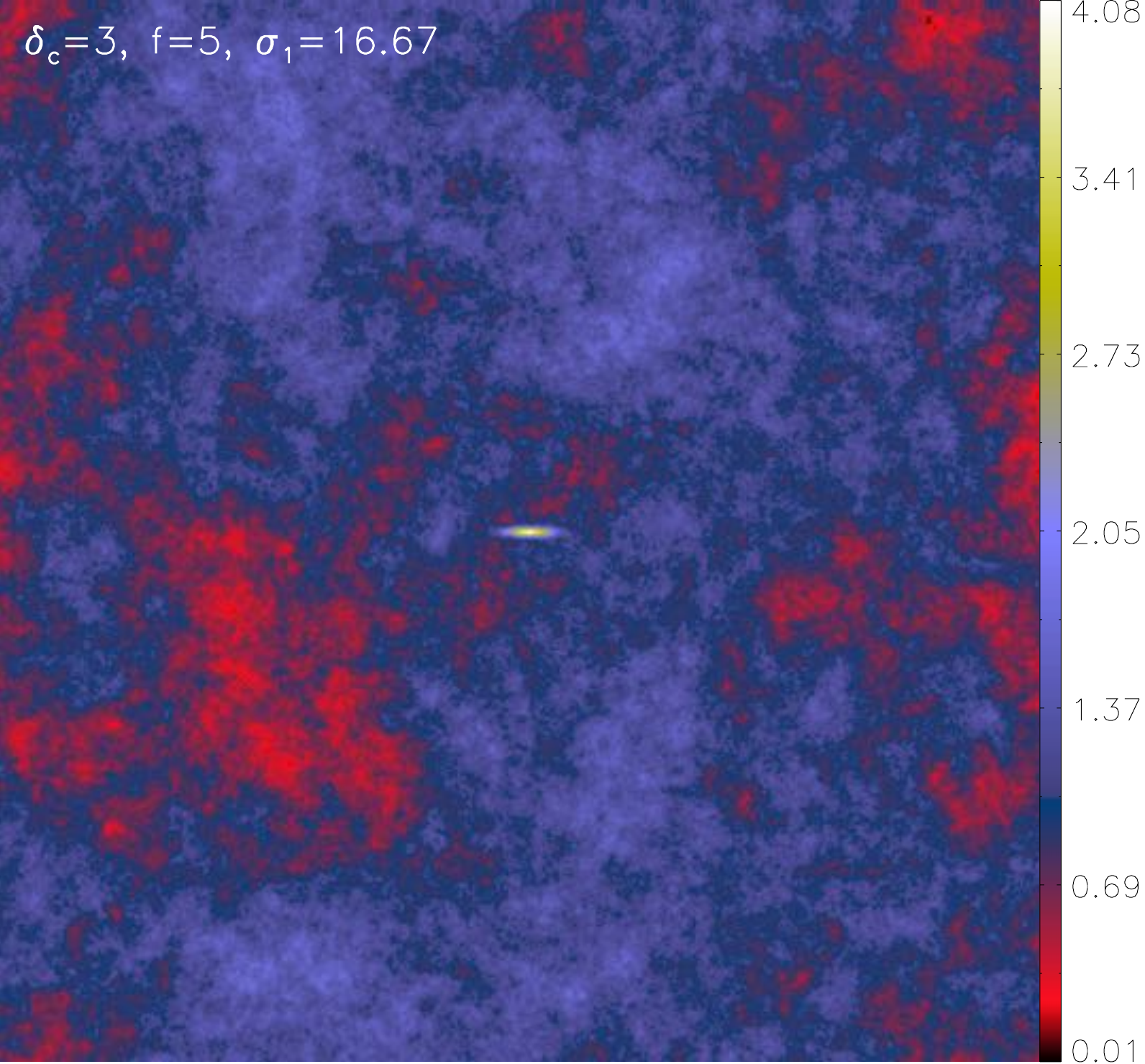}
\hspace{0.2cm} 
\includegraphics[width=0.47\columnwidth] {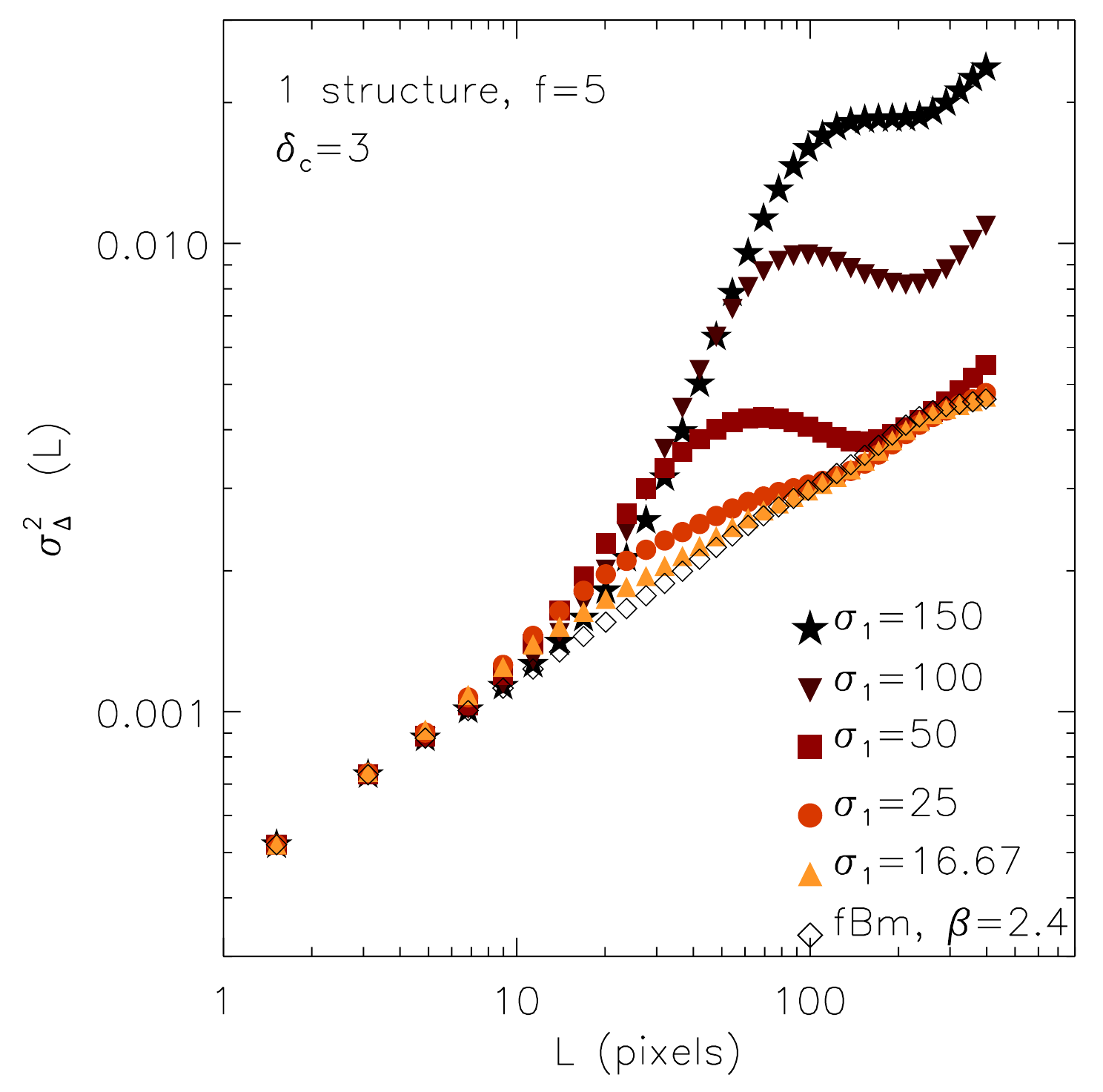}
\caption{2D Gaussian structures injected on top of an fBm image with $\beta=2.4$. The Gaussian functions all have an aspect ratio $f=0.2$, a column density contrast between the peak of the 2D Gaussian and the mean value of the fBm, $\delta_{c}=3$, and a value of $\sigma_{1}$ that is varied between $150$ and $16.67$ pixels. The bottom-right figure displays the corresponding $\Delta$-variance functions calculated for each case, and these are compared to the $\Delta$-variance function calculated for the underlying fBm image}
\label{fig9}
\end{figure}

In Fig.~\ref{fig8}, we investigate the effect of the surface area by increasing the number of structures that are superimposed onto the underlying fBm image. In this figure, one or several similar structures (all with $\delta_{c}=3$, $f=5$, and $\sigma_{1}=50$ pixels) are superimposed onto the fBm images (all with 1000$\times$1000 pixels resolution and $\beta=2.4$). The lower-right panel in Fig.~\ref{fig8} shows that increasing the surface area of the injected structures has a significant impact (i.e., linear with the number) on the increase of the $\sigma_{\Delta}^{2}$ on spatial scales that are on the order of the size of the structures. We also note an increase in the width of the excess of the $\Delta$-variance spectrum up to scales of $250$ pixels (i.e., larger than the sizes of the individual structures themselves) when there are several structures. This is the direct signature of the $250$ pixel separation between the structures. We also explore the effect of changing the size of the structure with respect to the image size while fixing the aspect ratio and column density contrast. Figure~\ref{fig9} displays five realizations with $f=5$, $\delta_{c}=3$ but where $\sigma_{1}$ is varied between $150$ and $16.67$ pixels. The $\Delta$-variance functions for these realizations are displayed in the lower-right panel of Fig.~\ref{fig9}. Here as well, the increment in $\sigma_{\Delta}^{2}$ with respect to the underlying fBm is maximized on scales that are on the order of the equivalent diameter of the structure, $D_{eq} \approx 4 \sqrt{\sigma_{1} \sigma_{2}}$. 
 
\begin{figure}
\centering
\includegraphics[width=0.497\textwidth]{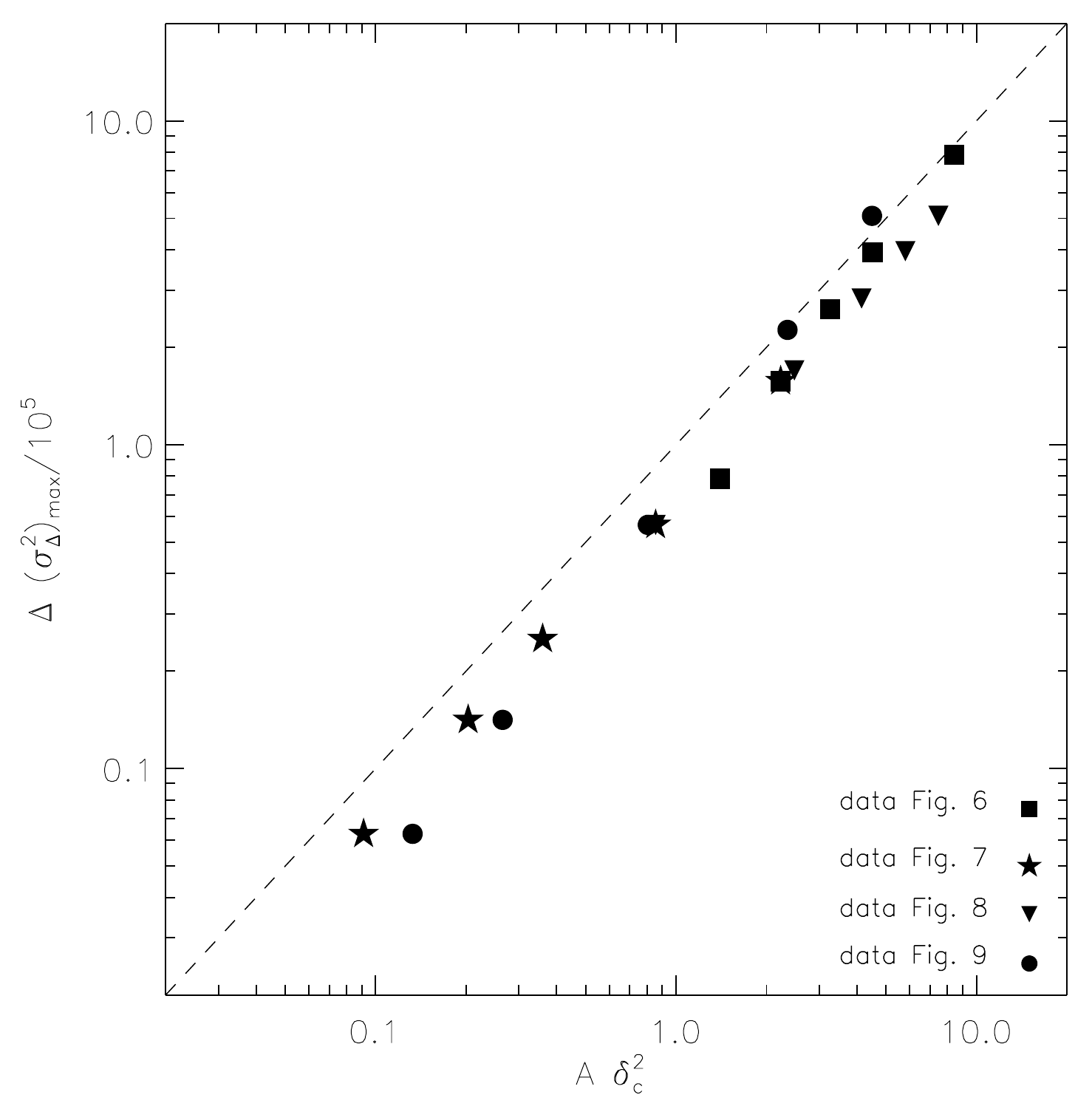}
\caption{Maximum deviation of the $\Delta$-variance function in the presence of structures from that of a pure fBm as a function of the quantity $A \delta_{c}^{2}$, where $A$ is the area covered by the discrete structure(s) and $\delta_{c}$ is the column density contrast between the peak of the structure and the mean value of the underlying fBm. The dashed line has a slope of one.}
\label{fig10}
\end{figure}

Given the expression of the $\Delta$-variance in Eq.~\ref{eq3}, one thus expects that the amplitude of the maximum deviation of $\sigma_{\Delta}^{2}$ in the presence of structures from the $\sigma_{\Delta}^{2}$ of an fBm (defined hereafter as $\Delta(\sigma_{\Delta}^{2})_{max}$) and which occurs on spatial scales that are equal to the equivalent diameter of the structures, to scale with $A \delta_{c}^{2}$, where $A$ is the total area covered by the structures. We verify whether this scaling holds for all cases displayed in Figs.~\ref{fig6}-\ref{fig9}. We calculate the area as being $A=N_{s} \pi 2 \sigma_1 \sigma_{2}$, where $N_{s}$ is the number of structures present on the map. Figure \ref{fig10} displays the value of $\Delta(\sigma_{\Delta}^{2})_{max}$ as a function of $A \delta_{c}^{2}$. A linear scaling between these quantities is found, even though we observe a small deviation from linearity for smaller values of $A \delta_{c}^{2}$. This is due to the fact that when structures are small, there are larger uncertainties associated with the determination of their surface. 

\subsection{Toward more realistic configurations}\label{realistic}

\begin{figure*}
\centering
\includegraphics[width=0.32\textwidth] {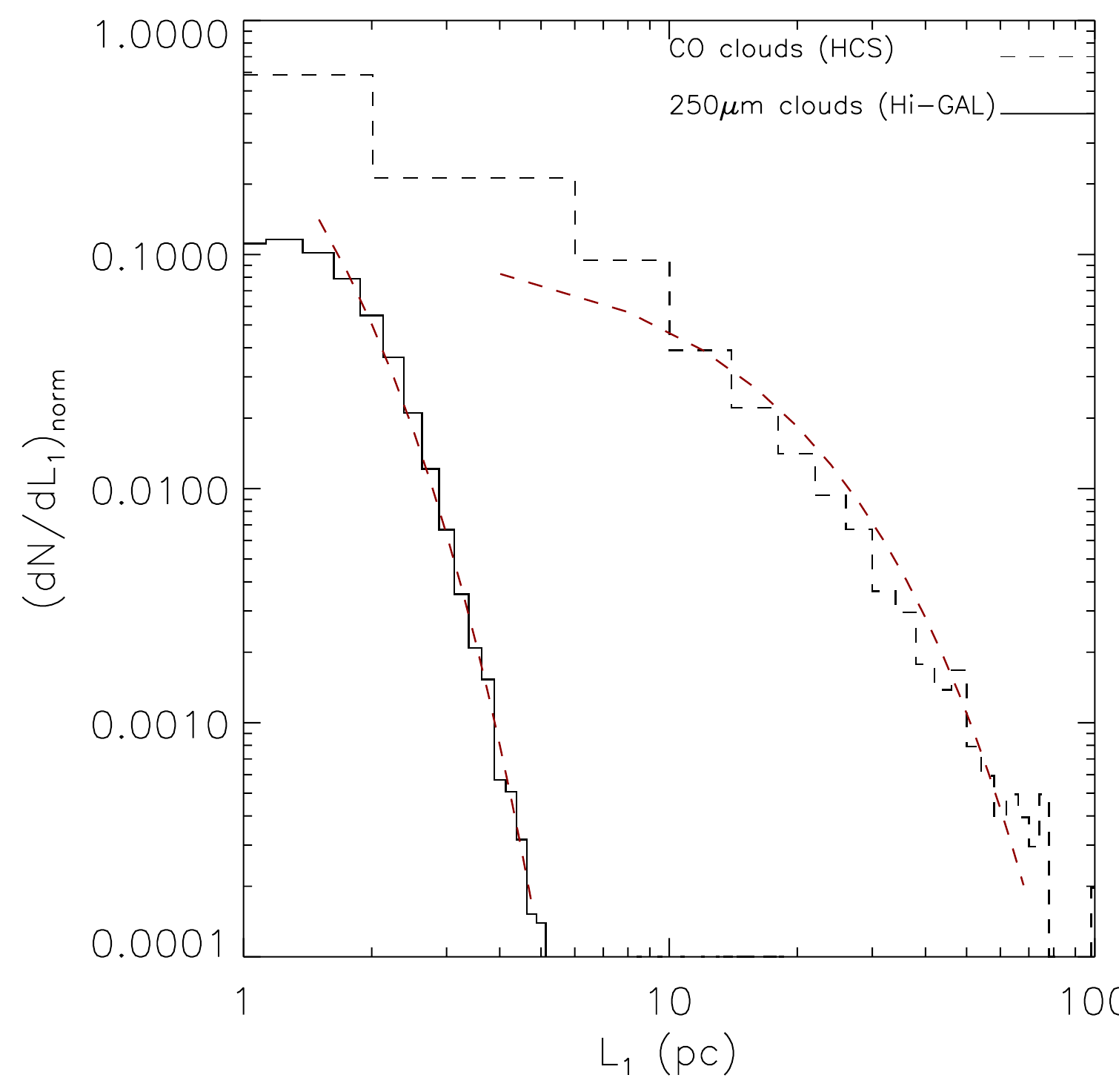}
\includegraphics[width=0.32\textwidth] {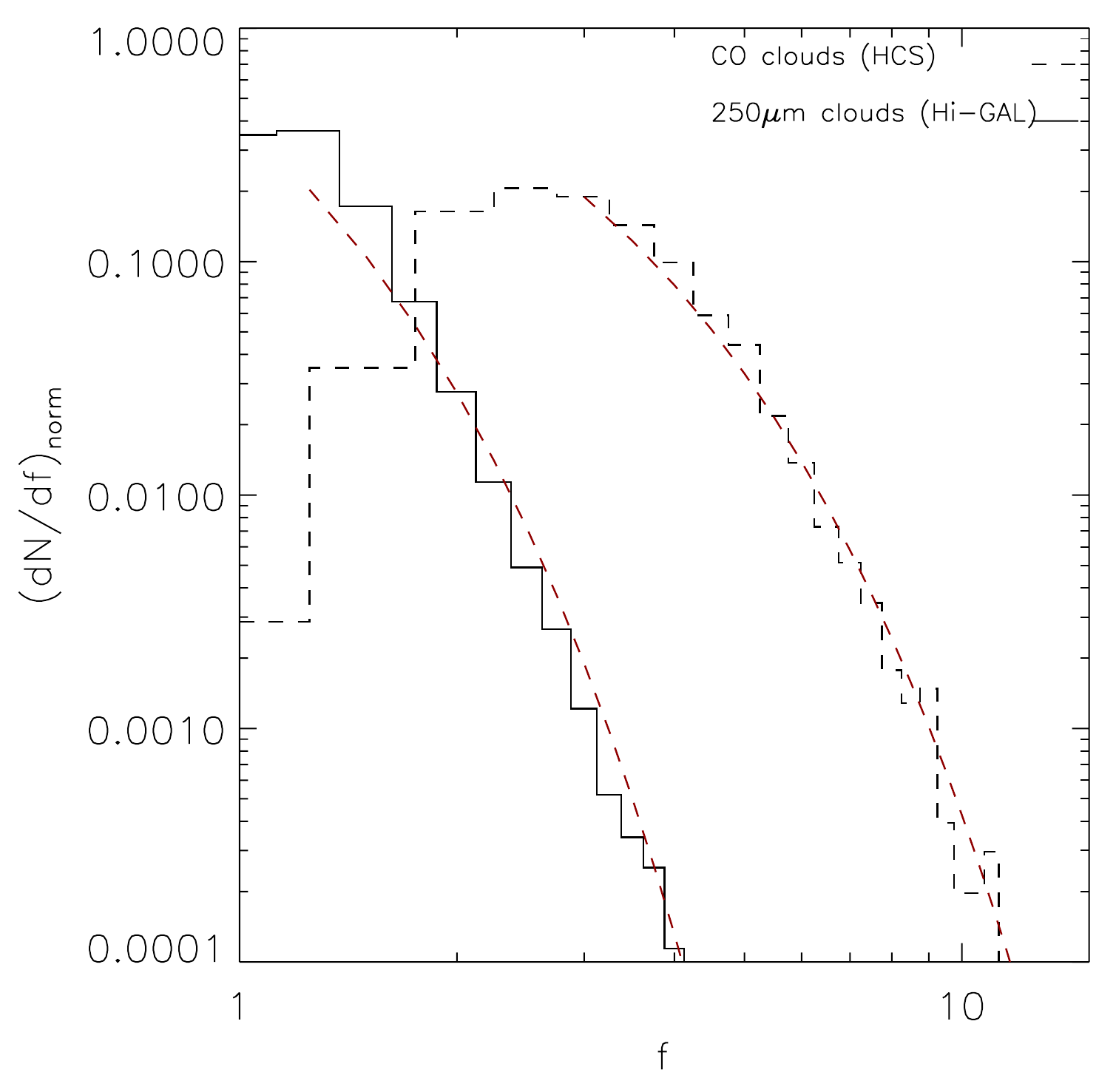}
\includegraphics[width=0.32\textwidth] {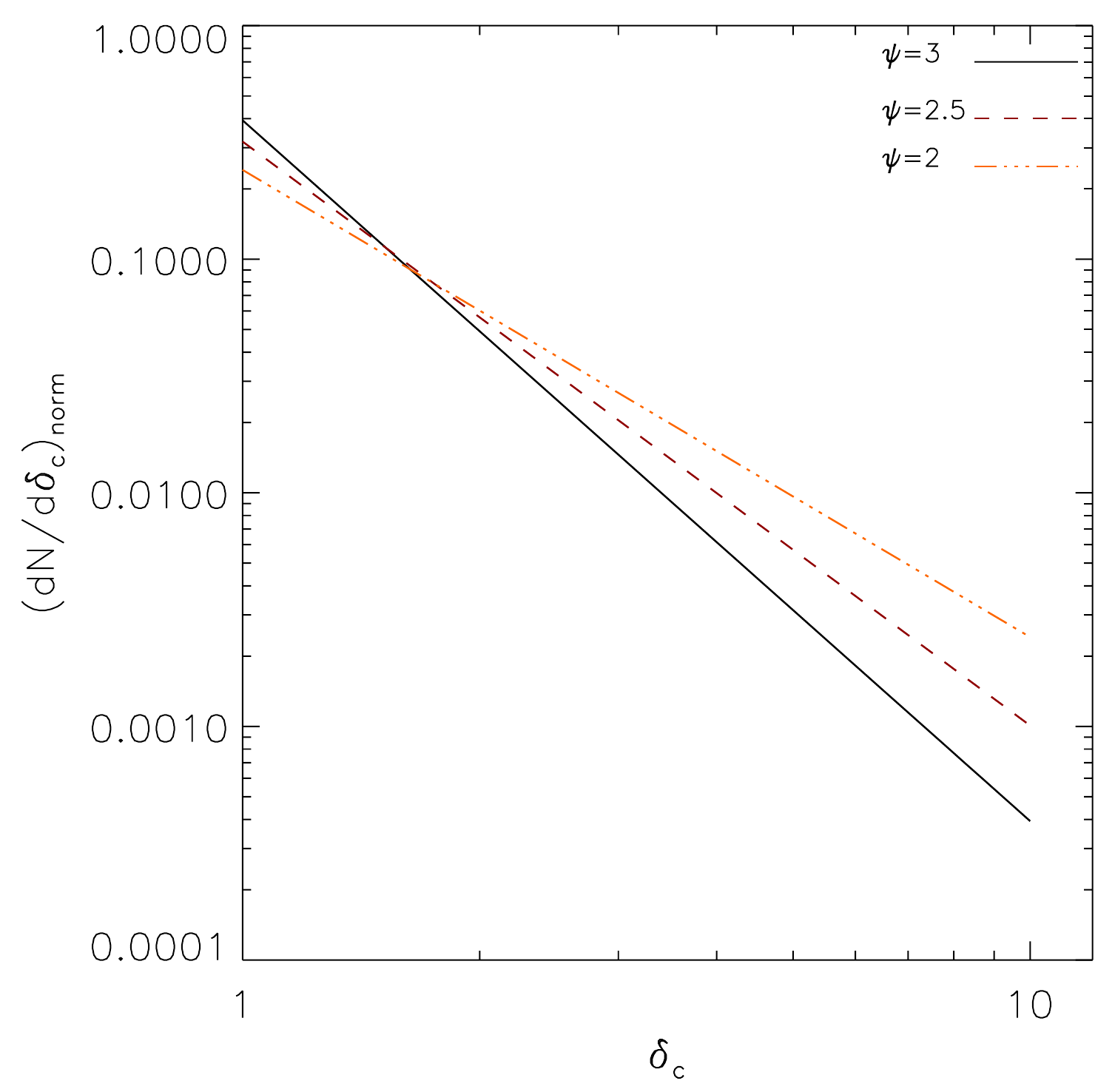}\\
\caption{Fractional probability distribution functions of the size of the major axis (left) and aspect ratio of clouds (middle) found in the $250 \mu$m maps of the Hi-GAL survey (Molinari et al. 2016; Elia et al. 2017; full line) and in the $^{12}$CO FCRAO HCS survey (Heyer et al 2001; Dib et al. 2009; dashed line). The distribution of column density contrast is assumed to be a power law. The distributions of $L_{1}$ and $f$ are fitted with parameterized functions. The values of the parameters of the fit are reported in the main text.}
\label{fig11}
\end{figure*}

In principle, the generation of realistic column density maps could rely on numerical simulations of turbulent and self-gravitating molecular clouds. However, the parameter space can be very large, namely, models with or without gravity, with various magnetic field strengths, and with various driving schemes, Mach numbers, and turbulence driving scales. When gravity is included, the extracted information will also unavoidably depend on the time evolution of the simulated clouds. This remains a valuable approach that has in fact been explored to a certain extent (e.g., Ossenkopf et al. 2001;2002) and deserves to be explored further with more refined models. In this work, we prefer to generate models whose parameters can be easily controlled and for which we can easily understand and disentangle the effects on the $\Delta$-variance function. As in Sect.~\ref{fbmplus}, we superimposed 2D Gaussians on top of predefined fBm images. However, instead of including individual structures or structures that are set apart from each other, we now include 2D Gaussians with specific distribution functions that characterize their properties. The parameters we varied are the number of 2D Gaussian structures, $N_{s}$, the distribution function of the size of the major axis $\left(dN/dL_{1}\right)$, the distribution function of the aspect ratios $\left(dN/df\right)$, and the distribution function of the structures column density contrast $\left(dN/d\delta_{c}\right)$. Each structure is assigned a randomly drawn orientation on the map, and the structures are allowed to overlap. Keeping in mind that clumps, cores, and filaments such as those found in the column density map of Cygnus-X North may have a more complex internal structure than 2D Gaussian functions, we aim to understand which combination of the parameters leads to $\Delta$-variance functions that are similar to that of the Cygnus-X North region. More broadly, our aim is also to understand the sensitivity of the $\Delta$-variance to the choice of the distribution functions that characterize the statistical properties of the structures. 

The distribution of sizes and aspect ratios of cores and clumps in molecular clouds is likely to depend on the density tracer as well as on the clump identification algorithm. To illustrate this, in Fig.~\ref{fig11} we compare the size (i.e., major axis; $L_{1}$, left panel) and aspect ratio ($f$, middle panel) distributions of structures found in the {\it Herschel infrared Galactic Plane survey} (Hi-GAL; Molinari et al. 2016; Elia et al. 2017) and in the {\it Five College Radio Astronomy Observatory (FCRAO) CO survey of the outer Galaxy} (HCS; Heyer et al. 1998; Dib et al. 2009). Structures in the Hi-GAL survey are extracted from $250 \mu$m emission maps (78952 objects in total), whereas the HCS survey is based on the (1-0) transition in $^{12}$CO molecular line observations, in which 10156 discrete structures were identified. The clouds and clumps reported in the Hi-GAL survey are ostensibly more roundish than the ones detected in molecular line observations\footnote{The Hi-GAL sources were extracted with CuTEx (Molinari et al. 2016), which is designed to identify relatively roundish sources. In principle, during the detection step, it keeps only structures with both minor and major axes ranging from 1 to 3 instrumental point spread functions. Subsequently, starting from this initial guess, the 2D Gaussian fit that is used to determine the flux as well as the final estimate of the two axes has an additional tolerance to adjust itself on the source profile, so that one can find one (or both) of the two axes shorter than 1 RMS or longer than 3 RMS. Usually, no large differences are found between the two axes, so that the ratio is never larger than $\approx 4$. On the contrary, algorithms used to extract sources from CO surveys, such as the one used in Heyer et al. (2001) that is based on a friend-of-friend approach in position-position-velocity space, do not have any constraint on source size. If the CO emission is kinematically connected over an elongated area, such a structure might be identified as a single source with a large aspect ratio.}. The distribution functions in Fig.~\ref{fig11} are normalized and are thus transformed into probability distribution functions. 

For the aspect ratio distributions of the Hi-GAL and FCRAO HCS clouds and clumps, $(dN/df)_{norm}$ (Fig.~\ref{fig11}, middle panel), we find that these distributions are best approximated by the following function:

\begin{equation}
{\rm log}\left(\frac{dN}{df}\right)_{norm} = \eta~f+A_{f}.
\label{eq9}
\end{equation}

Fitting the distributions of aspect ratios for the Hi-GAL (for $f > 1$) and HCS (for $f > 2.5$) clouds yields [$\eta=-1.15\pm0.04, A_{f}=0.75\pm0.13$] and [$\eta=-0.37\pm0.01,A_{f}=0.41\pm0.10$], respectively. The results of the corresponding fits are shown in Fig.~\ref{fig11} (dashed red lines, middle panel). For any other chosen value of $\eta$, the corresponding value of $A_{f}$ can be calculated by requiring that $\int_{f_{min}}^{f_{max}} \left(dN/df\right)_{norm} df=1$, and where $f_{min}$ and $f_{max}$ are the lower and upper limits on $f$, respectively. In the same vain, we fitted the normalized distribution function of the size of the major axis, $\left(dN/dL_{1}\right)_{norm}$. Here again, we find that the data is best fitted with a function that is given by:

\begin{equation}
{\rm log}\left(\frac{dN}{dL_{1}}\right)_{norm} = \xi~L_{1}+A_{L_{1}}.
\label{eq10}
\end{equation} 

Using Eq. \ref{eq10} the fit to the data of the Hi-GAL clouds for values of $L_{1}$ in the range $1.5~{\rm pc} \geq L_{1} \geq 5~{\rm pc}$ and for the HCS clouds using values of $L_{1}$ in the range $4~{\rm pc}\geq L_{1} \geq 70~{\rm pc}$ yields values of the parameters $\xi$ and $A_{L_{1}}$ of [$\xi=-0.89\pm 0.02,A_{L_{1}}=0.50\pm0.06$] and [$\xi=-0.04\pm0.002,A_{L_{1}}=-0.91\pm0.09$], respectively. The fit functions are displayed with the dashed red lines in Fig.~\ref{fig11} (left-hand panel). For any other value of $\xi$, the corresponding values of $A_{L_{1}}$ can be obtained by requiring that $\int_{L_{1,min}}^{L_{1,max}} \left(dN/dL_{1}\right)_{norm} dL_{1}=1$, where $L_{1,min}$ and $L_{1,max}$ are the lower and upper limits on $L_{1}$.   

The distribution of column density contrasts of dense structures in nearby molecular clouds is not yet fully established. Recent work by Arzoumanian et al. (2019) derived the contrast between the average column density on filament crests and their local background for filaments detected in a number of nearby molecular clouds. Roy et al. (2019)  constructed the distribution function of the contrast between filaments and their local background and found that it scales as $\left(dN/d{\rm log}\delta_{c}\right) \propto \delta_{c}^{-1.5}$ for $\delta_{c} > 1$. The exact scaling found in Arzoumanian et al. and Roy et al. may not apply directly to our synthetic models since we define the contrast as being the one between the peak column density of the structure and the mean value over the entire map. With this in mind, we parameterized the distribution of column density contrasts as being a power law of the form:

\begin{equation}
\frac{dN}{d\delta_{c}} = A_{c} \delta_{c}^{-\psi},  
\label{eq11}
\end{equation} 

\begin{figure*}
\centering
\hspace{0.5cm}
\includegraphics[width=0.25\textwidth] {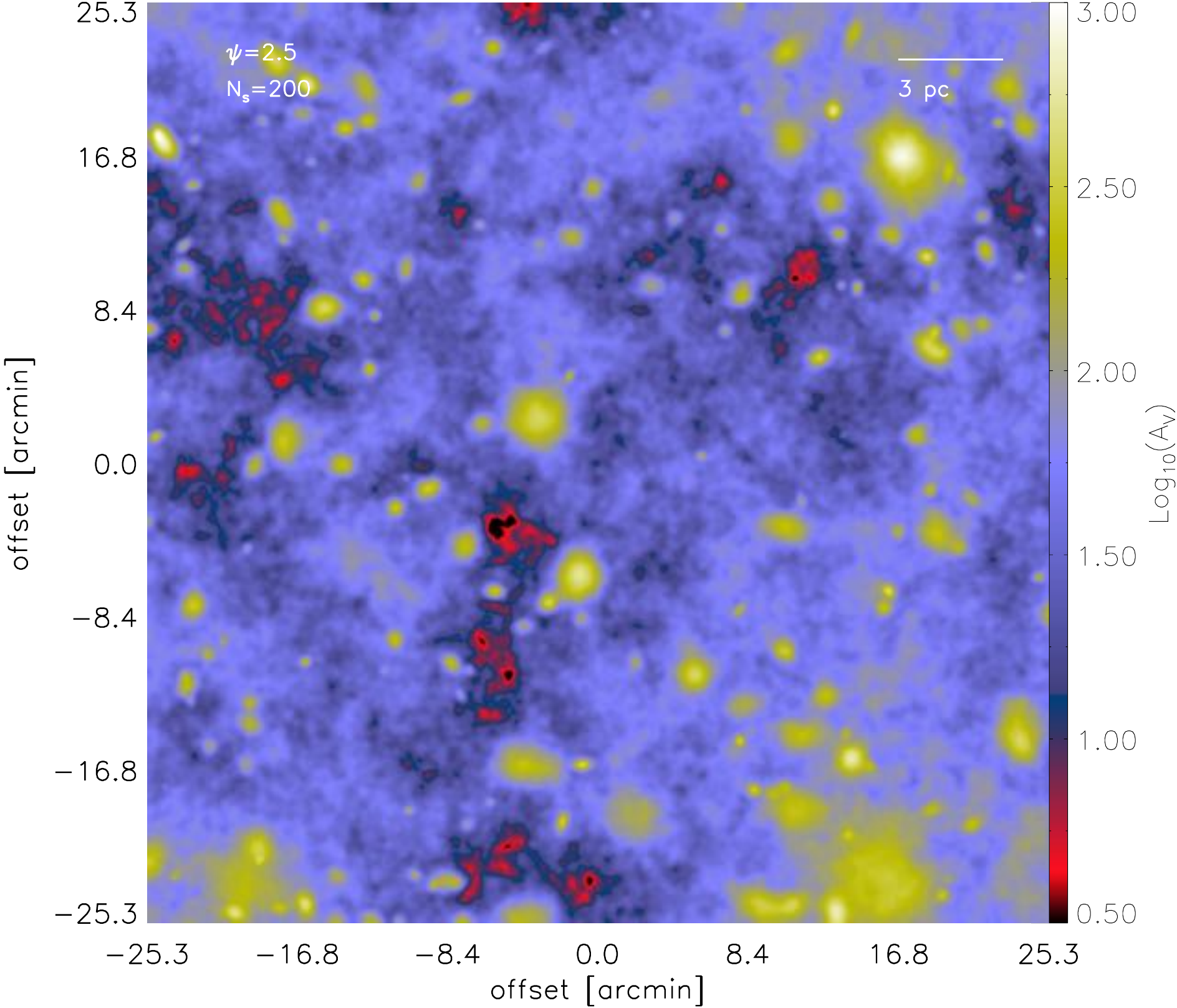}
\hspace{1.5cm}
\includegraphics[width=0.25\textwidth] {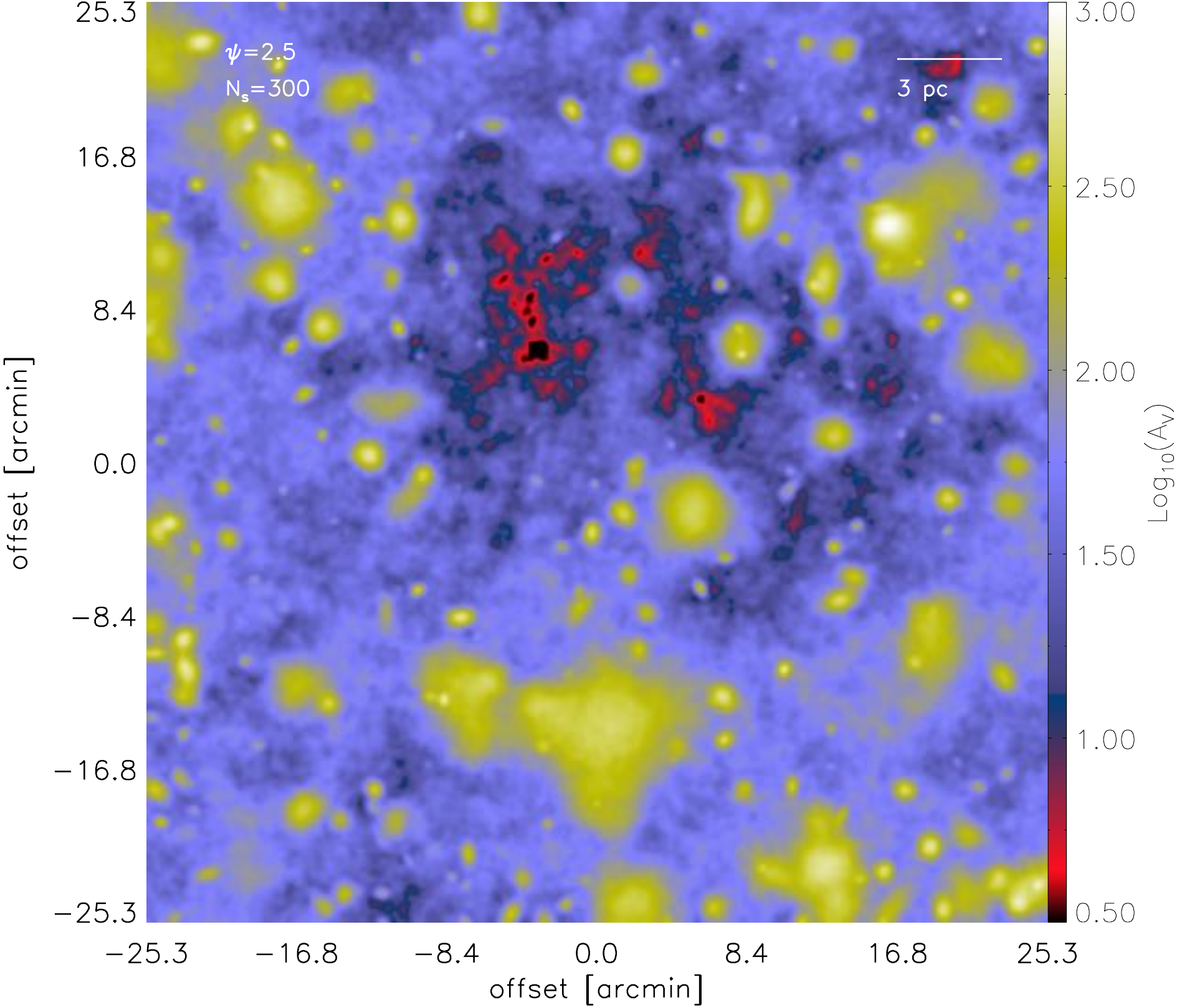}
\hspace{1.5cm}
\includegraphics[width=0.25\textwidth] {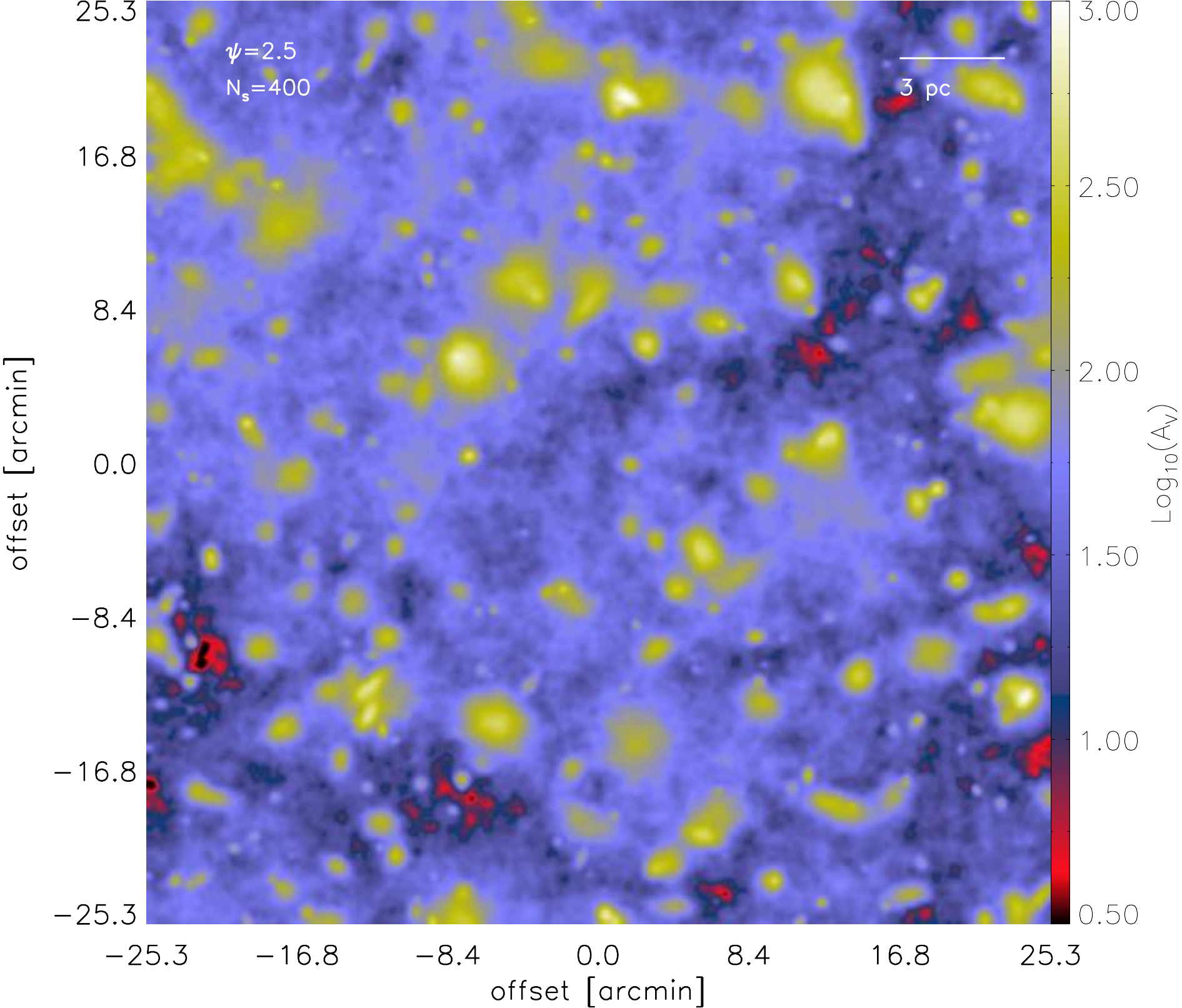}\\
\vspace{0.5cm}
\includegraphics[width=0.32\textwidth] {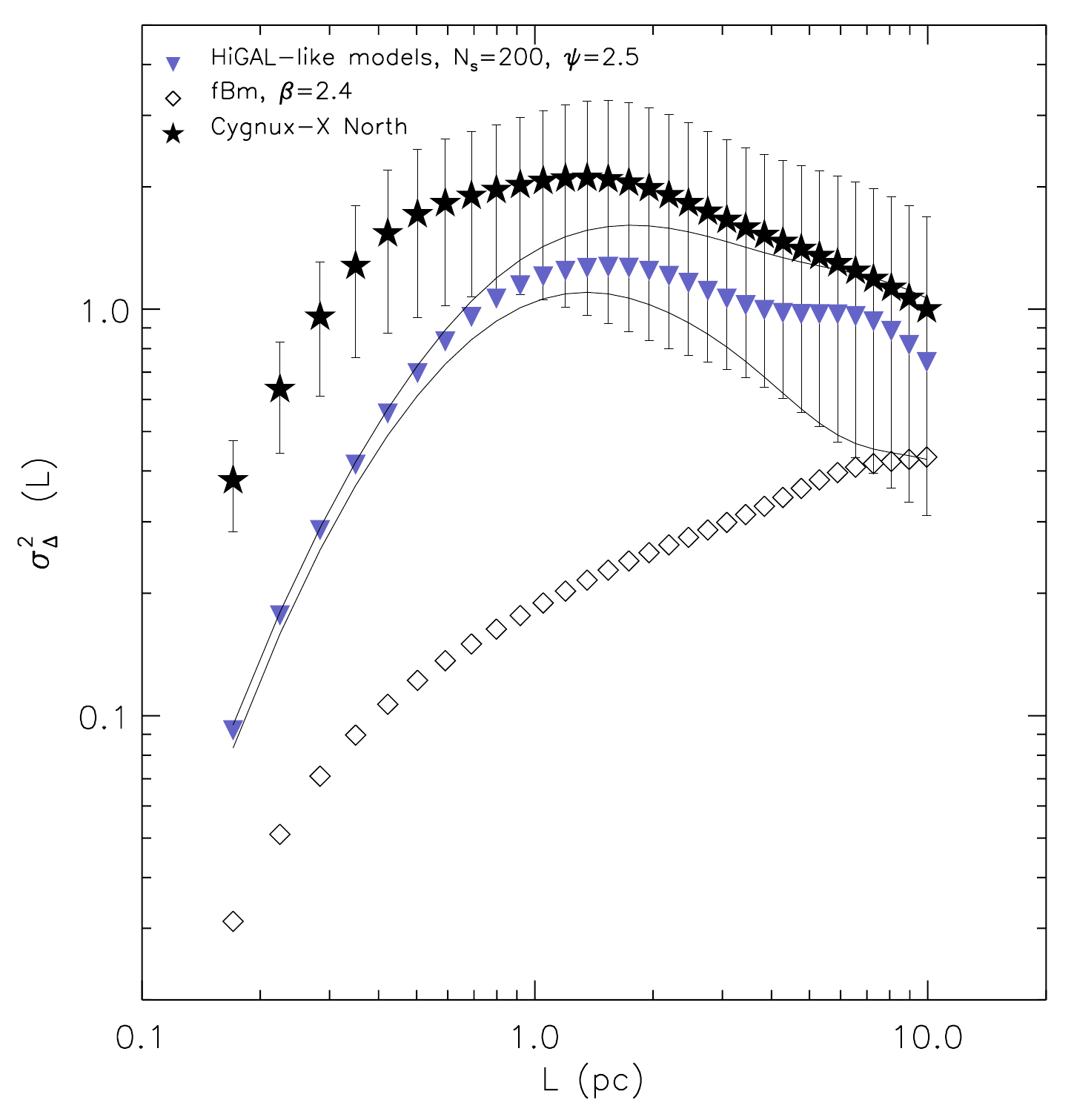}
\hspace{0.2cm}
\includegraphics[width=0.32\textwidth] {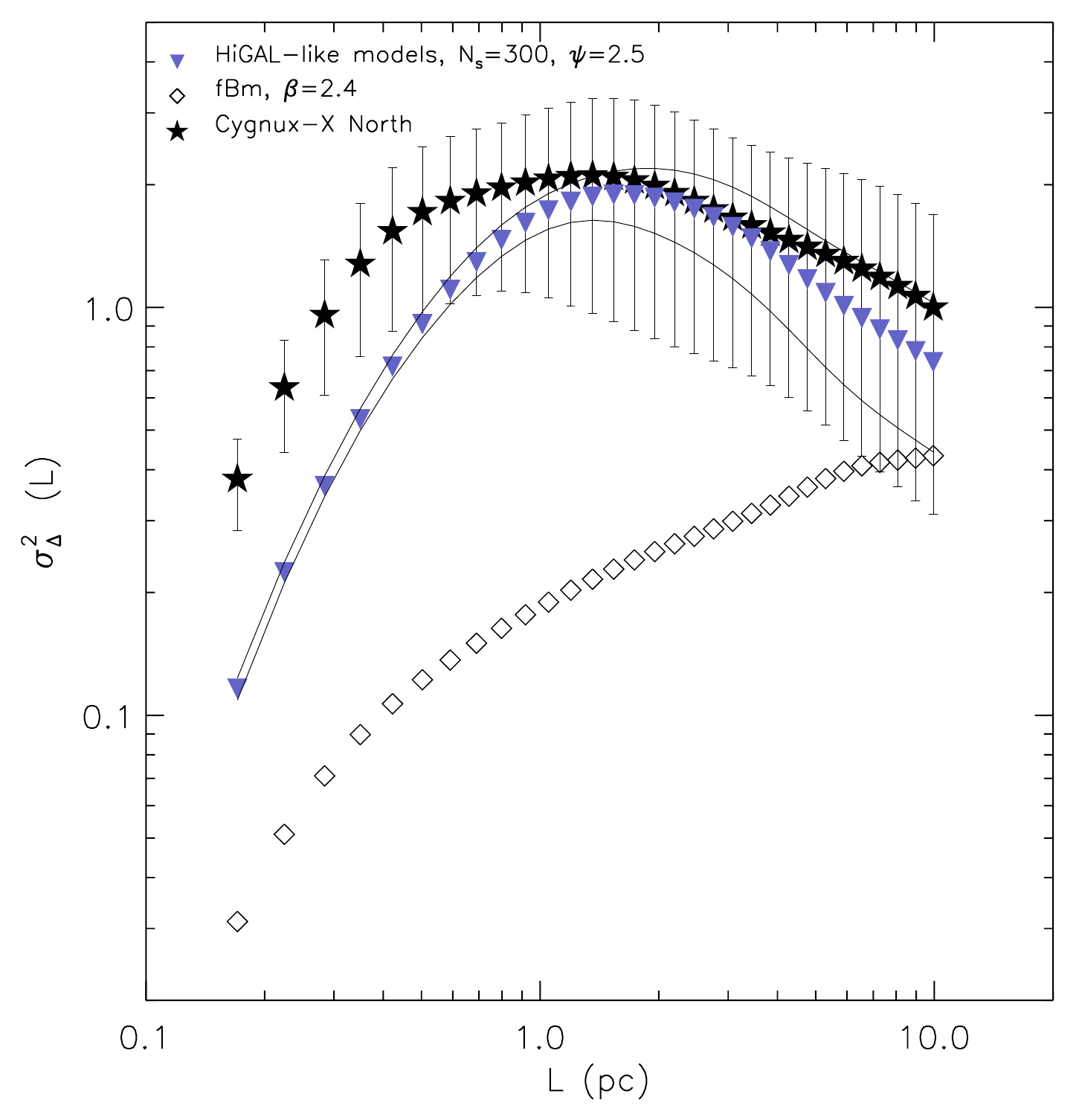}
\hspace{0.2cm} 
\includegraphics[width=0.32\textwidth] {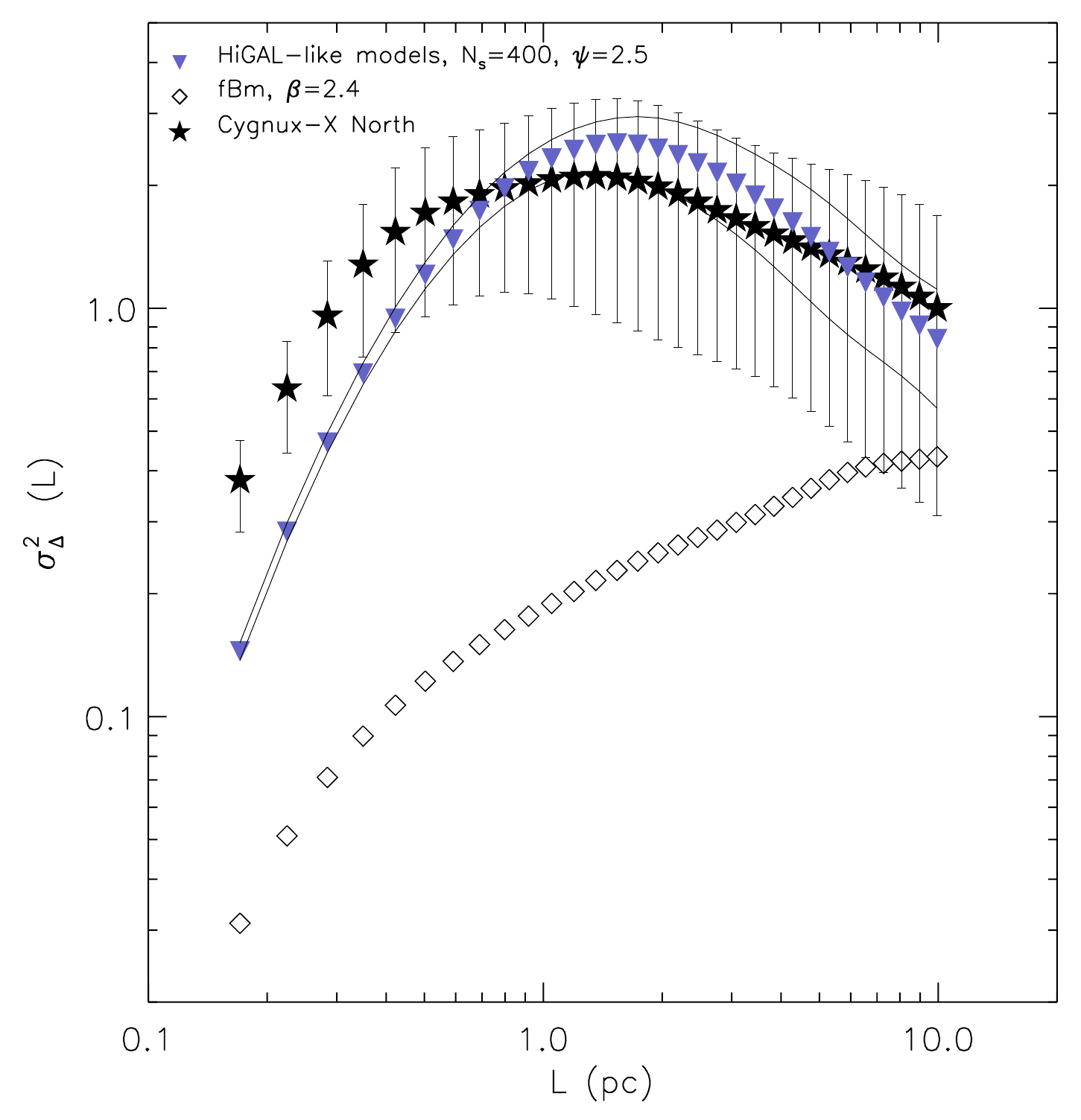}
\caption{Top: Examples of 200 (left), 300 (middle), and 400 (right) 2D Gaussian structures injected on top of an fBm image with $\beta=2.4$. The Gaussian structures are randomly sampled using the distribution functions of the major axis size and aspect ratio distributions of the Hi-GAL clumps. The column density contrasts are randomly sampled from a distribution with $\psi=2.5$. All synthetic maps are convolved with a beam whose FWHM=$18.2\arcsec$. Bottom: $\Delta$-variance spectrum of the synthetic models for cases with $N_{s}=200$ (left), $N_{s}=300$ (middle), and $N_{s}=400$ (right) injected structures. Each synthetic spectrum is the average over 25 realizations, and the full lines represent the $1 \sigma$ dispersion around the mean. The synthetic $\Delta$-variance spectra are compared to those of the Cygnus-X North region and an fBm with $\beta=2.4$.}
\label{fig12}
\end{figure*}

\begin{figure*}
\centering
\hspace{0.5cm}
\includegraphics[width=0.25\textwidth] {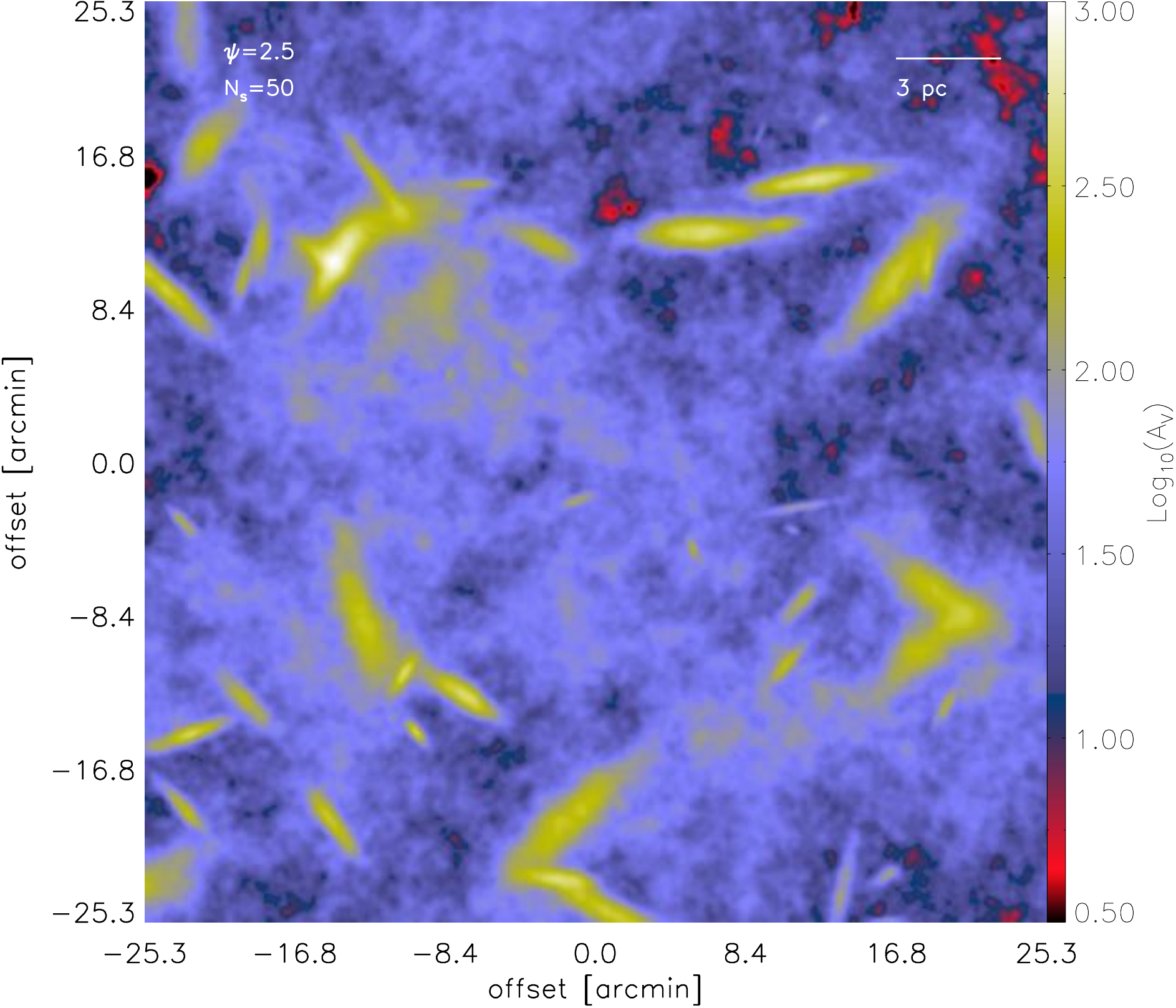}
\hspace{1.5cm}
\includegraphics[width=0.25\textwidth] {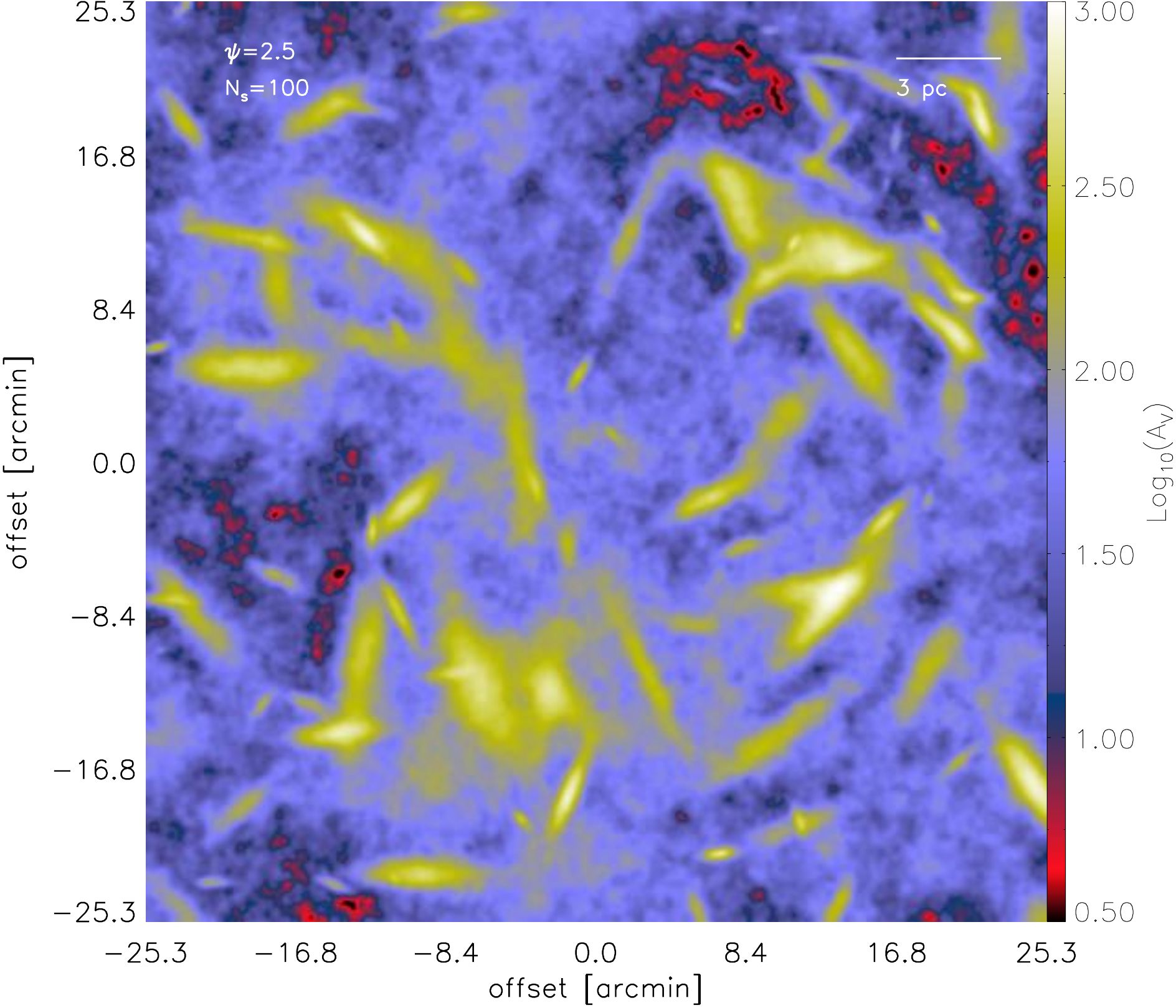}
\hspace{1.5cm}
\includegraphics[width=0.25\textwidth] {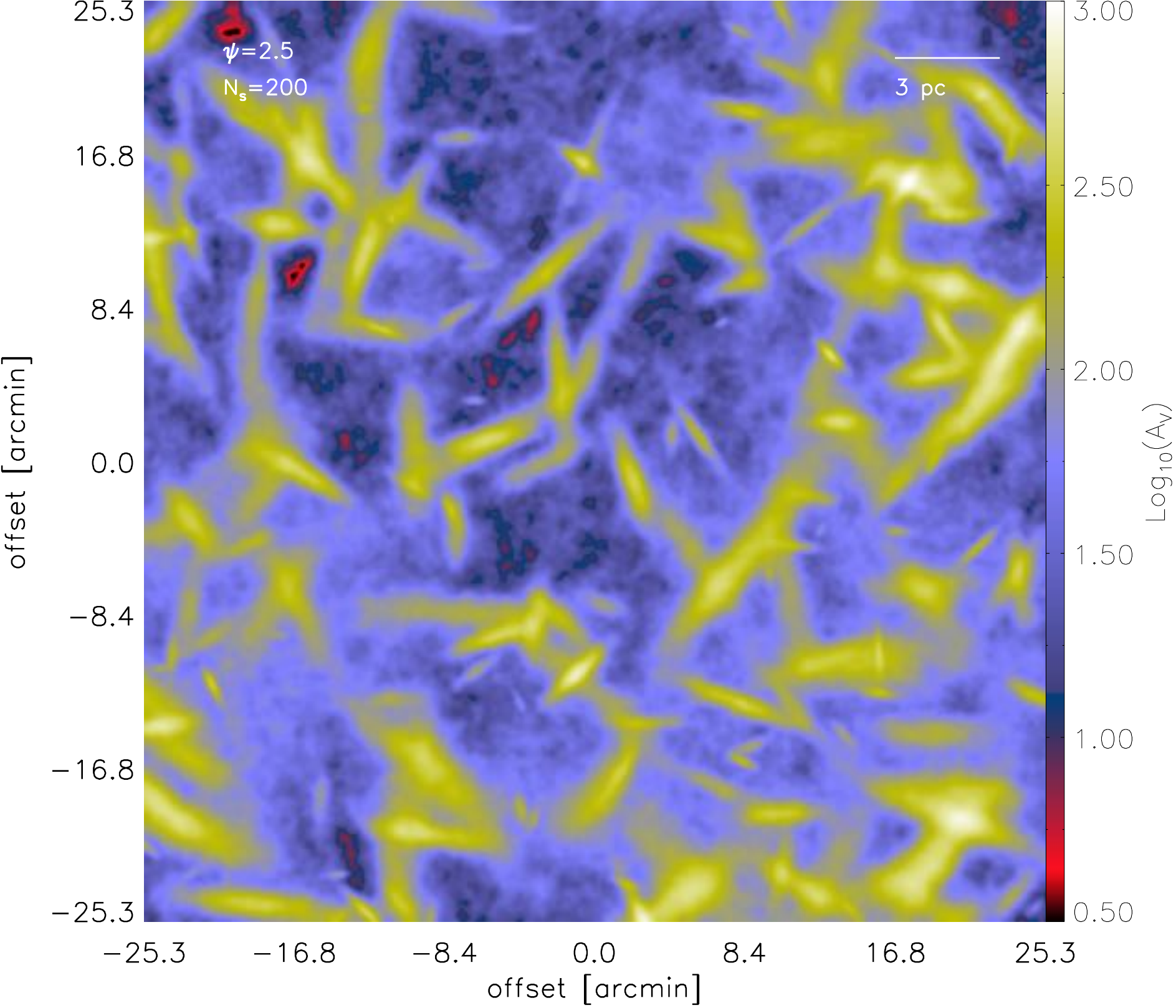}\\
\vspace{0.5cm}
\includegraphics[width=0.32\textwidth] {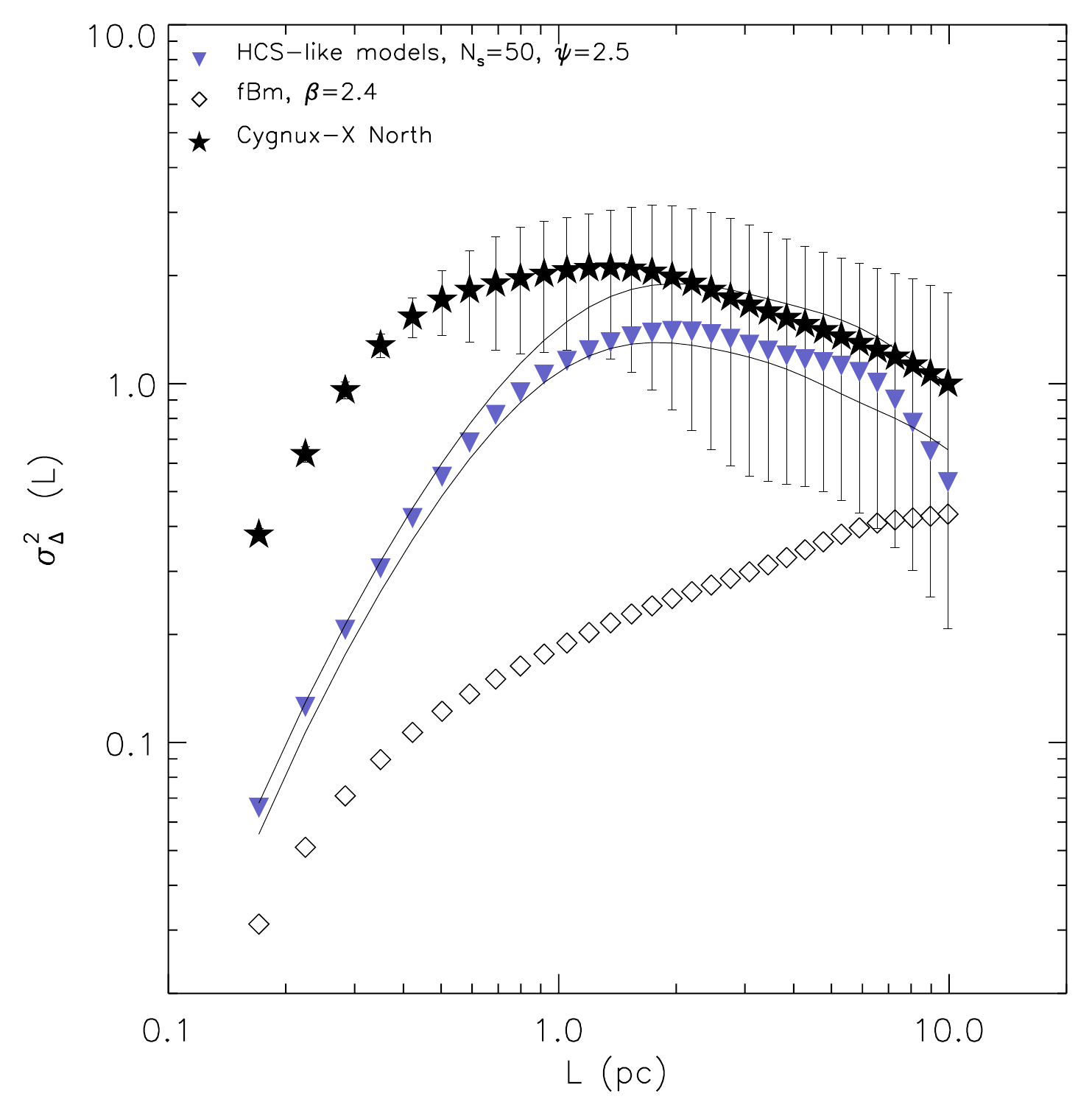}
\hspace{0.2cm}
\includegraphics[width=0.32\textwidth] {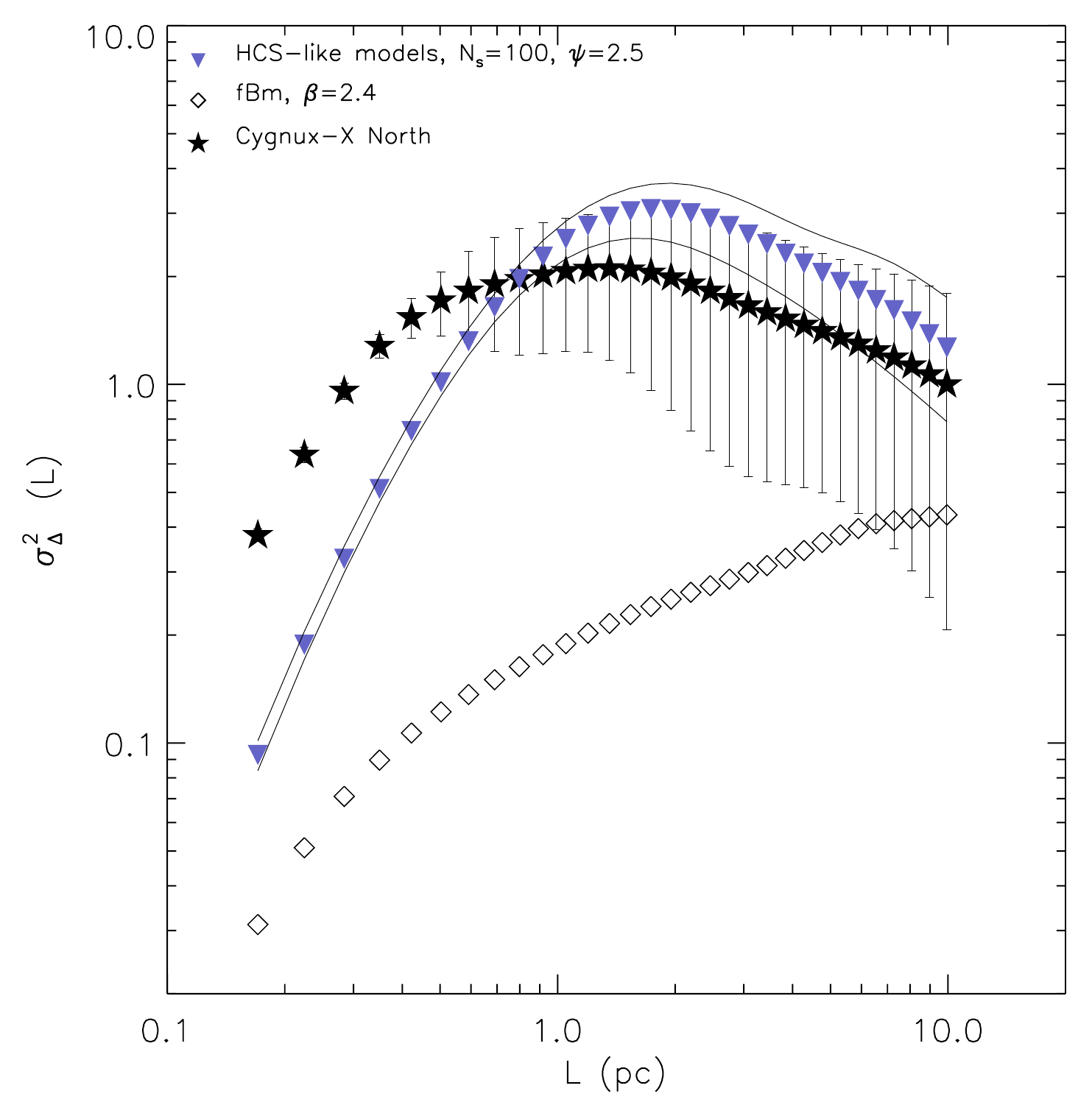}
\hspace{0.2cm} 
\includegraphics[width=0.32\textwidth] {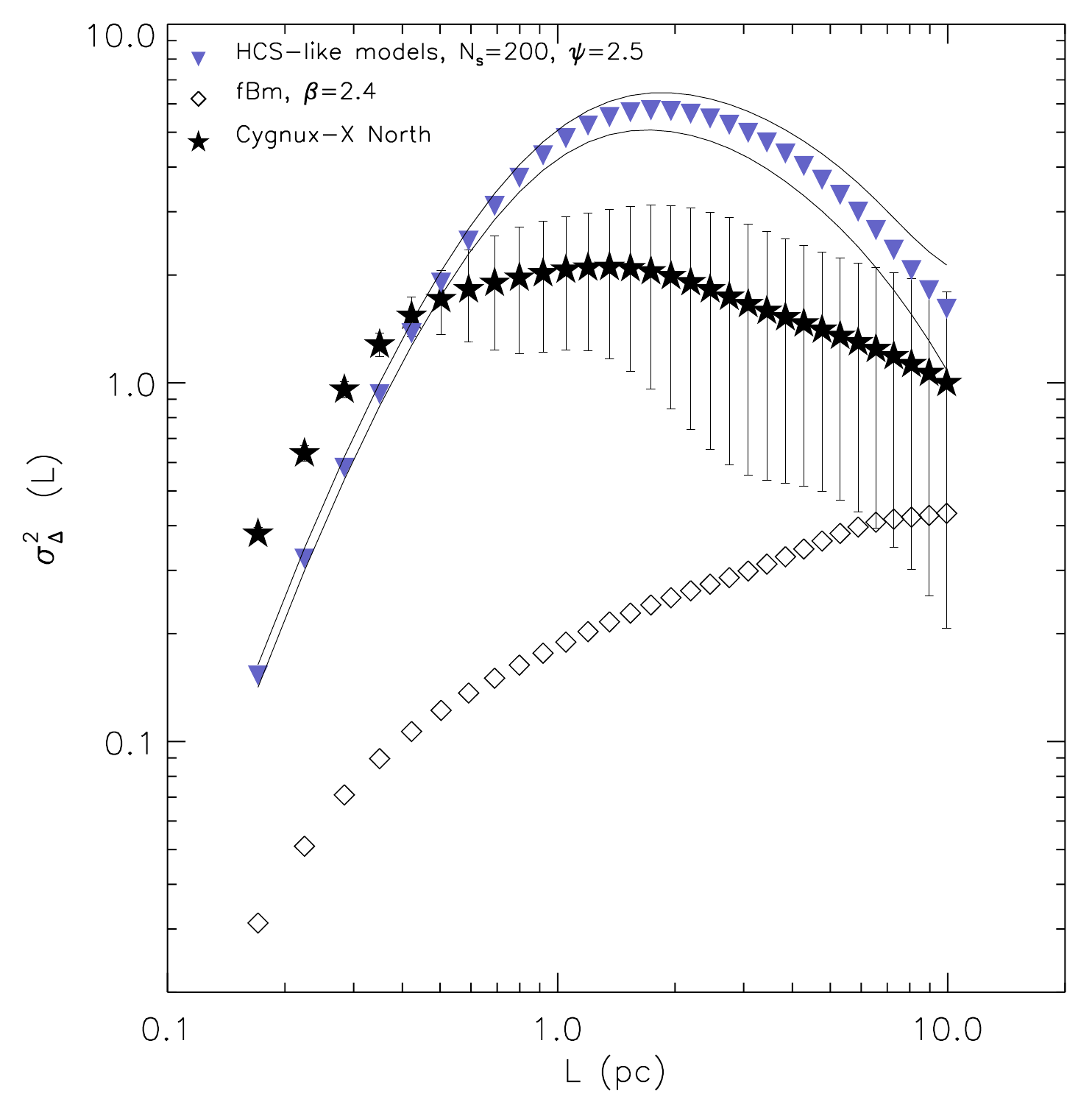}
\caption{Top: Examples of 50 (left), 100 (middle), and 200 (right) 2D Gaussian structures injected on top of an fBm image with $\beta=2.4$. The Gaussian structures are randomly sampled using the distribution functions of the major axis size and aspect ratio distributions of the HCS clouds. The column density contrasts are randomly sampled from a distribution with $\psi=2.5$. All synthetic maps are convolved with a beam whose FWHM=$18.2\arcsec$. Bottom: $\Delta$-variance spectrum of the synthetic models for cases with $N_{s}=50$ (left), $N_{s}=100$ (middle), and $N_{s}=200$ (right) injected structures. Each synthetic $\Delta$-variance spectrum is the average over 25 realizations and the full lines represent the $1 \sigma$ dispersion around the mean. The synthetic $\Delta$-variance spectra are compared to those of the Cygnus-X North region and an fBm with $\beta=2.4$.}
\label{fig13}
\end{figure*} 

\noindent where $A_{c}$ is a normalization coefficient that is given by $\int_{\delta_{c,min}}^{\delta_{c,max}} \left(dN/d\delta_{c}\right) d\delta_{c}=1$, and $\delta_{c,min}$ and $\delta_{c,max}$ are the lower and upper limits on $\delta_{c}$. We took $\delta_{c,min}=1$ in all cases and varied $\delta_{c,max}$ between 3 and 10. This is consistent with the range of values found by Arzoumanian et al. (2019) for filaments and with the range of column densities that are present in the high density tail of the N-PDF of the Cygnus-X North cloud. We chose three values of $\psi$, of $2$, $2.5$, and $3$, which, as an extrapolation of the results presented in Arzoumanian et al. (2019), should cover both variations due to differences in the clouds environmental conditions and variations due to temporal evolution (i.e., self-gravitating structures will have, statistically, higher column density contrasts at time goes by).

As stated above, we would like to understand the sensitivity of the $\Delta$-variance spectrum in relation to the underlying distribution functions of the different parameters, and, as a by-product, understand which particular set of parameters can help generate a $\Delta$-variance spectrum that resembles the one found in Cygnus-X North. In principle, the parameter space is relatively large with four free parameters to probe ($N_{s}, \eta, \xi$, and $\psi$), and even larger if the lower and upper limits on $f$, $L_{1}$, and $\delta_{c}$ are also varied. As a first step, we explore below models of synthetic clouds whose properties are inspired from the Hi-GAL and the HCS samples. In a second step, we expand this "forward modeling" approach and present a broader parameter study where we vary, in a more systematic fashion, the parameters of the distributions functions. 
 
For any given choice of ($N_{s}, \eta, \xi, \psi$), and owing to the fact that the orientations and positions of the injected structures are random, it is important that, for each choice of the parameters, a statistically significant number of realizations is performed in order to capture the mean behavior and standard deviation around the mean of the $\Delta$-variance spectrum. We chose to perform $25$ realizations with any given set of the parameters. Furthermore, and owing to computational limitations, we performed the synthetic calculations on maps with $1000\times1000$ pixels, whereas the map of Cygnus-X North has $5740\times5740$ pixels. We assigned the same physical size to each pixel in the synthetic maps as in the observations ($\approx 0.025$ pc), and we matched the mean $A_{V}$ in the synthetic maps to the mean $A_{V}$ of the Cygnus-X North cloud. Because we are comparing the results from the synthetic maps to the observations of the Cygnus-X North region, all synthetic maps (i.e., both the reference fBm and the fBm plus structures maps) were convolved with a beam similar to that of the observations. The beam is represented by a Gaussian function whose FWHM=$18.2\arcsec$. We compared the $\Delta$-variance spectra of the synthetic maps and the observations on scales that are larger than the beam size, namely scales that are $\geq 0.15$ pc.  

\subsubsection{Hi-GAL- and HCS-like clumps}

\begin{figure*}
\centering
\includegraphics[width=0.2\textwidth] {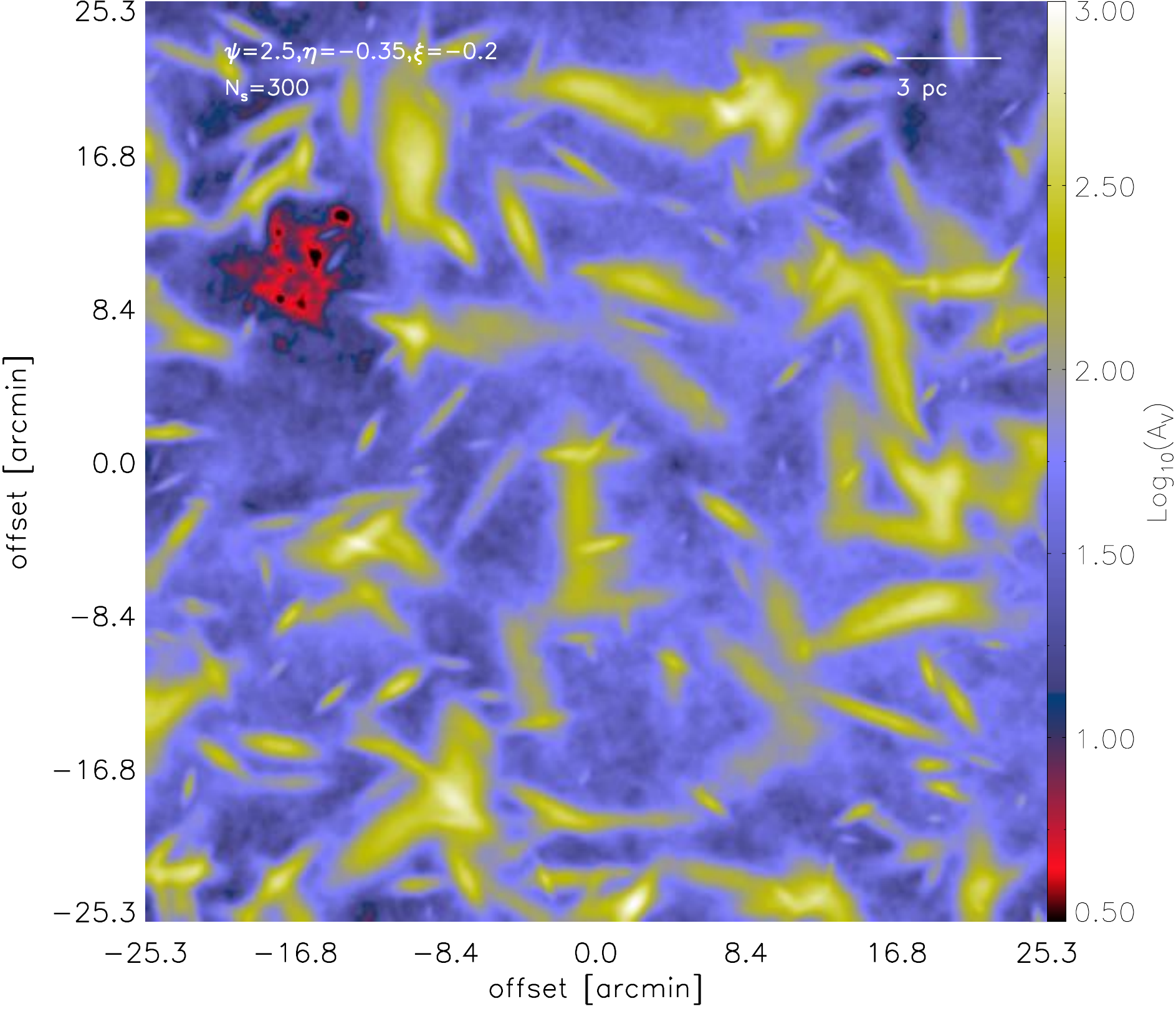}
\hspace{1cm}
\includegraphics[width=0.2\textwidth] {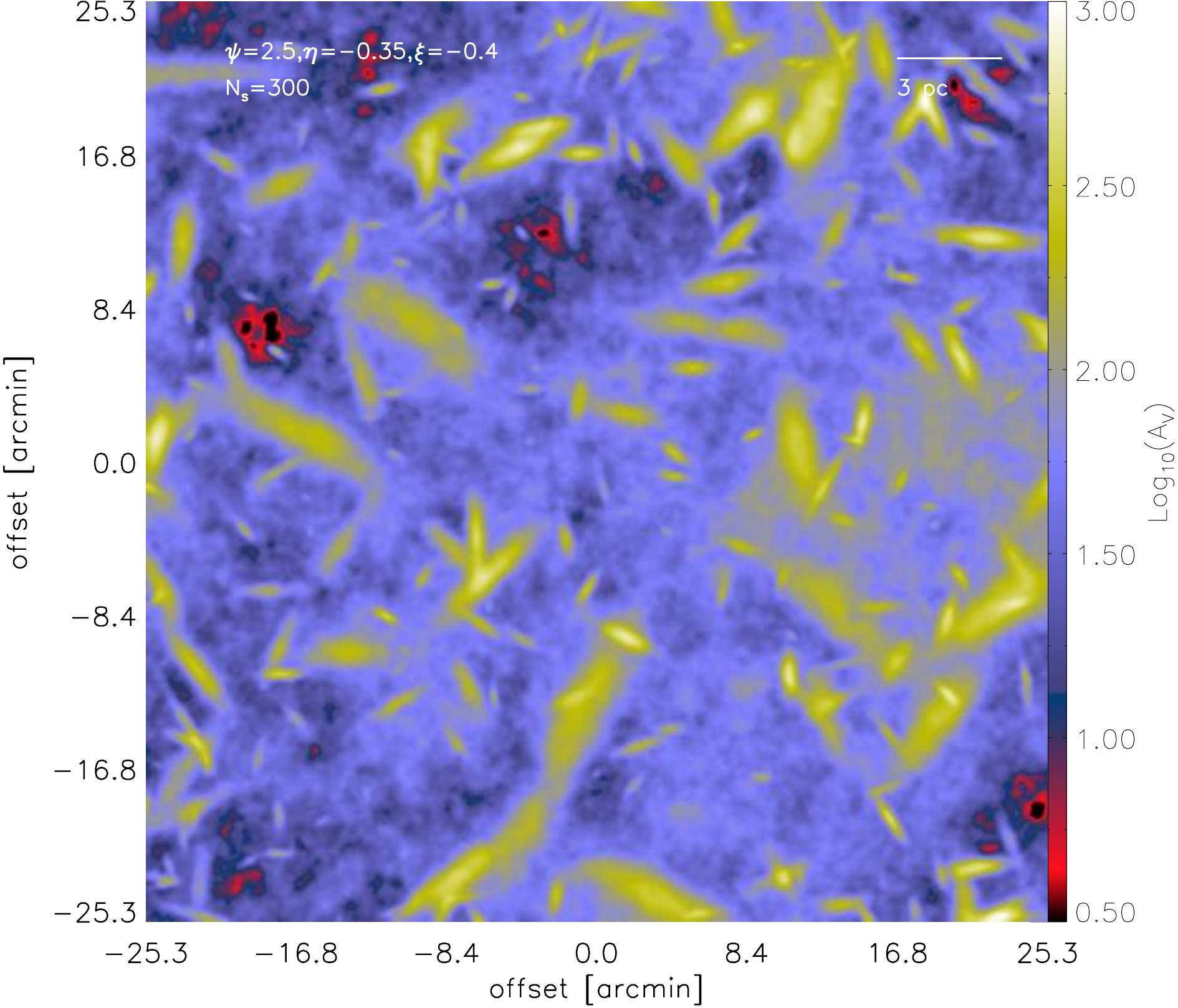}
\hspace{1cm}
\includegraphics[width=0.2\textwidth] {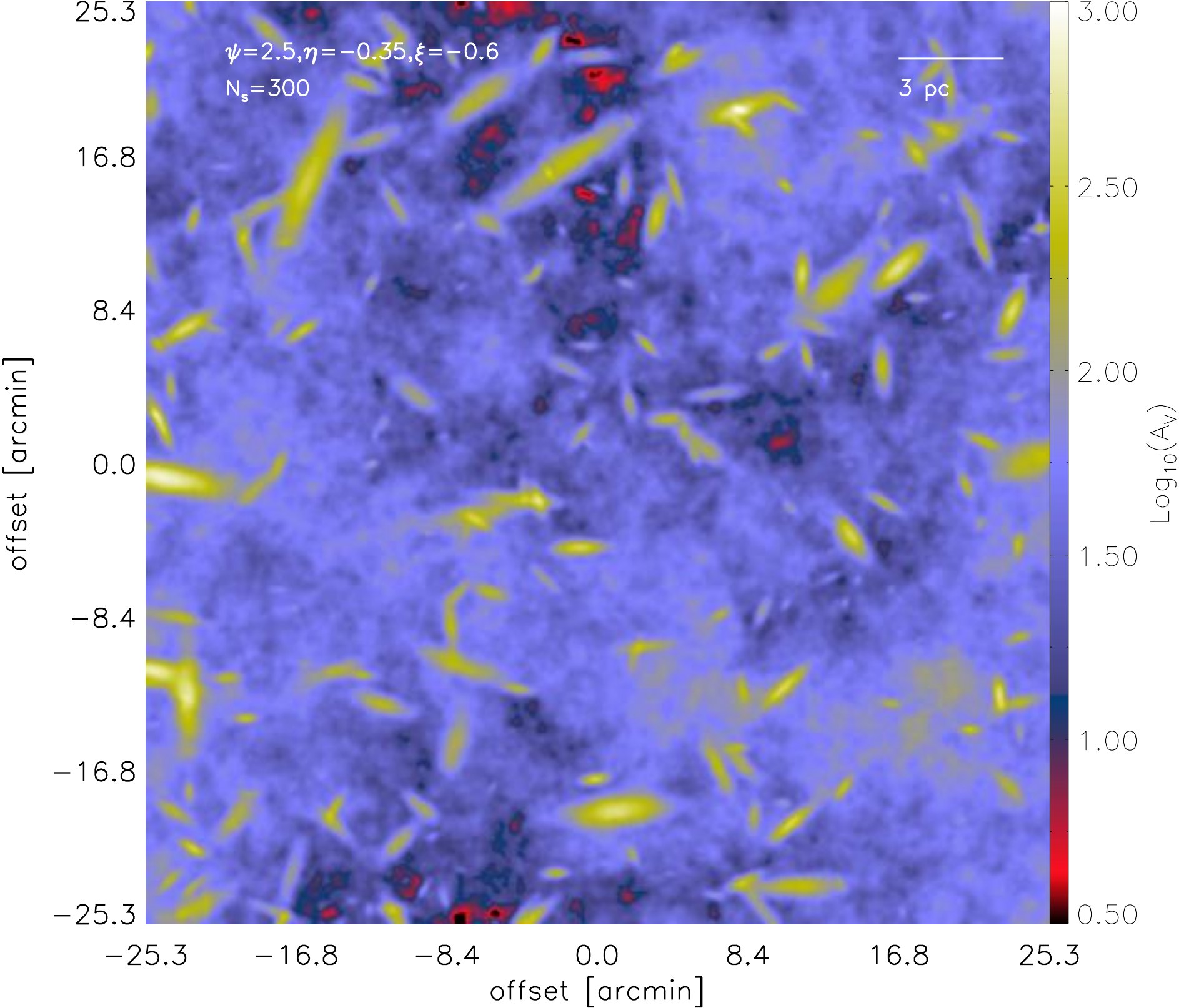}
\hspace{1cm}
\includegraphics[width=0.2\textwidth] {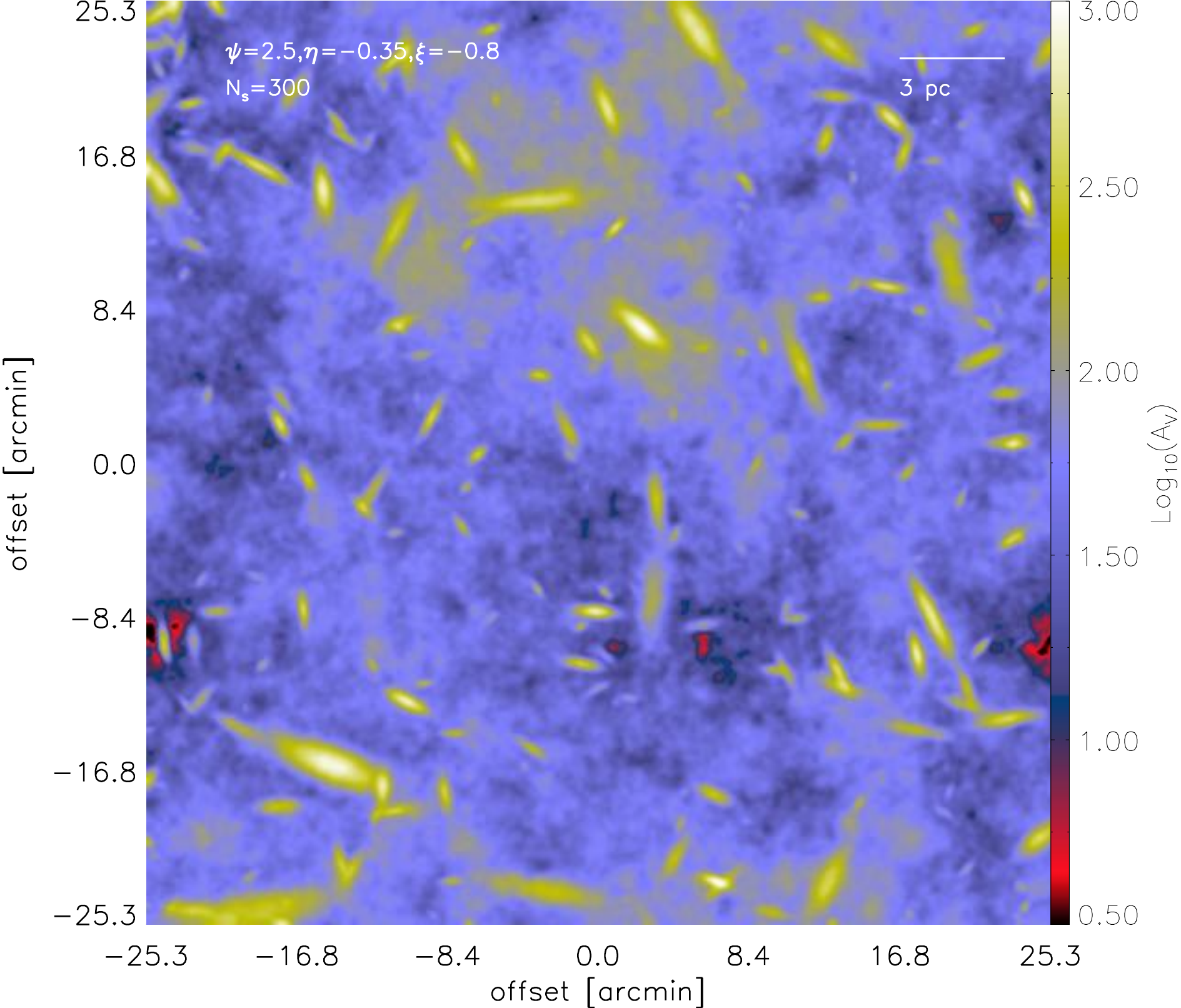}\\
\vspace{0.5cm}
\includegraphics[width=0.2\textwidth] {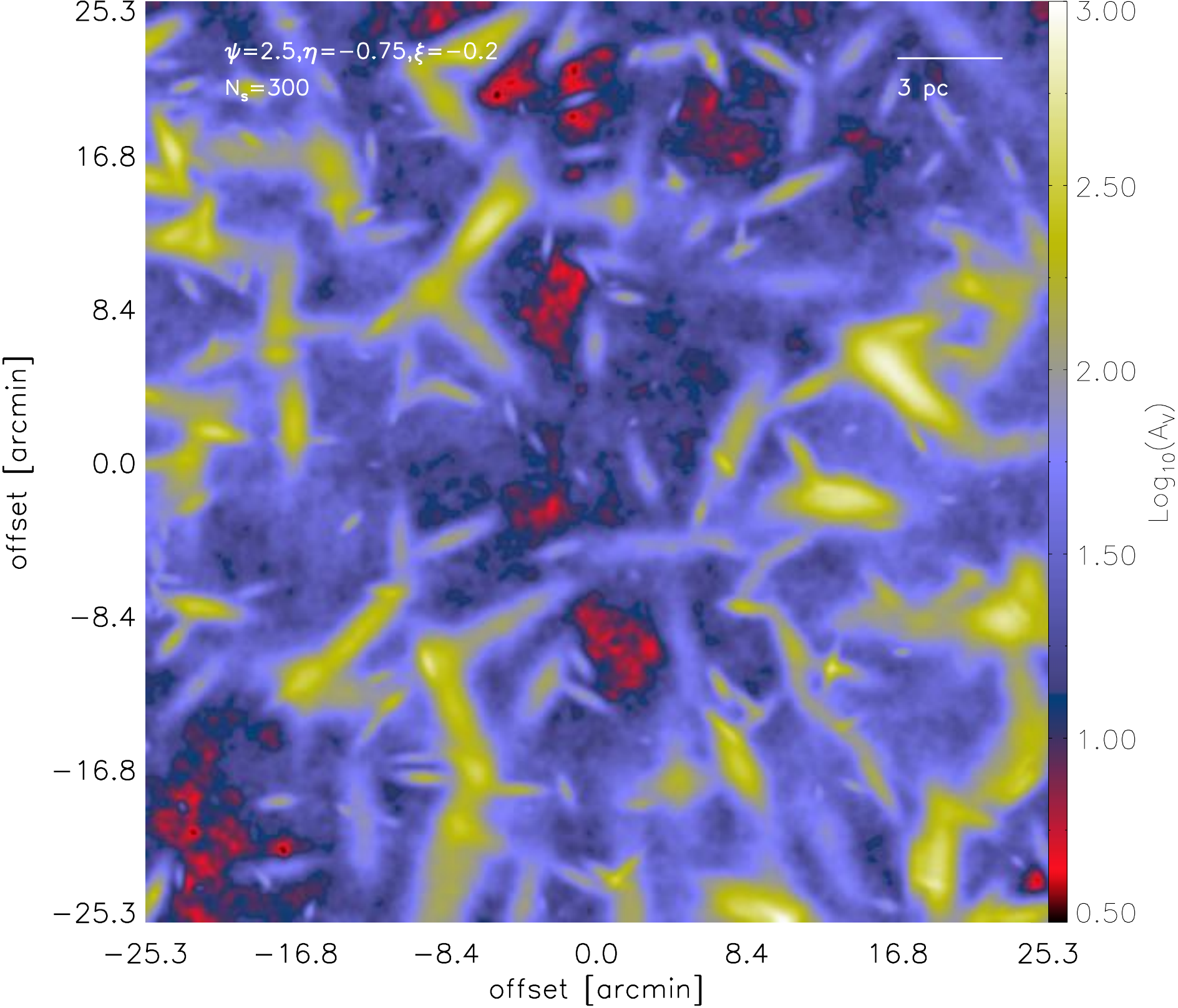}
\hspace{1cm} 
\includegraphics[width=0.2\textwidth] {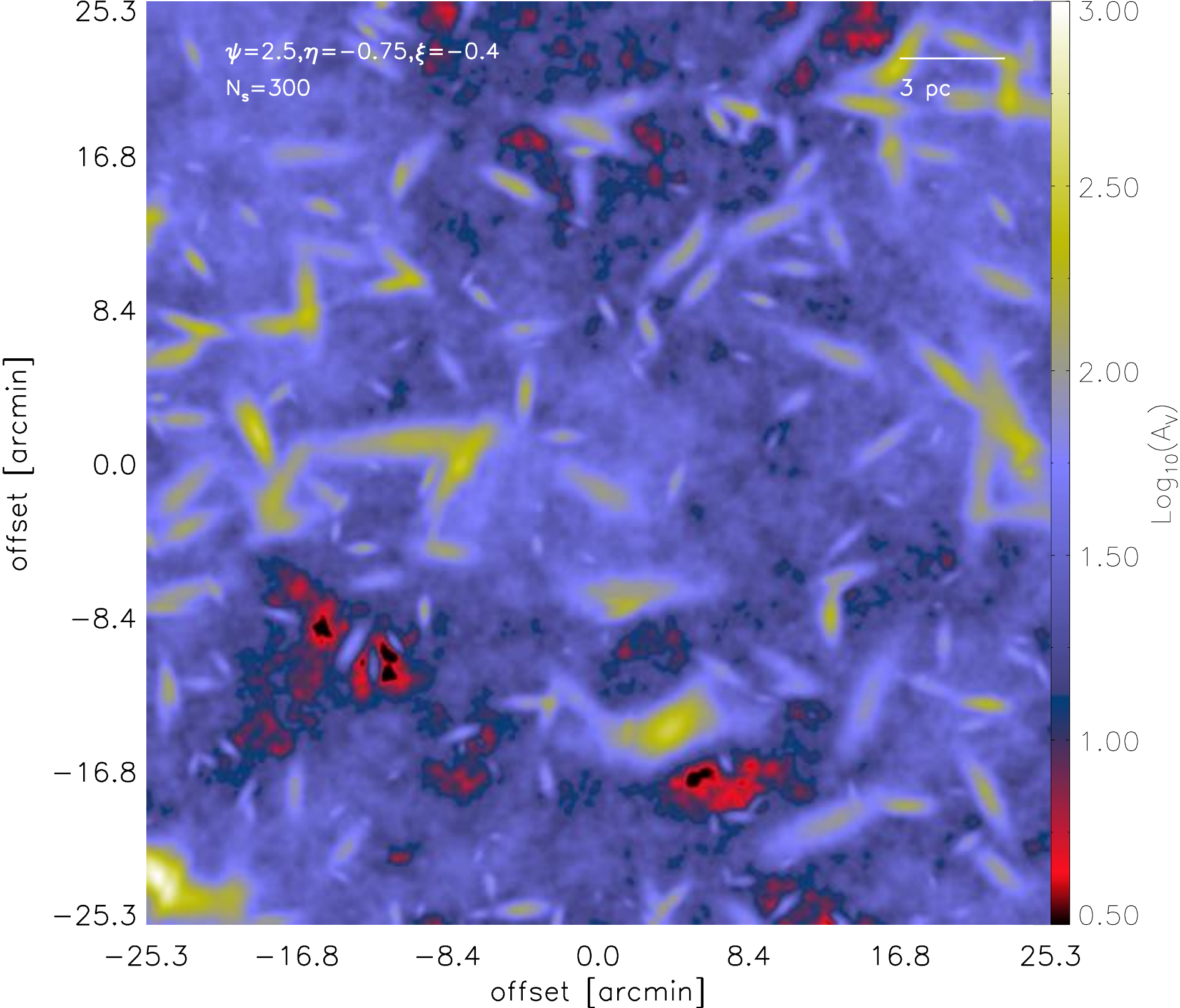}
\hspace{1cm}
\includegraphics[width=0.2\textwidth] {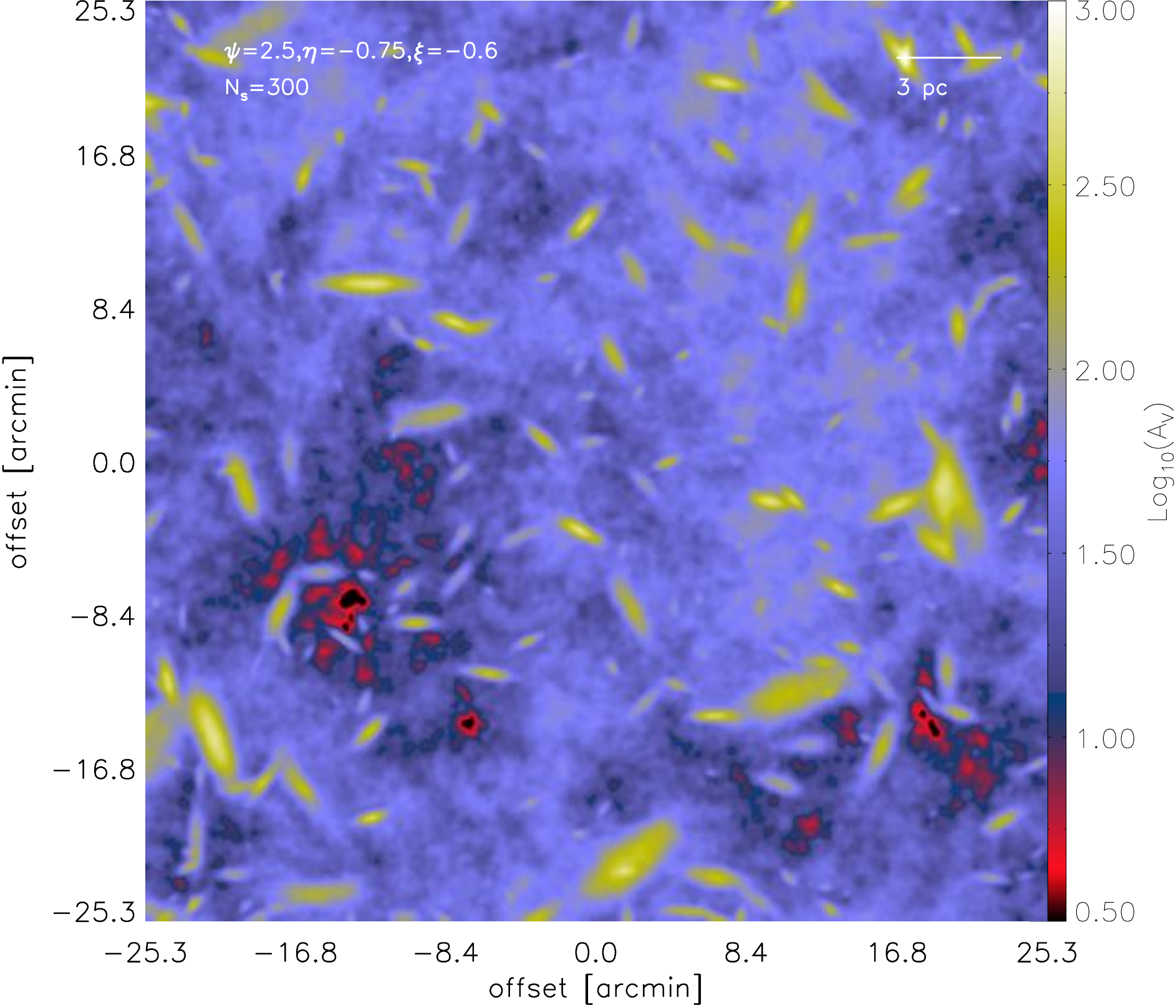}
\hspace{1cm} 
\includegraphics[width=0.2\textwidth] {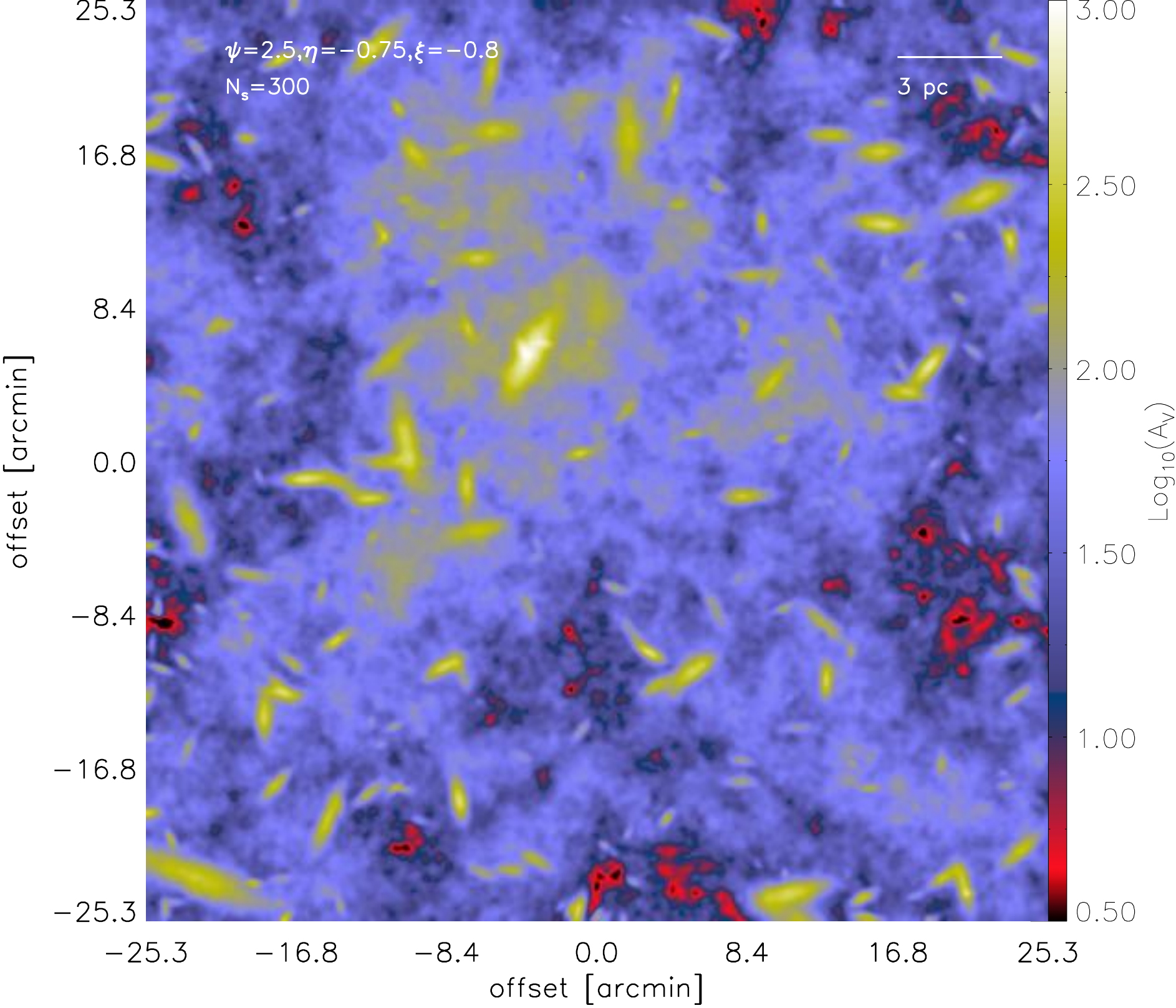}\\
\vspace{0.5cm}
\includegraphics[width=0.2\textwidth] {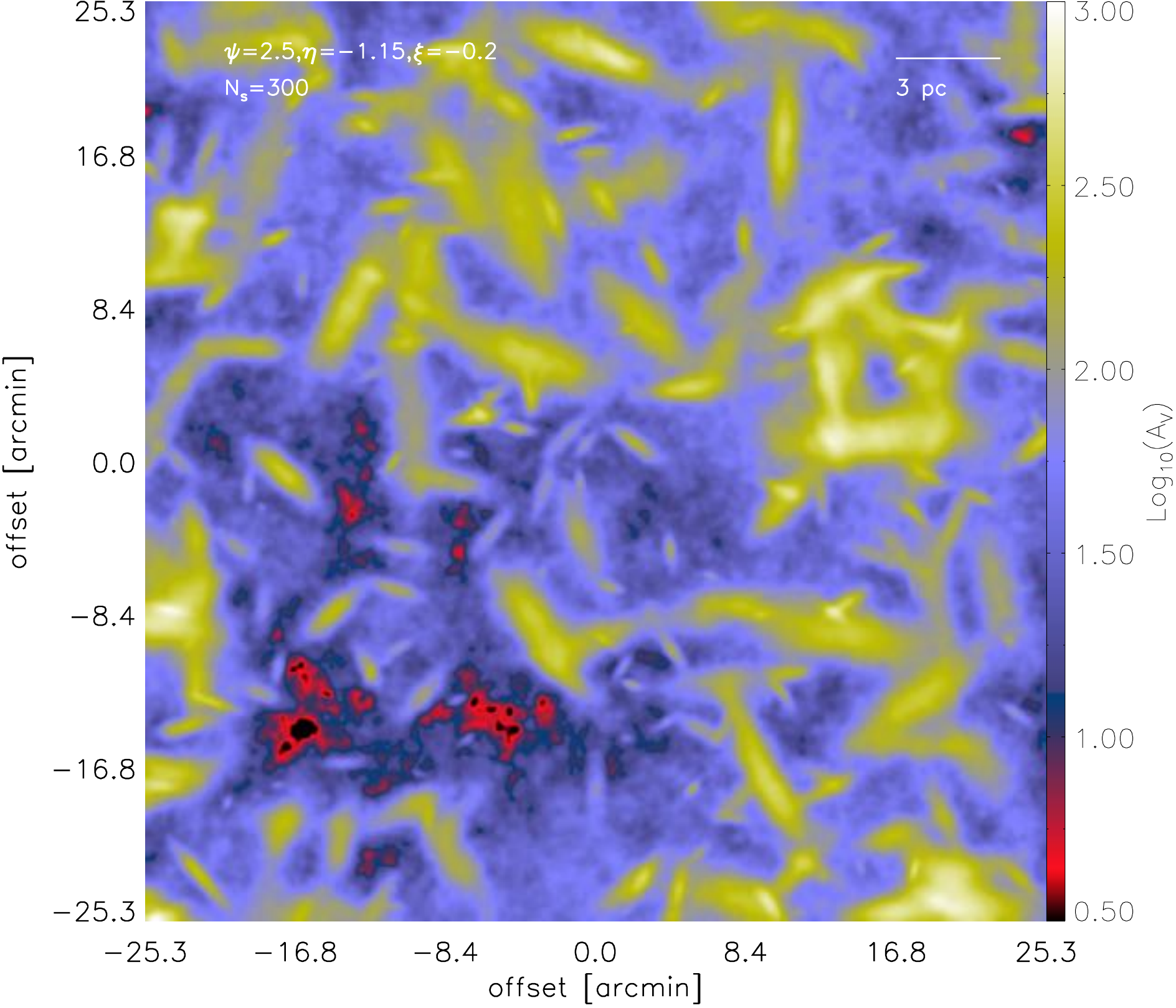}
\hspace{1cm} 
\includegraphics[width=0.2\textwidth] {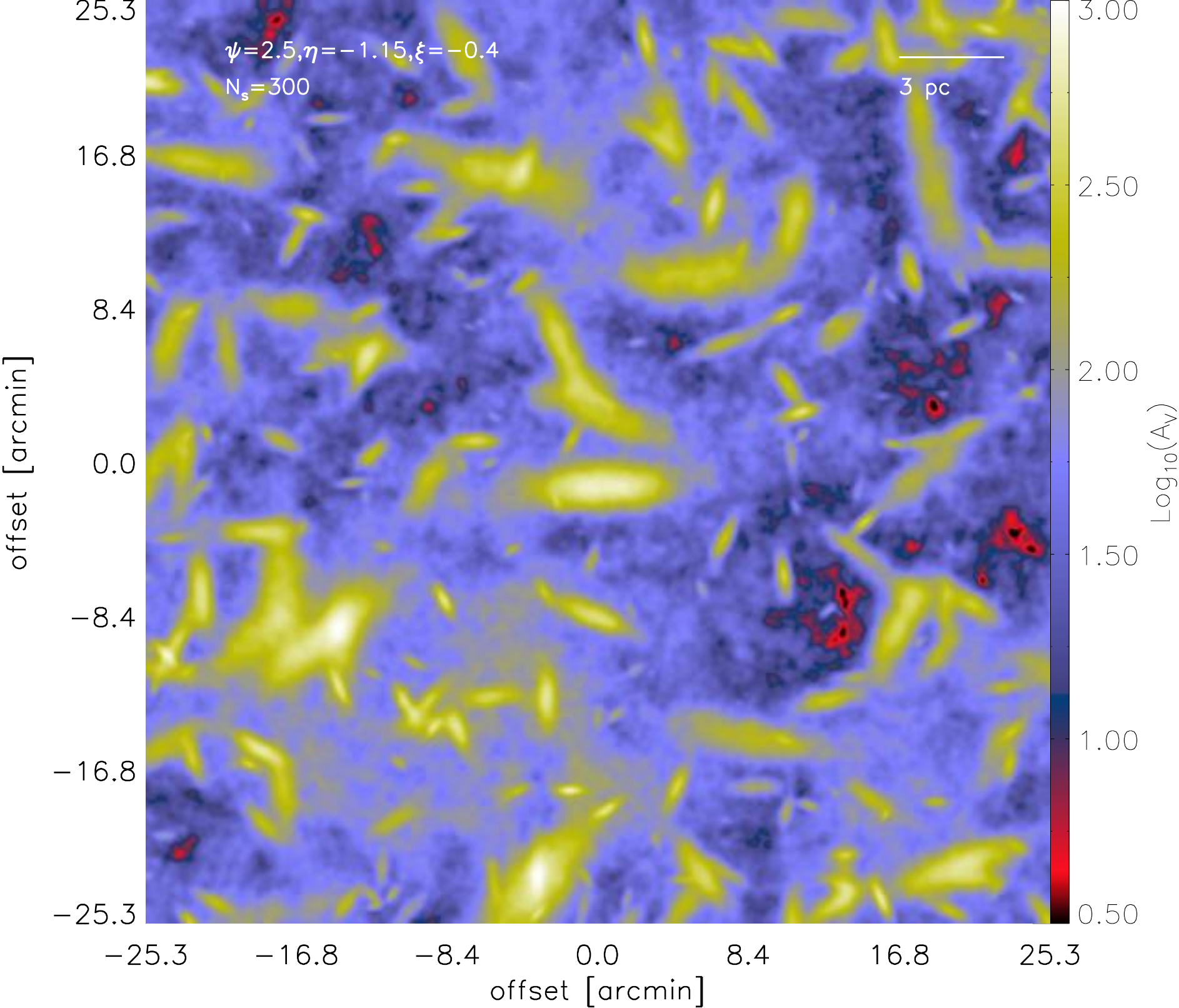}
\hspace{1cm}
\includegraphics[width=0.2\textwidth] {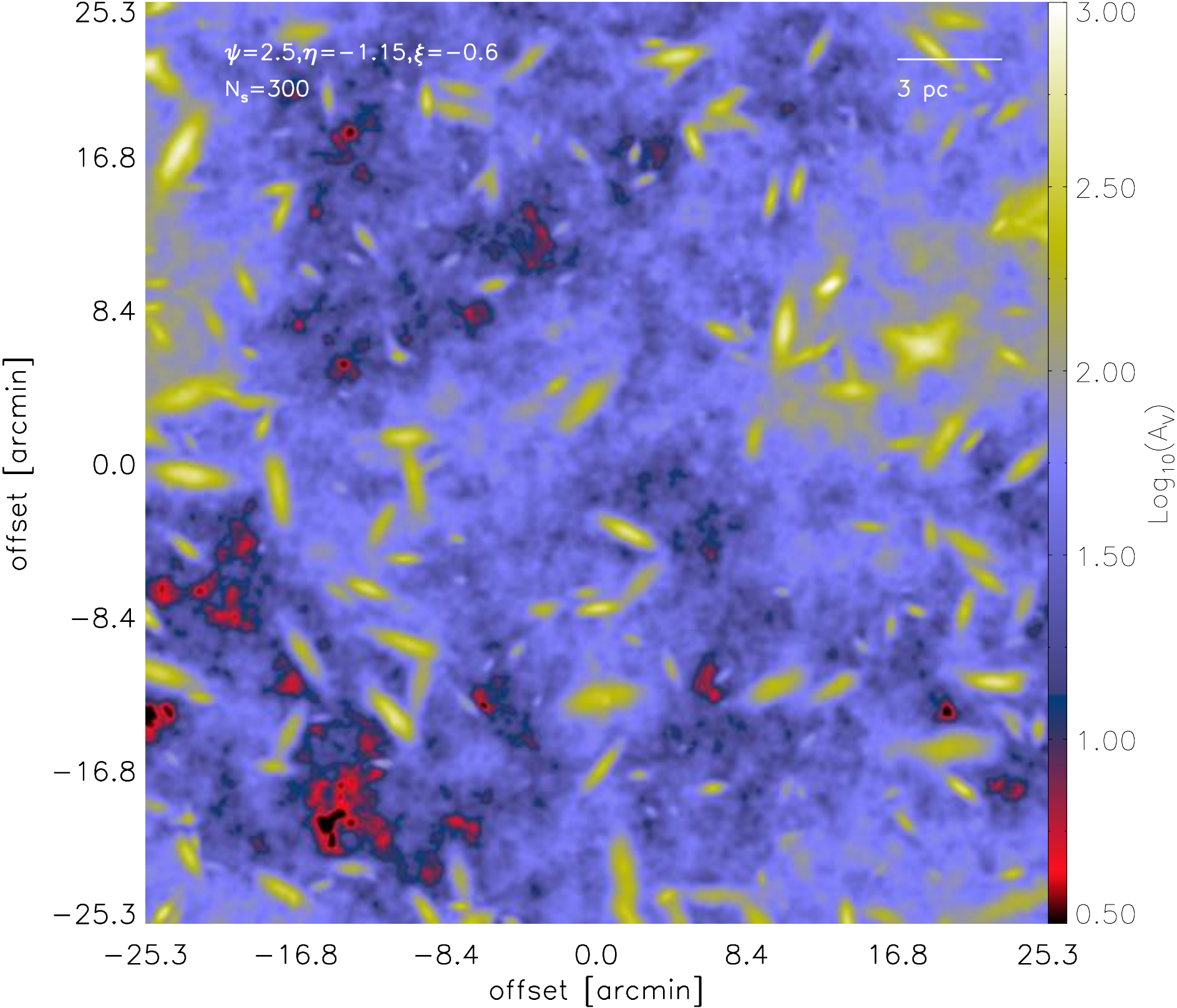}
\hspace{1cm} 
\includegraphics[width=0.2\textwidth] {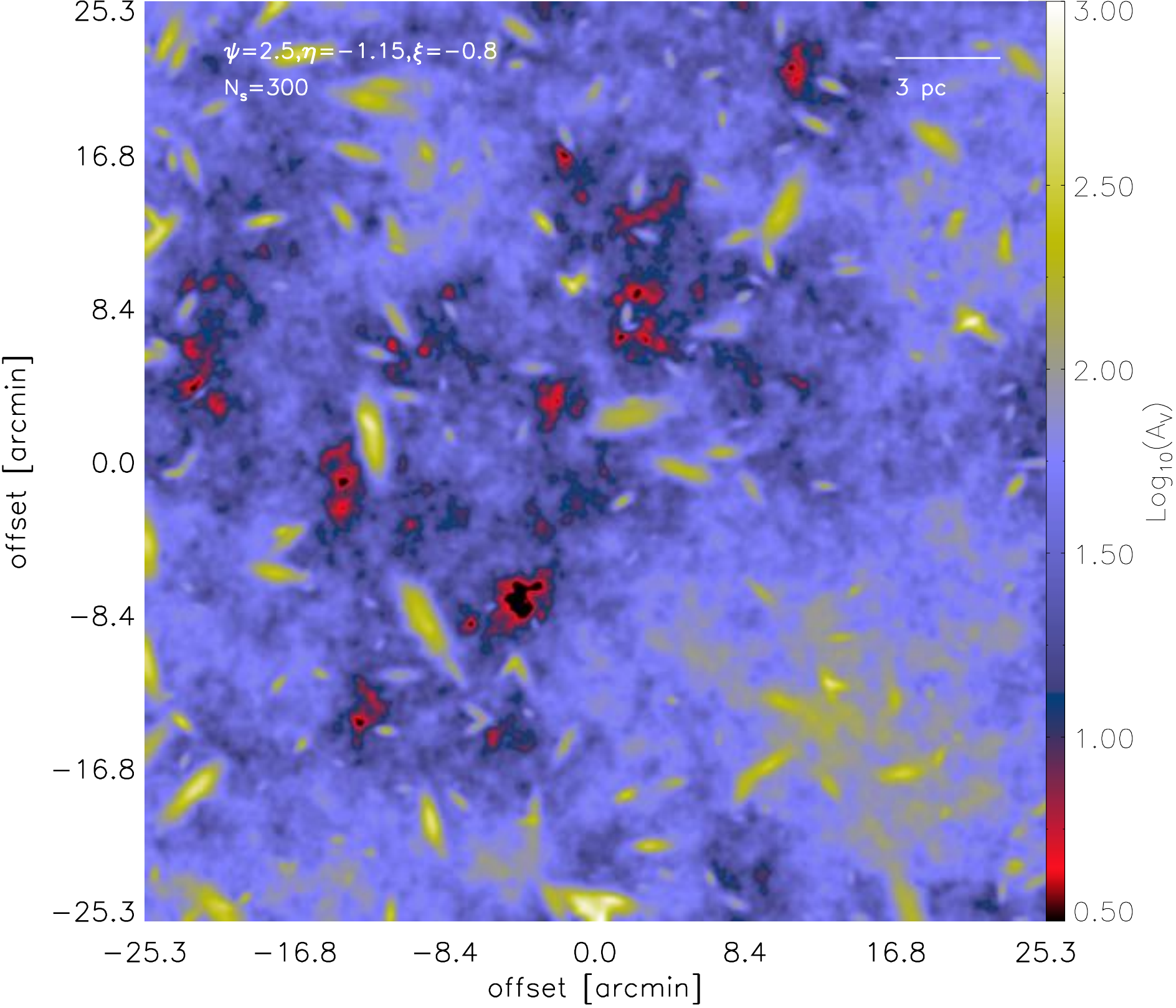}
\vspace{0.5cm}
\caption{Examples of synthetic maps generated by overlaying structures (2D Gaussians, $N_{s}=300$) on top of an fBm image with $\beta=2.4$. The properties of the structures are randomly sampled from distribution functions of the aspect ratio (Eq.~\ref{eq9}), size of the major axis (Eq.~\ref{eq10}), and column density contrast (Eq.~\ref{eq11}). All models shown here share the same values of $\delta_{c}=2.5$, ($L_{1,min}=0.025~{\rm pc},L_{1,max}=5~{\rm pc}$), ($f_{min}=3,f_{max}=12$), and ($\delta_{c,min}=1,\delta_{c,max}=3$). All synthetic maps are convolved with a beam whose FWHM=$18.2\arcsec$.}
\label{fig14}
\end{figure*} 

\begin{figure*}
\centering
\includegraphics[width=0.94\textwidth]{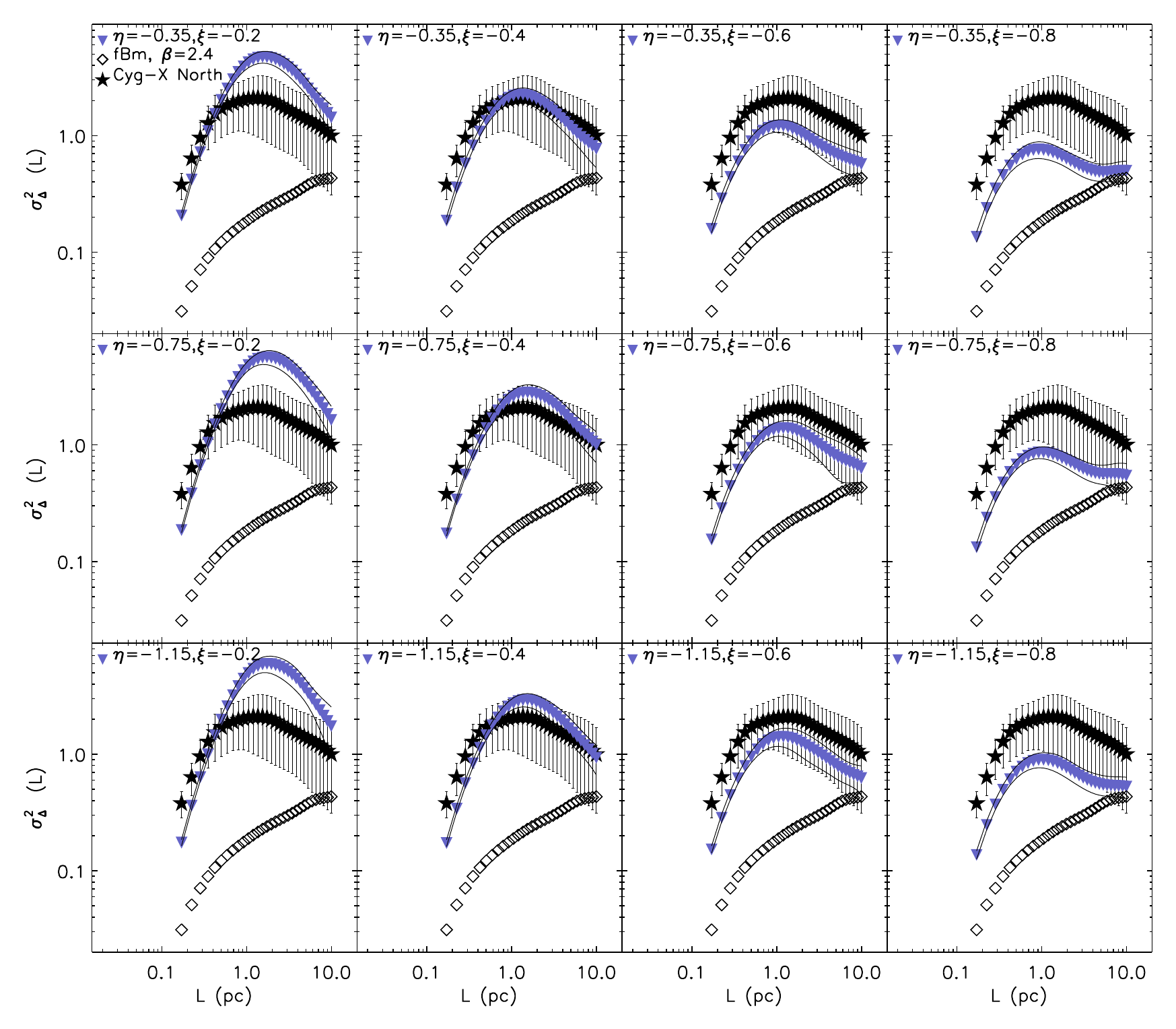}
\vspace{0.8cm}
\caption{$\Delta$-variance spectra related to the models presented in Fig.~\ref{fig14}. Each synthetic spectrum (blue triangles) is the average over 25 realizations of the maps with the same set of parameters. In all models, $N_{s}=300$ and $\psi=2.5$. The full line is the $1 \sigma$ dispersion around the mean. The synthetic $\Delta$-variance spectra are compared to that of the Cygnus-X North region and that of an fBm image with $\beta=2.4$. In the grid of models, the best fit to the data of Cygnus North is for the case with $\xi=-0.4$ and $\eta=-0.35$.}
\label{fig15}
\end{figure*}

\begin{figure}
\centering
\includegraphics[width=\columnwidth]{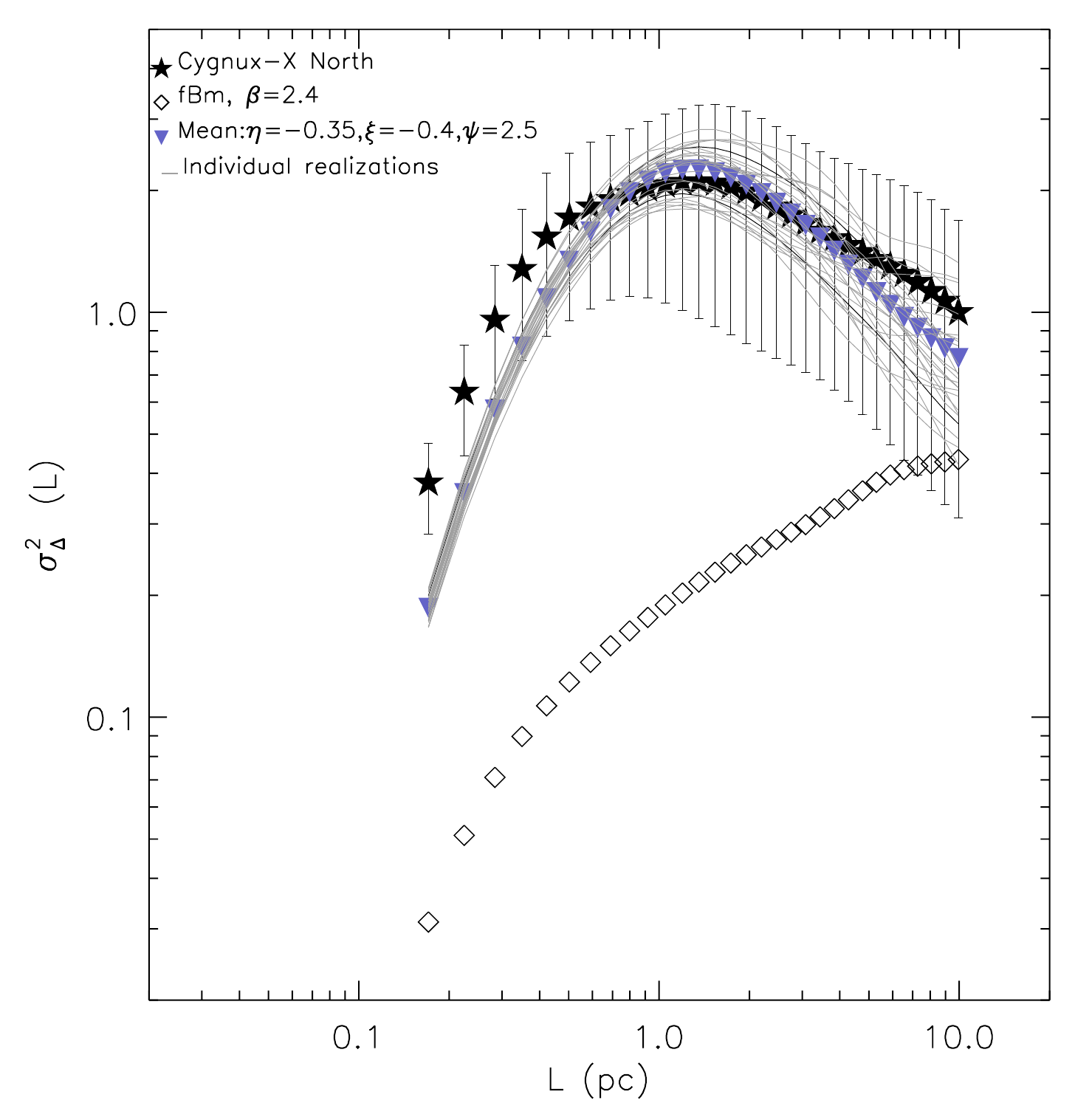}
\caption{Same as the figure with $\xi=-0.4$, $\eta=-0.35$, $\psi=2.5$, and $N_{s}=300$ displayed in Fig.~\ref{fig15} but additionally showing the $\Delta$-variance spectra of individual realizations with this set of parameters.}
\label{fig16}
\end{figure}

As illustrative examples, we first explored the resulting $\Delta$-variance spectra corresponding to population of clumps and cores similar to the ones found in the Hi-GAL (submm) and HCS (CO) surveys. We recall that the values of $\xi$ and $\eta$ are [$-0.04,-0.37$] and [-0.89,-1.15] for the Hi-GAL clouds and the CO clouds, respectively, which implies that the CO structures in the HCS survey are both larger and more elongated than structures detected in the Hi-GAL survey. The lower and upper limits on the aspect ratios for the submm-like and CO-like clouds are taken to be ($f_{min}=1,f_{max}=4$) and ($f_{min}=3,f_{max}=12$), as per the observational constraints (Fig.~\ref{fig11}, middle panel). For the lower limits on the major axis size, we extrapolated the major axis size distributions for both the submm-like and CO-like clouds down to the resolution limit such that $L_{1,min}=0.025$ pc\footnote{In Cygnus-X North, the real lower limit on the sizes of the structures and the true shape of the size distribution are very uncertain in the regime of small sizes. The choice of $L_{1,min}=0.025$ pc basically means no lower limit (as it is the smallest resolved structure), while all other higher values would be questionable.}. For the upper limit on $L_{1}$, we adopted a common value of $L_{1,max}=5$ pc for both the Hi-GAL-like and  HCS-like clouds. Adopting a larger value of $L_{1,max}$ (up to $\approx 70$ pc) for the HCS-like clouds would be excessive on the grounds that Cygnus-X North does not contain any such large structures, and this is further motivated by the fact that we are considering a region that is $5.74$ smaller, in each direction, than the real map. In the absence of additional information from the surveys, we imposed, in both cases, a value of $\psi=2.5$ and lower and upper bounds on $\delta_{c}$ of 1 and 3, respectively. 

The top panel in Fig.~\ref{fig12} displays three examples of the Hi-GAL-like maps generated with $N_{s}=200$ (left), 300 (middle), and 400 structures (right) out of a total of 25 realizations performed for each case. The size of the major axis, aspect ratio, and column density contrast for each individual structure are randomly sampled from the corresponding distribution functions, and the structures are assigned random positions and orientations and overlaid on top of an fBm image with $\beta=2.4$. The bottom panel in Fig.~\ref{fig12} displays the corresponding $\Delta$-variance spectra, which are calculated, in each case, as the mean spectrum from the $25$ realizations (blue triangles). The $\Delta$-variance spectra for the synthetic maps are compared to the spectrum of the underlying fBm (open diamonds) and to that of the Cygnus-X North cloud (filled stars). As observed earlier in the case where individual structures are injected (in Sect.~\ref{fbmplus}), the $\Delta$-variance spectrum in the presence of structures shows a departure from that of the underlying fBm, and, in the case of an entire population of structures, the point of maximum departure from the underlying fBm case corresponds to the characteristic scale of the ensemble of structures that are injected onto the map. Figure~\ref{fig12} shows that a better agreement is obtained for $N_{s}=300$. We followed the same procedure and generated structures similar to those found in the HCS survey. Three examples of such maps with $N_{s}=50$, $100$, and $200$ are displayed in Fig.~\ref{fig13} (top panel). The HCS-like structures have a shallower spectrum of major axis sizes, and the corresponding $\Delta$-variance spectrum peaks at higher spatial scales than their Hi-GAL-like counterparts (Fig.~\ref{fig13}, bottom panel). Figures~\ref{fig12} and \ref{fig13} show that while the Hi-GAL-like structures provide a better match to the observations of Cygnus-X, neither of these two cloud samples fit the observations of Cygnus-X North well. However, it is useful to compare these two cases to Cygnus-X in order to highlight how structures with fundamentally different statistical properties impact the $\Delta$-variance spectrum. These cases also provide a starting point for a more detailed exploration of the parameter space. 

\subsubsection{Parameter study} 

\begin{figure*}
\centering
\hspace{0.5cm}
\includegraphics[width=0.25\textwidth] {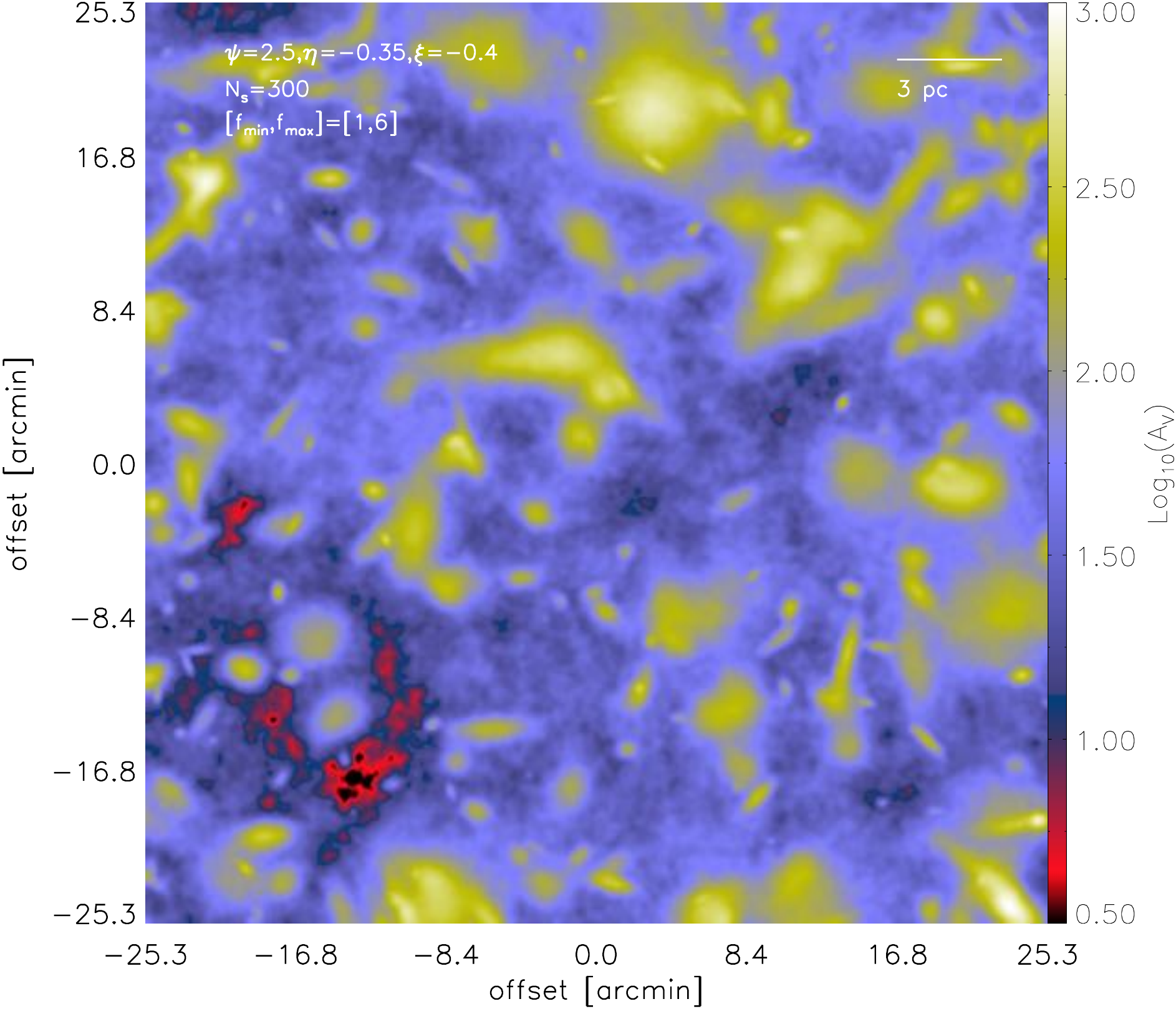}
\hspace{1.5cm}
\includegraphics[width=0.25\textwidth] {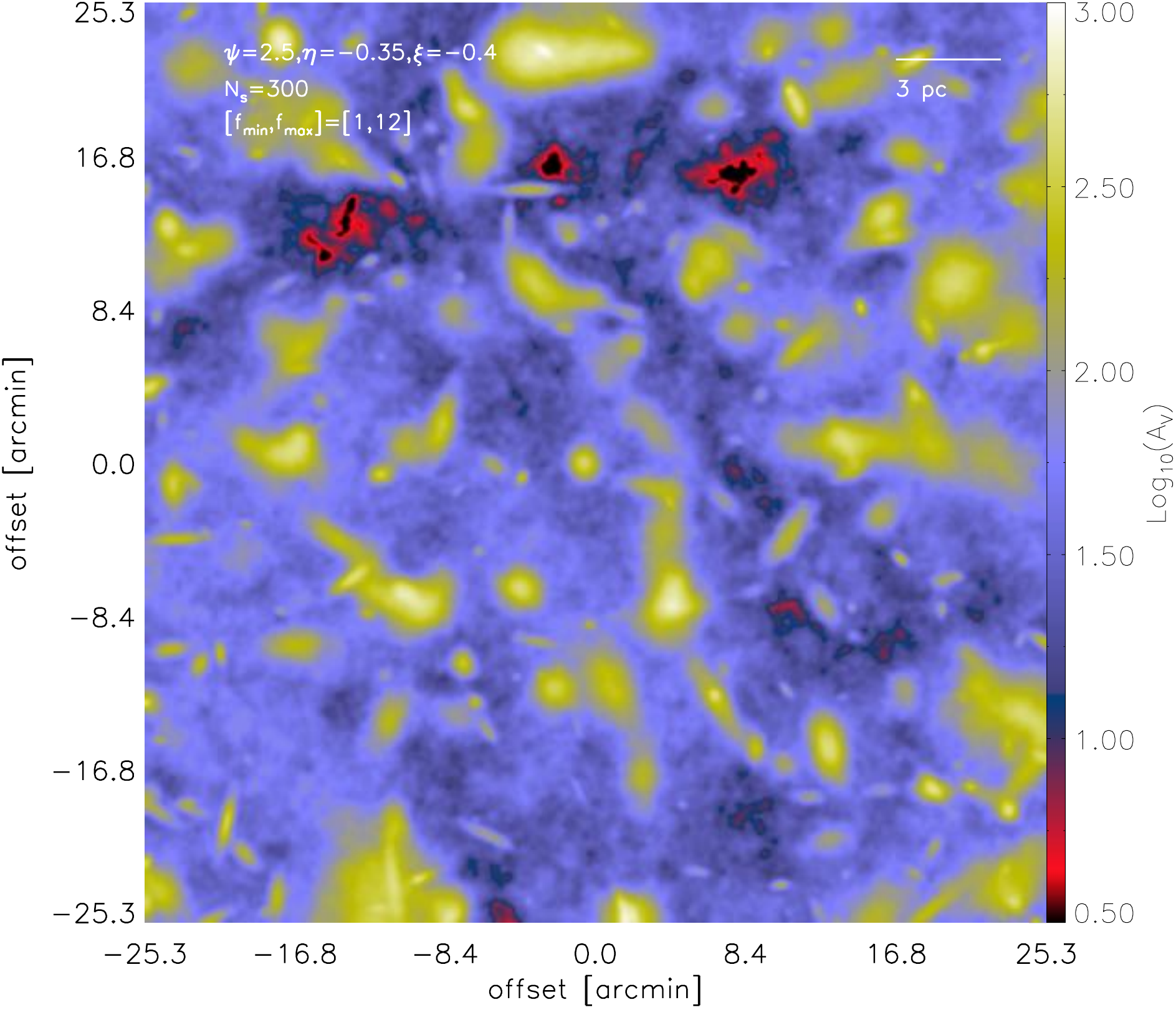}
\hspace{1.5cm}
\includegraphics[width=0.25\textwidth] {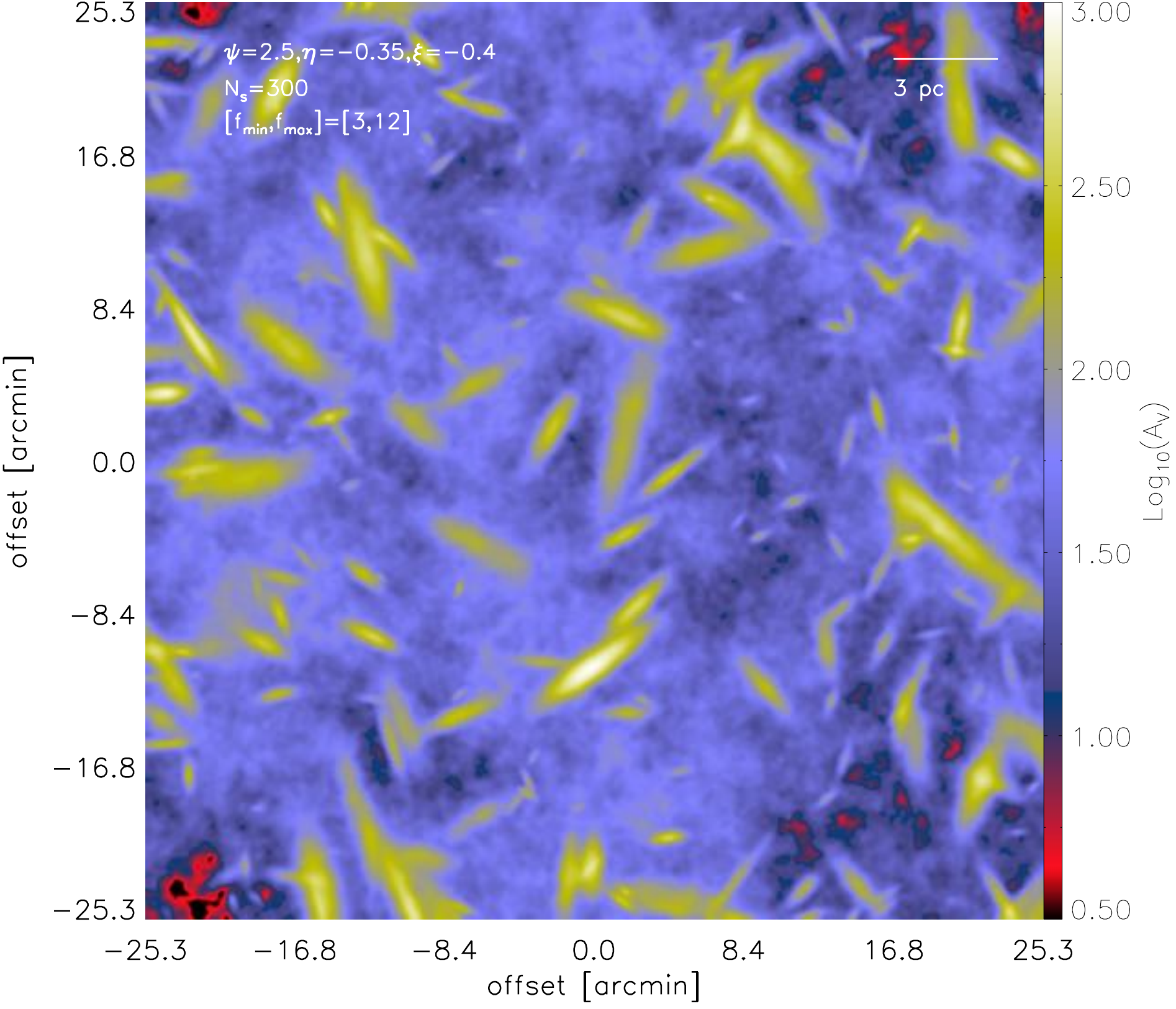}\\
\vspace{0.5cm}
\includegraphics[width=0.32\textwidth] {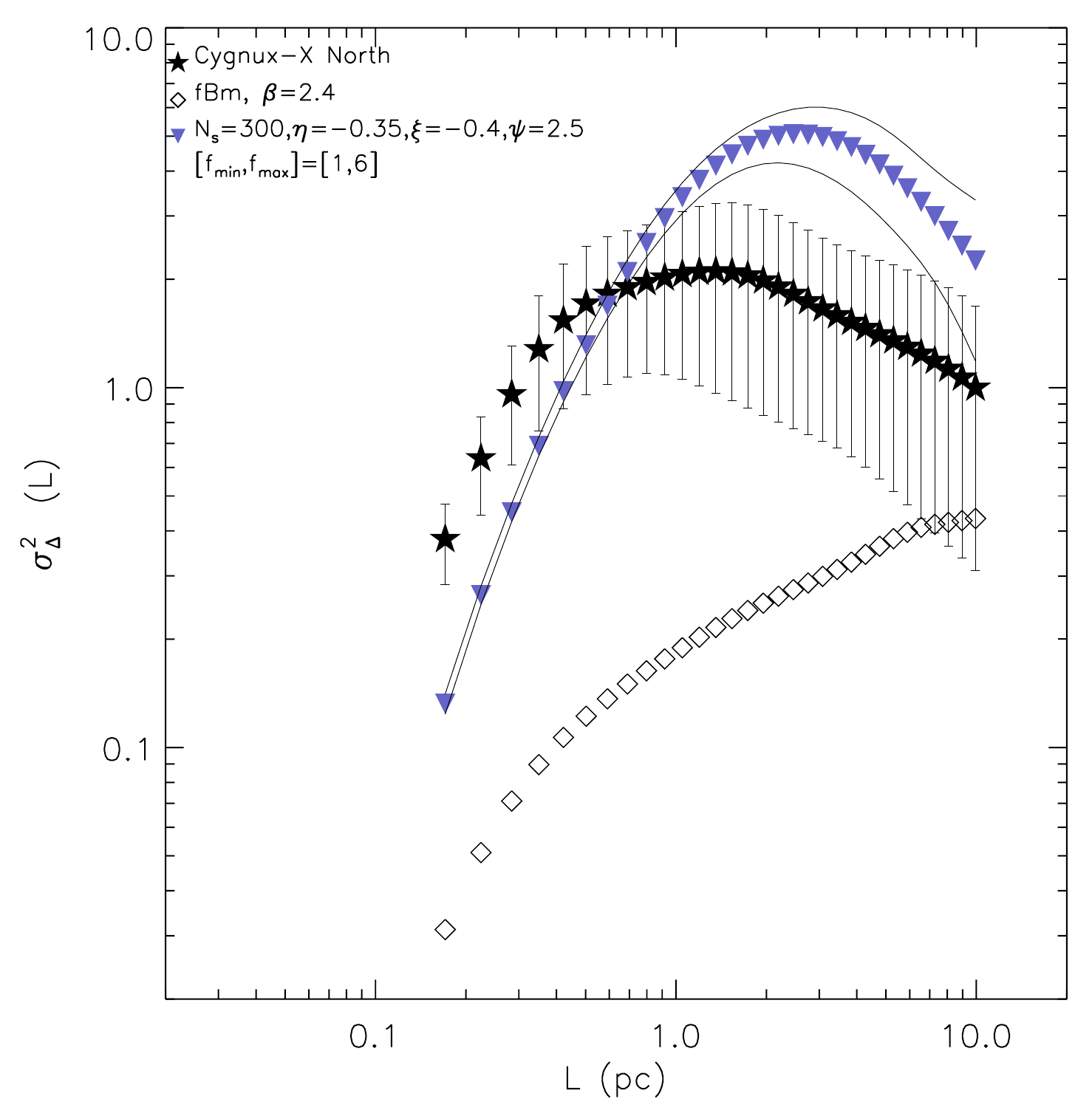}
\hspace{0.2cm}
\includegraphics[width=0.32\textwidth] {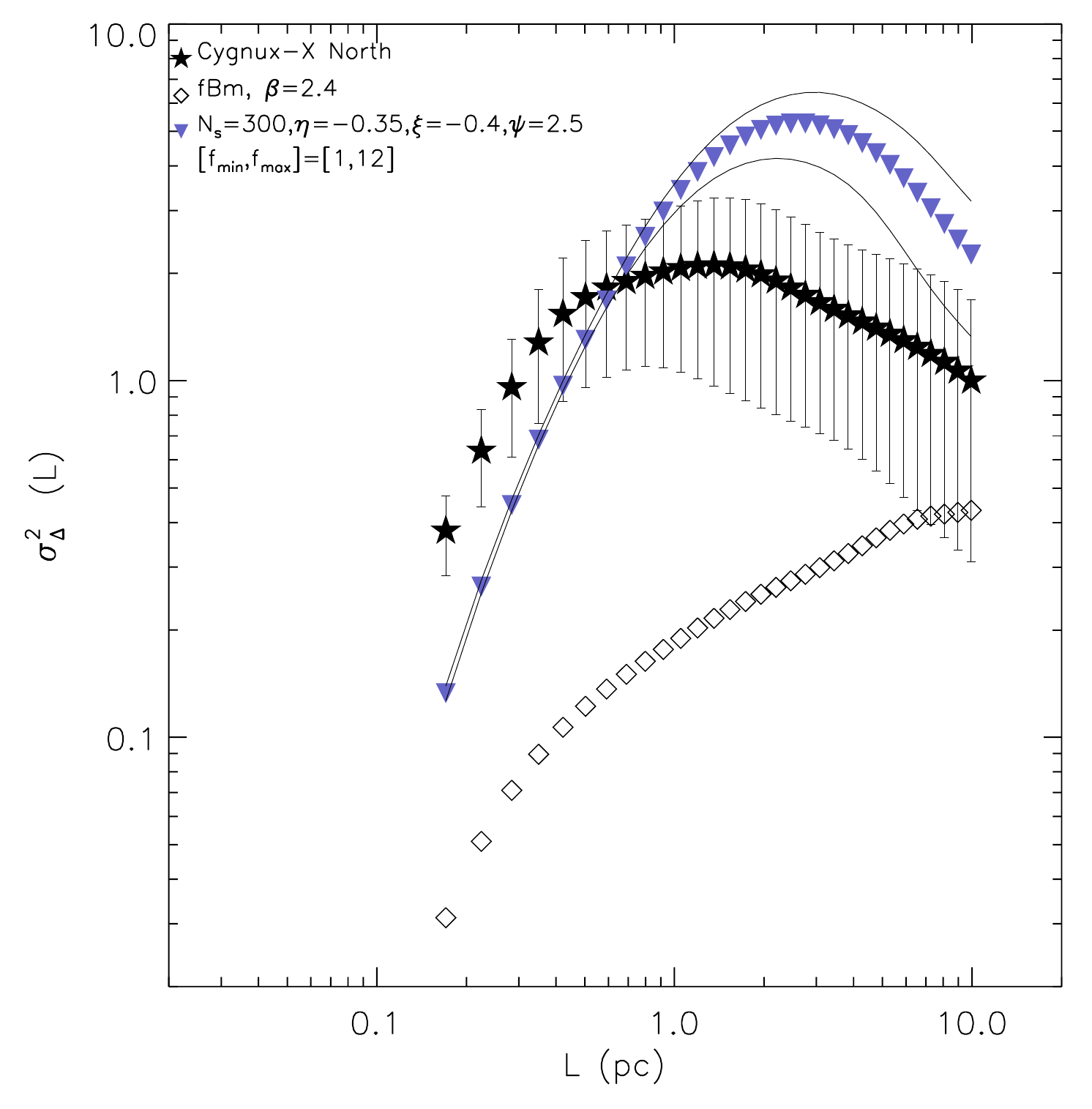}
\hspace{0.2cm} 
\includegraphics[width=0.32\textwidth] {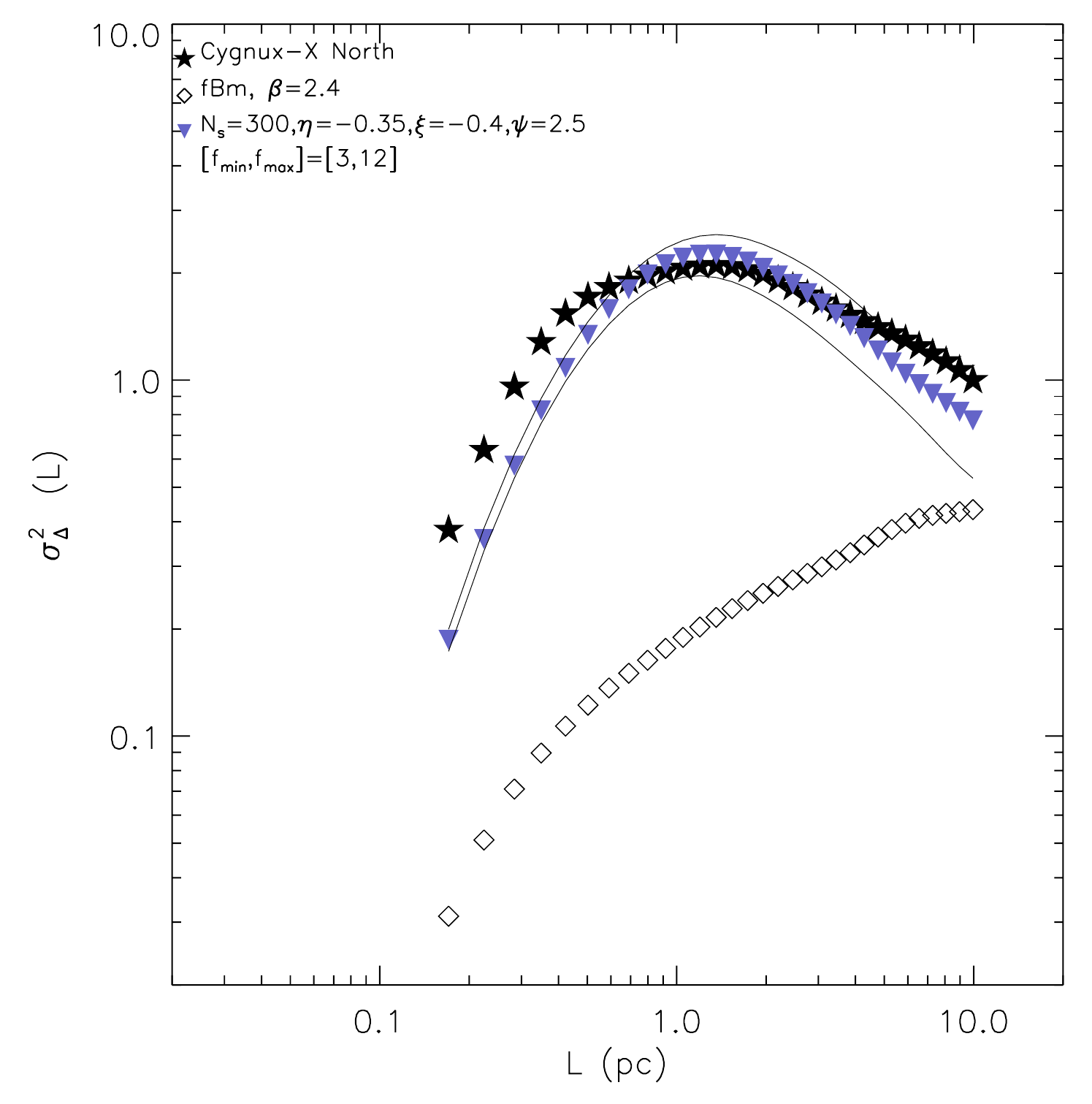}
\caption{Top: Examples of 2D Gaussian structures injected on top of an fBm image with $\beta=2.4$. The maps only differ in the values of the lower and upper bounds of the aspect ratios, $f_{min}$ and $f_{max}$, respectively. All other parameters have the same values (see text for details). All synthetic maps are convolved with a beam whose FWHM=$18.2\arcsec$. Bottom: $\Delta$-variance spectrum of the synthetic models for the three cases with the considered sets of $f_{min}$ and $f_{max}$. Each synthetic spectrum is the average over 25 realizations, and the full lines represent the $1 \sigma$ dispersion around the mean. The synthetic $\Delta$-variance spectra are compared to that of the Cygnus-X North region and to that of an fBm image with $\beta=2.4$.}
\label{fig17}
\end{figure*}

\begin{figure*}
\centering
\hspace{0.5cm}
\includegraphics[width=0.25\textwidth] {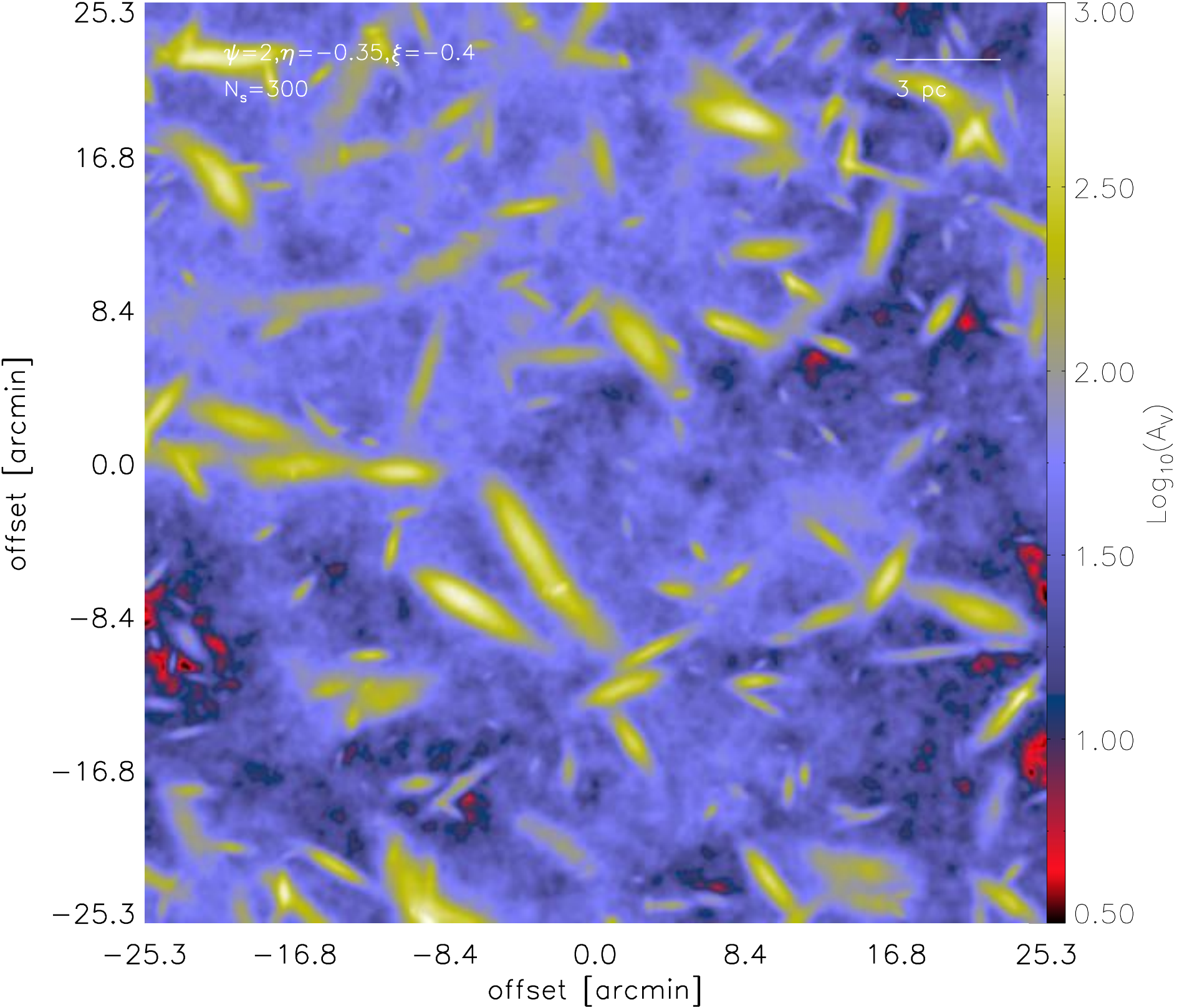}
\hspace{1.5cm}
\includegraphics[width=0.25\textwidth] {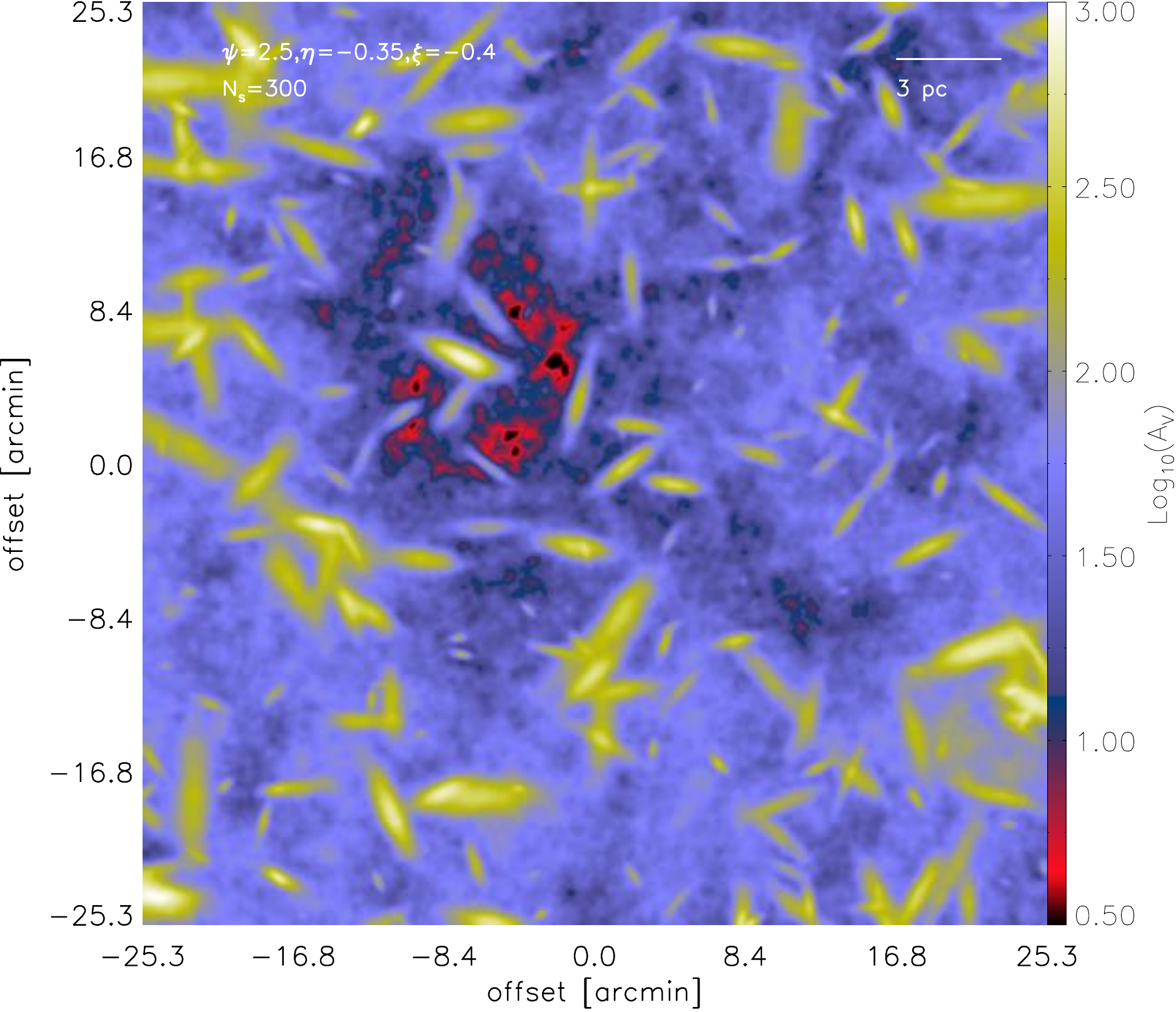}
\hspace{1.5cm}
\includegraphics[width=0.25\textwidth] {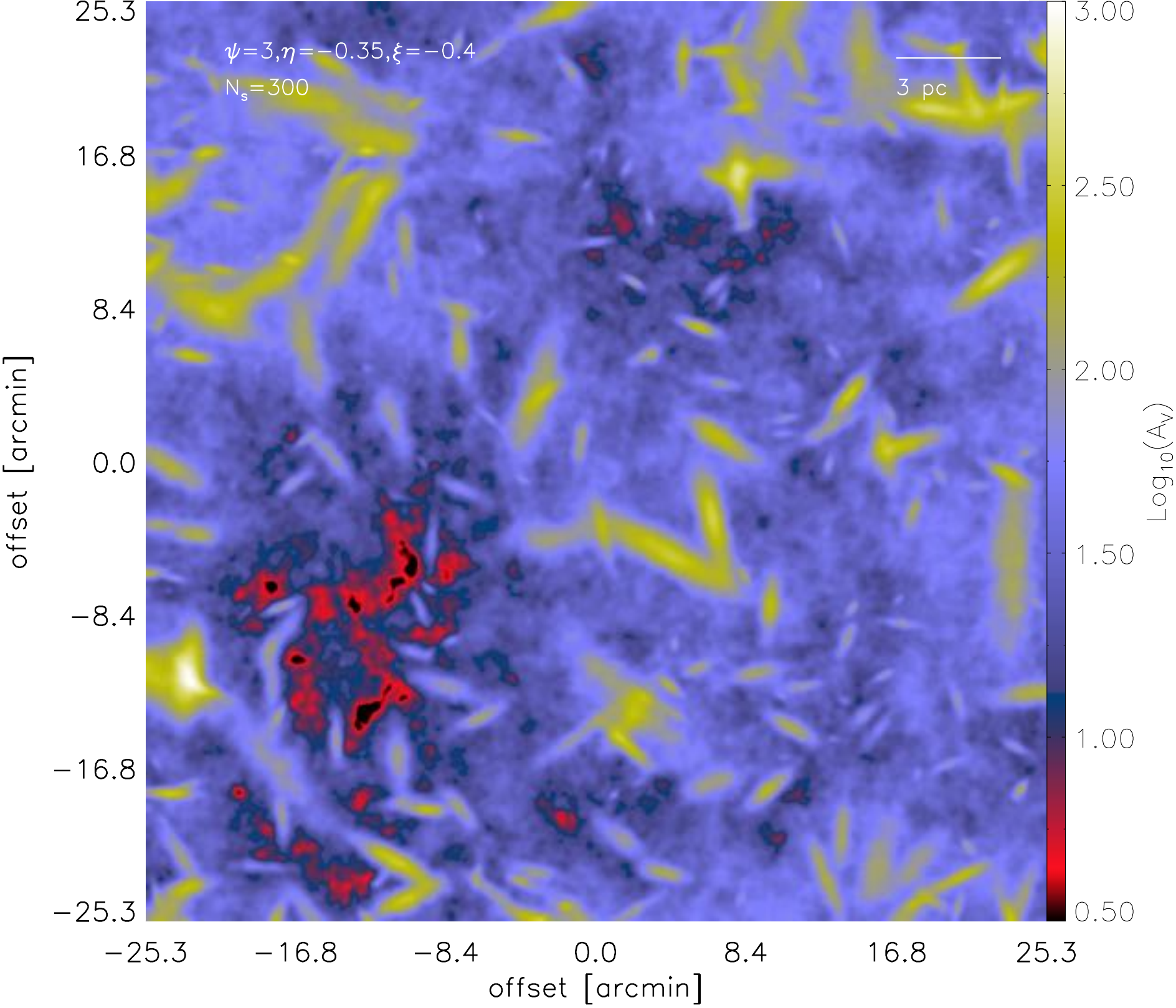}\\
\vspace{0.5cm}
\includegraphics[width=0.32\textwidth] {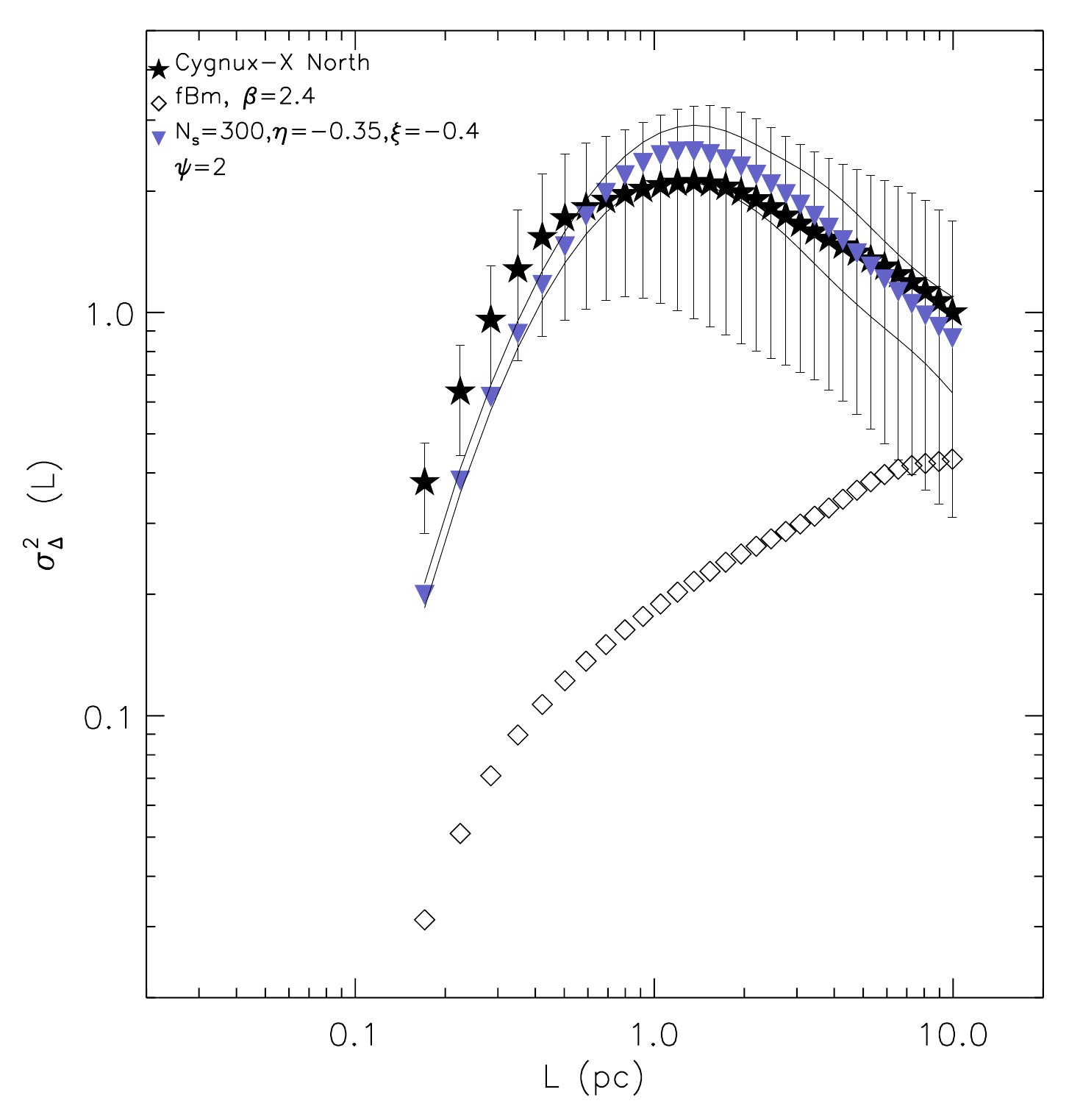}
\hspace{0.2cm}
\includegraphics[width=0.32\textwidth] {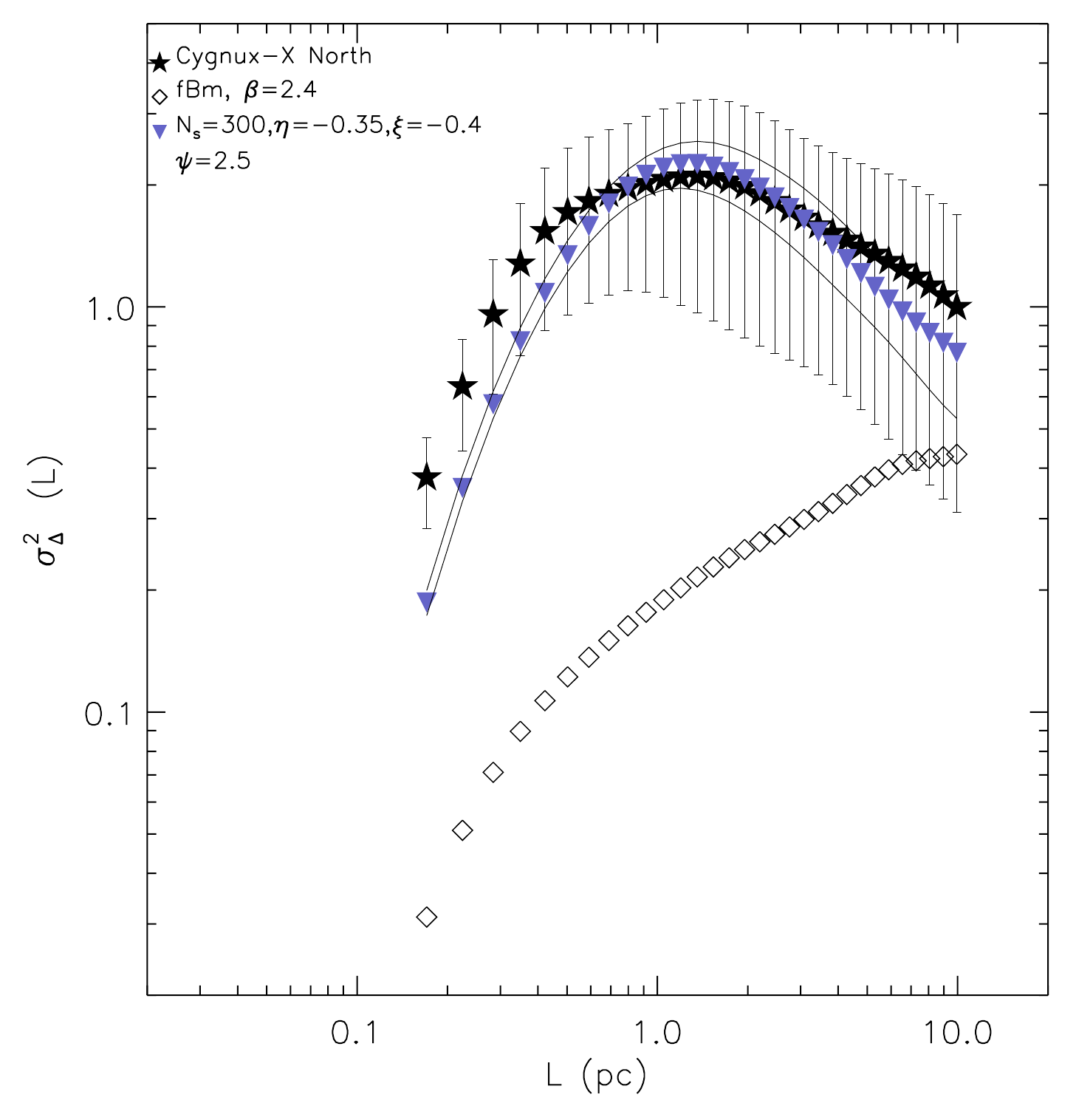}
\hspace{0.2cm} 
\includegraphics[width=0.32\textwidth] {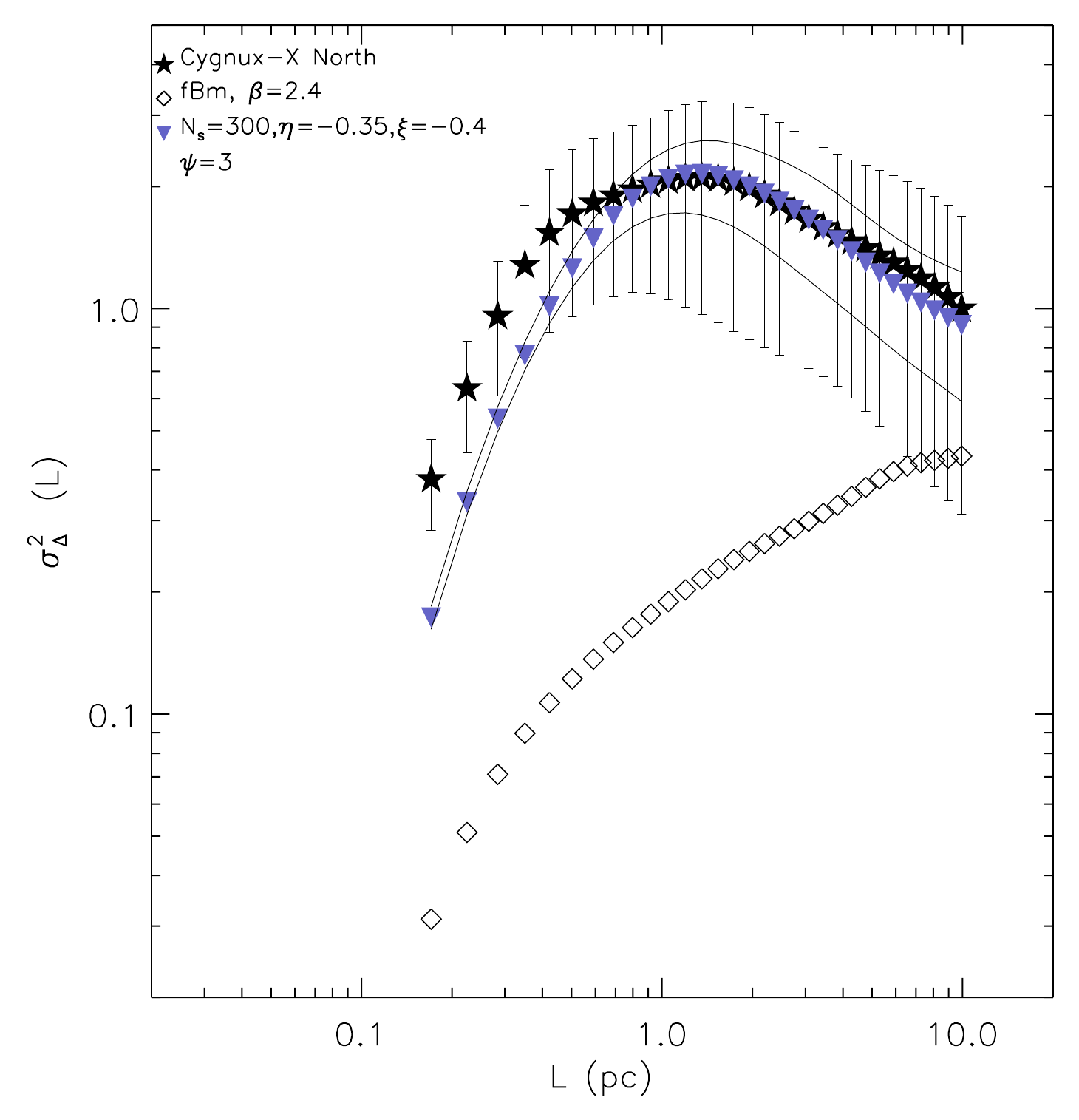}
\caption{Top: Examples of 2D Gaussian structures injected on top of an fBm image with $\beta=2.4$. The maps only differ in the values of the exponent $\psi$. All other parameters have the same values (see text for details). All synthetic maps are convolved with a beam whose FWHM=$18.2\arcsec$. Bottom: $\Delta$-variance spectrum spectra of the synthetic models for the three cases with $\psi=2$, $2.5$, and $3$. Each synthetic spectrum is the average over 25 realizations, and the full lines represent the $1 \sigma$ dispersion around the mean. The synthetic $\Delta$-variance spectra are compared to that of the Cygnus-X North region and that of an fBm image with $\beta=2.4$.}
\label{fig18}
\end{figure*}

 \begin{figure*}
\centering
\hspace{0.5cm}
\includegraphics[width=0.25\textwidth] {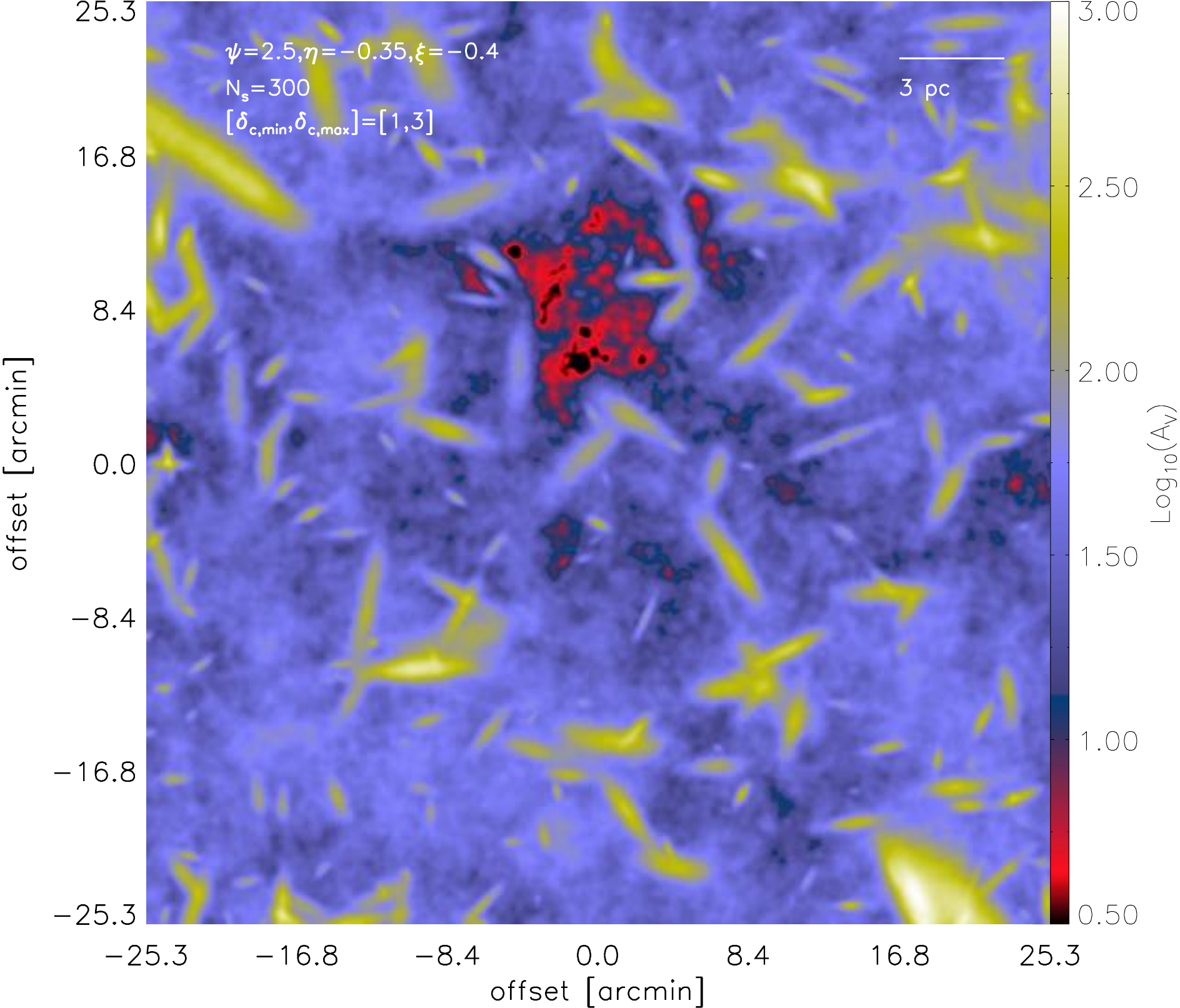}
\hspace{1.5cm}
\includegraphics[width=0.25\textwidth] {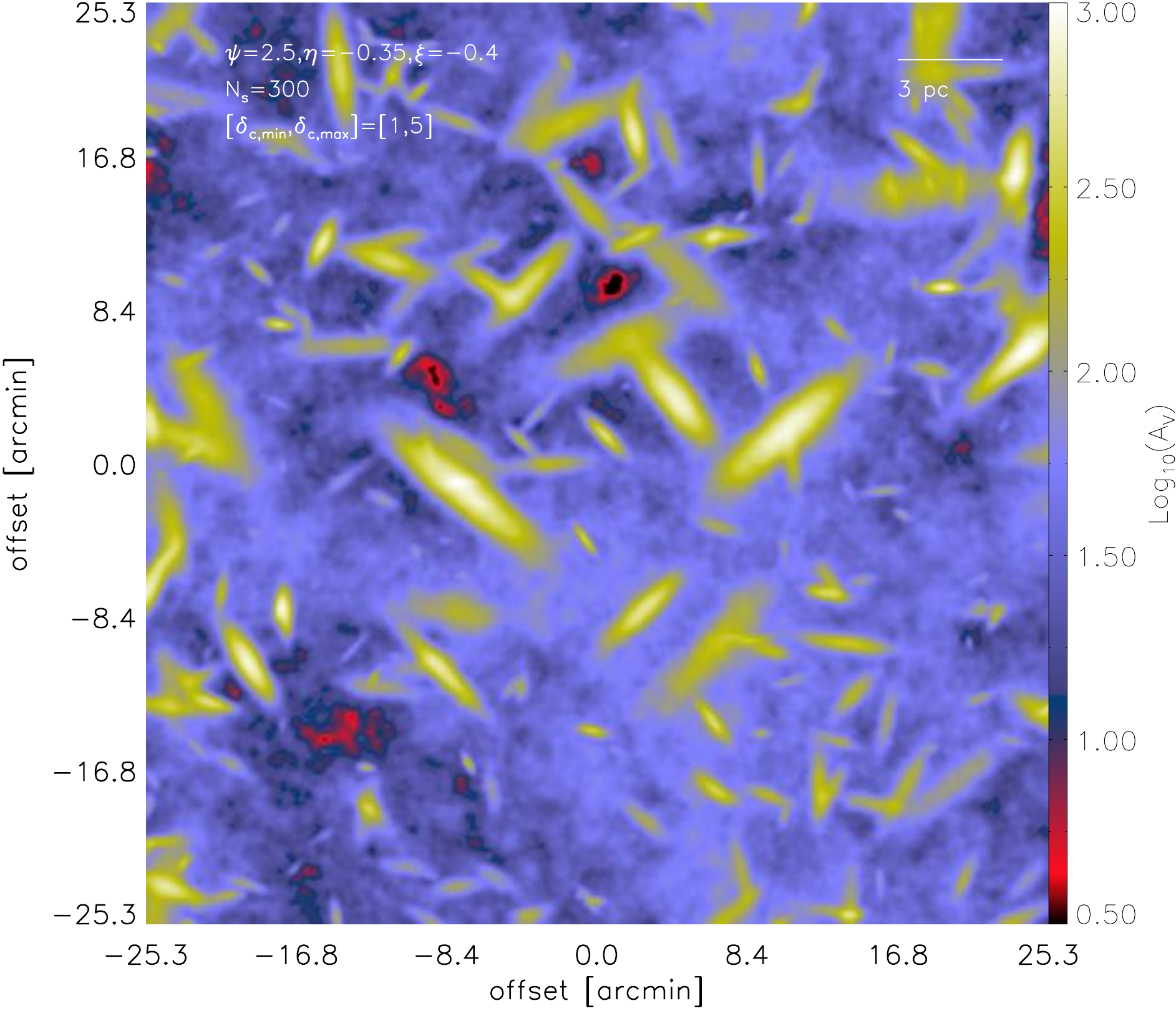}
\hspace{1.5cm}
\includegraphics[width=0.25\textwidth] {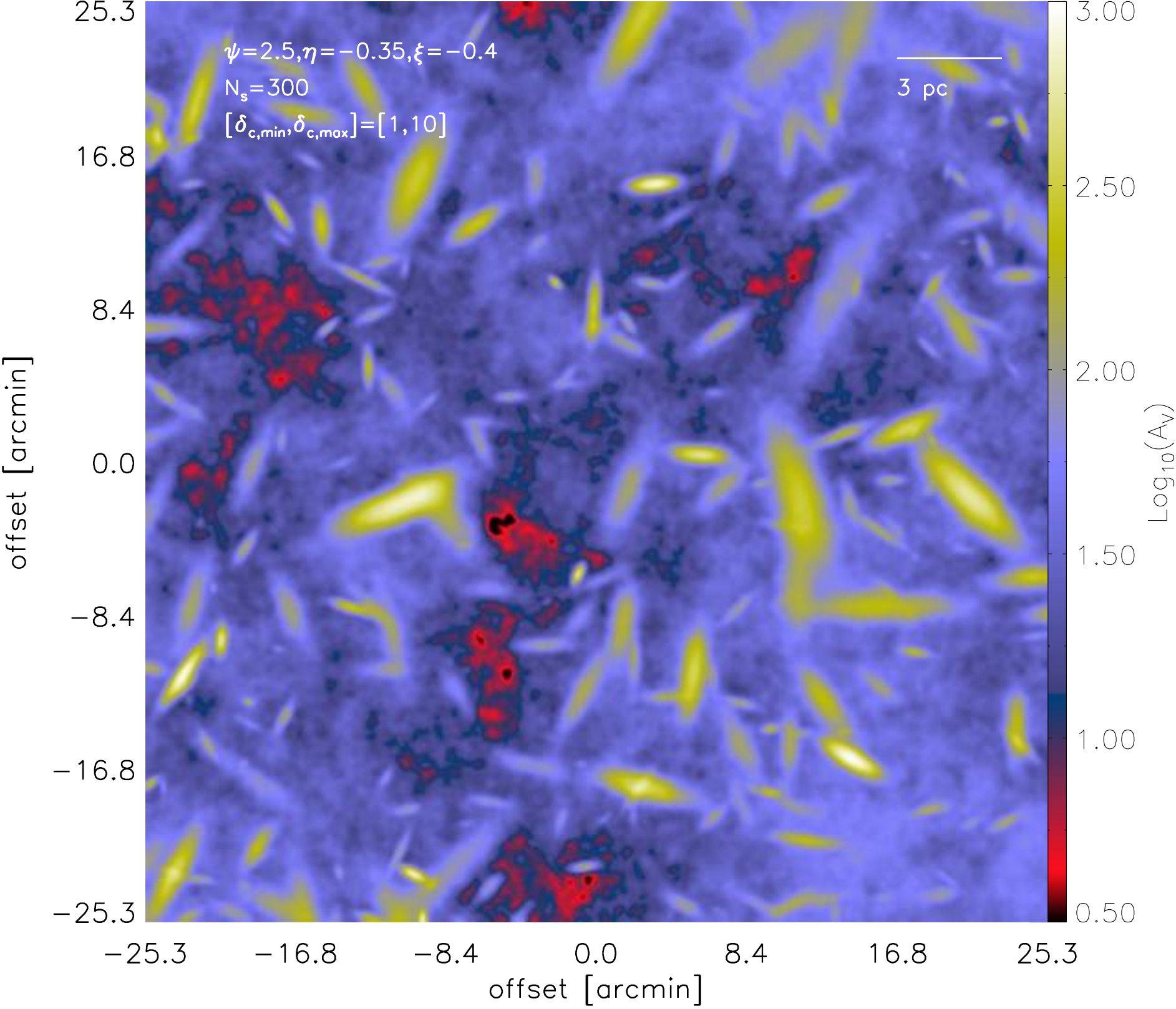}\\
\vspace{0.5cm}
\includegraphics[width=0.32\textwidth] {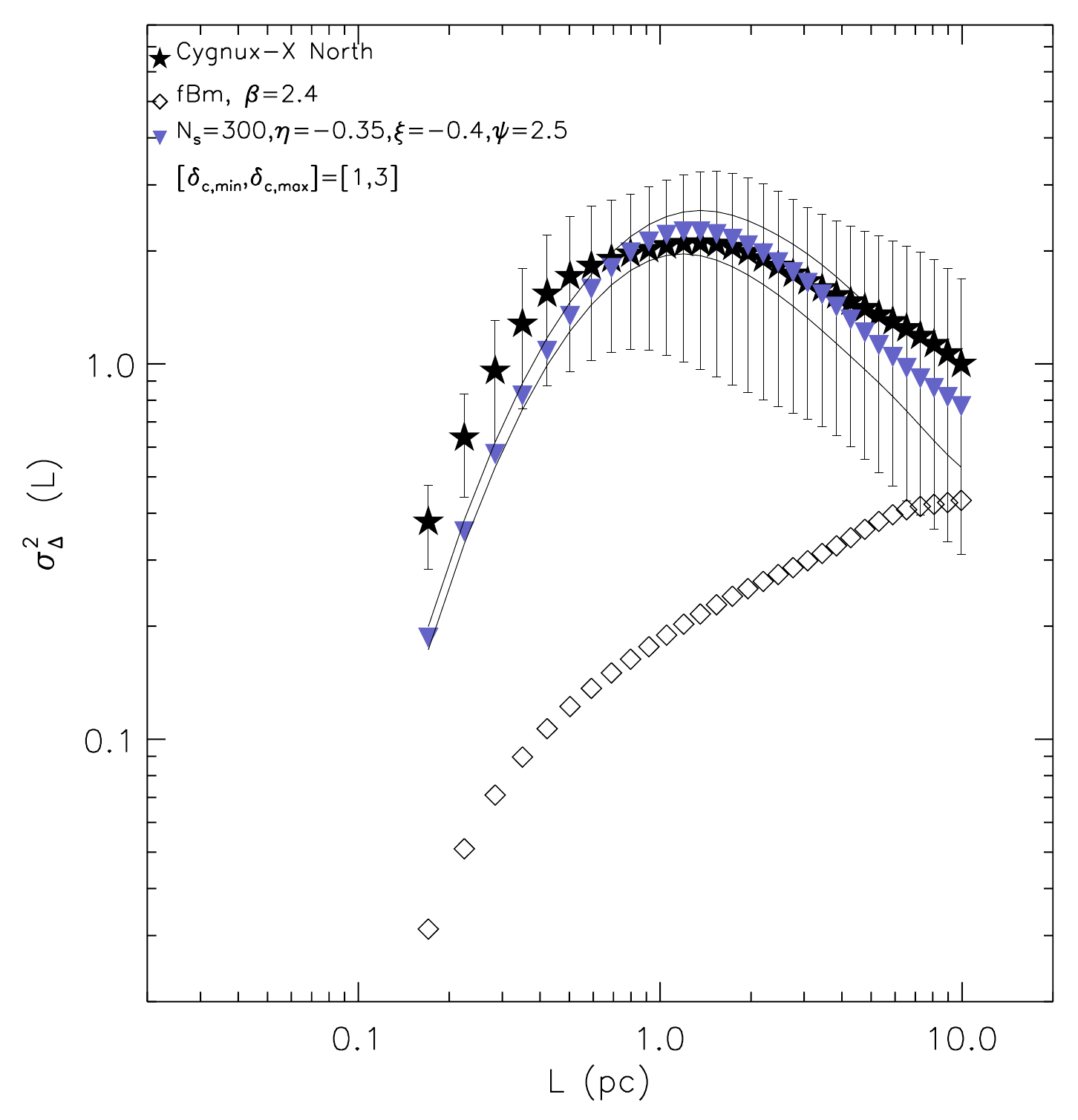}
\hspace{0.2cm}
\includegraphics[width=0.32\textwidth] {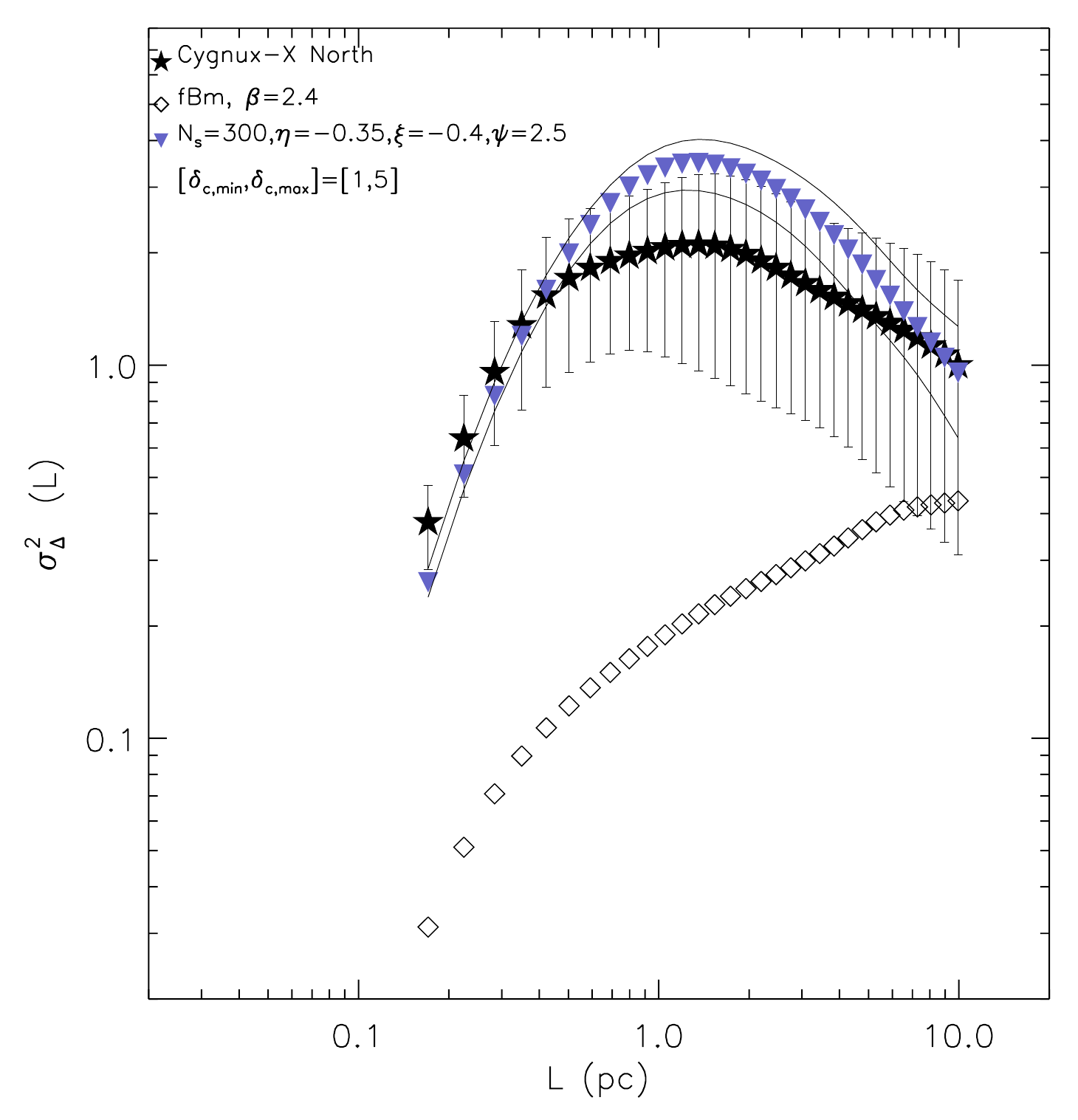}
\hspace{0.2cm} 
\includegraphics[width=0.32\textwidth] {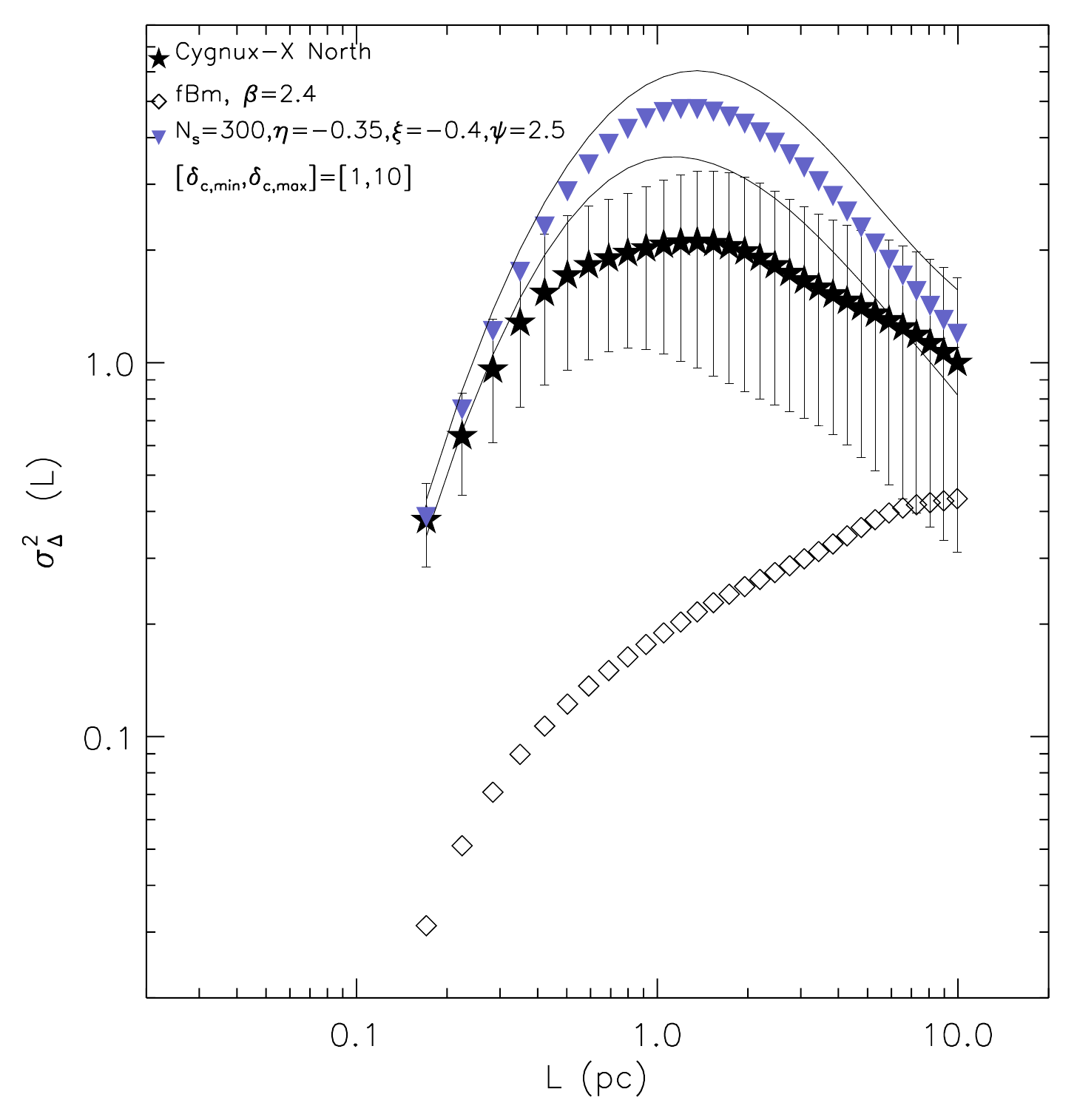}
\caption{Top: Examples of 2D Gaussian structures injected on top of an fBm image with $\beta=2.4$. The maps only differ in the values of the lower and upper bounds of column density contrast $\delta_{c,min}$ and $\delta_{c,max}$, respectively. All other parameters have the same values (see text for details). All synthetic maps are convolved with a beam whose FWHM=$18.2\arcsec$. Bottom: $\Delta$-variance spectra of the synthetic models for the three cases with the considered sets of $\delta_{c,min}$ and $\delta_{c,max}$. Each synthetic spectrum is the average over 25 realizations, and the full lines represent the $1 \sigma$ dispersion around the mean. The synthetic $\Delta$-variance spectra are compared to that of the Cygnus-X North region and to that of an fBm image with $\beta=2.4$.}
\label{fig19}
\end{figure*}

In this section, we perform a broader parameter study and investigate how the $\Delta$-variance spectrum is affected by variations in the distribution functions of the structure sizes, aspect ratios, and column density contrasts. It is nearly impossible to cover the entire parameter space for the four parameters ($N_{s}, \eta, \xi, \psi$). We therefore fixed $N_{s}=300$, and adopted, in the first instance, a value of $\psi=2.5$. We varied the shapes of the major axis and aspect ratio distribution functions and considered values of $\xi=[-0.8,-0.6,-0.4,-0.2]$ and $\eta=[-1.15,-0.75,-0.35]$. Furthermore, we fixed here the lower and upper bounds of the structures sizes (i.e., size of the major axis) to $L_{1,min}=0.025$ pc and $L_{1,max}=5$ pc, respectively. The lower and upper bounds on the aspect ratios are fixed in all cases to the values of $f_{min}=3$ and $f_{max}=12$, and the lower and upper limits on the column density contrasts are $\delta_{c,min}=1$ and $\delta_{c,max}=3$. Examples of maps generated with each permutation of these parameters are displayed in Fig.~\ref{fig14}. The $\Delta$-variance spectra for all of these cases are displayed in Fig.~\ref{fig15}. Here again, we performed 25 realizations with each set of parameters and computed the mean value and standard deviation on each spatial scale. What Fig.~\ref{fig15} reveals is that the $\Delta$-variance spectrum is more sensitive to the shape of the distribution of the major axis sizes, characterized here by the parameter $\xi,$ than to the distribution of aspect ratios (parameter $\eta$). Shallower distributions of the major axis ($\xi=-0.2$) lead to an over abundance of larger structures on the map and to a noticeable mismatch of the $\Delta$-variance for those cases with the $\Delta$-variance spectrum of the Cygnus-X North region, irrespective of the value of $\eta$. In contrast, steeper distribution functions of the major axis (e.g., $\xi=-0.8$) lead to significantly less variance than what is observed in Cygnus-X North on scales $\gtrsim 0.2$ pc. For intermediate values of $\xi$ in the range [-0.6,-0.4], there is a good agreement between the $\Delta$-variance spectrum of the synthetic models and the spectrum of Cygnus-X North. Models with $\xi=-0.4$ present the best fit, but one can reasonably argue that cases with $\xi=-0.6$ could still be considered a good fit to the data if higher values of $N_{s}$ were employed. From the grid of models shown in Fig.~\ref{fig15}, the case with $\xi=-0.4$ and $\eta=-0.35$ represents the best fit to the observations. This corresponds to a mean value of the size of the major axis of $\bar{L_{1}}=\int_{0.025}^{5}\left(dN/dL_{1}\right)L_{1} dL_{1}/\int_{0.025}^{5}\left(dN/dL_{1}\right)dL_{1} \approx 1.05$ pc and to a mean value of the aspect ratio of $\bar{f}=\int_{3}^{12} \left(dN/df\right) f df/\int_{3}^{12} \left(dN/df\right) df \approx 4.23$. This implies a mean value of the minor axis $\bar{L_{2}}=0.25$ pc, and, taking $\sigma_{1}=L_{1}/3$ and $\sigma_{2}=L_{2}/3$, this yields a value of the effective size $D_{eff}=4\sqrt{\sigma_{1}\sigma_{2}}\approx 0.70$ pc, which is very close to the mid position of the plateau in the $\Delta$-variance spectrum of the Cygnus-X North cloud.
 
 Figure~\ref{fig16} shows the individual $\Delta$-variance spectra for the 25 individual realizations (i.e., gray lines) when structures are randomly drawn from the distribution functions with the best fitting set of parameters, namely $N_{s}=300$, $\xi=-0.4$, $\eta=-0.35$, and $\psi=2.5$. We adopted these values of $N_{s}$, $\xi$, and $\eta$ and further investigated the effect of the remaining parameters.

In order to further explore the effect of the aspect ratio, $f$, we performed additional tests in which we varied its lower and upper bounds. In addition to the fiducial case in which $f_{min}=3$ and $f_{max}=12$, we considered models with $(f_{min}=1,f_{max}=6)$ and $(f_{min}=1,f_{max}=12)$. All other parameters were fixed to those of the best fitting model in Fig.~\ref{fig15}, namely $\xi=-0.4$, $\eta=-0.35$, $\psi=2.5$, and $(L_{1,min}=0.025~{\rm pc}, L_{1,max}=5~{\rm pc)}$, $(\delta_{c,min}=1,\delta_{c,max}=3)$. Figure~\ref{fig17} (top panel) displays examples of the maps for each one of the considered cases. The calculations of the $\Delta$-variance spectra in those cases (Fig.~\ref{fig17}, bottom panel) show that the existence of more roundish structures (i.e., larger structures for the same major axis size) results in $\Delta$-variance spectra that peak at higher spatial scales, and those cases present a poor fit to the observations of Cygnus-X North. 

We now explore the effect of varying the distribution function of the column density contrast. In addition to the fiducial case with $\psi=2.5$, we constructed synthetic maps with $\psi=2$ and $\psi=3$. We also generated additional maps in which the value of $\psi$ is fixed to $2.5$ and varied the values of the lower and upper limits on the column density contrast, $\delta_{c}$. We considered cases with $(\delta_{c,min}=1,\delta_{c,max}=3$; fiducial case shown earlier) and cases with $(\delta_{c,min}=1,\delta_{c,max}=5)$ and $(\delta_{c,min}=1,\delta_{c,max}=10)$. The remaining parameters were fixed to their fiducial values, namely $\xi=-0.4$, $\eta=-0.35$, $(L_{1,min}=0.025~{\rm pc}, L_{1,max}=5~{\rm pc)}$, and $(f_{min}=3,f_{max}=12)$. Figure~\ref{fig18} (top panel) displays selected realizations of maps generated with various values of $\psi$, and Fig.~\ref{fig19} (top panel) displays examples of maps generated with different values of $\delta_{c,max}$ and with a fixed value of $\psi=2.5$. The $\Delta$-variance spectra for various cases of $\psi$ are shown in Fig.~\ref{fig18} (bottom panel) and are always calculated as being the mean values from 25 realizations. Overall, the $\Delta$-variance spectrum is less impacted by variations in $\psi$, even though values of $\psi \ge 2.5$ lead to a better agreement with the observations. On the other hand, allowing for higher values of the maximum column density contrast, $\delta_{c,max}$, has an impact on the amplitude of the deviation from the $\Delta$-variance of the underlying fBm, but it has no effect on the position of the point of maximum deviation (Fig.~\ref{fig19}, bottom panel), in agreement with our findings in Sect.~\ref{fbmplus}. We find that a value of $\delta_{c,max}=3$ fits the observations better.     

In summary,  we are able to show that it is possible to reproduce $\Delta$-variance spectra that resemble that of the Cygnus-X North region under reasonable assumptions of the size distributions of structures, their aspect ratios, and column density contrasts. Broadly speaking, reproducing the $\Delta$-variance spectrum of the Cygnus-X North region requires a size distribution that is steeper than the size distribution of structures detected in CO surveys, such as the HCS survey, and shallower than the one inferred from the Hi-GAL submm survey. We also show that the observations are best fitted when structures are allowed to have aspect ratios that are predominantly $\gtrsim 3$. 

\begin{figure}
\centering
\includegraphics[width=0.47\columnwidth]{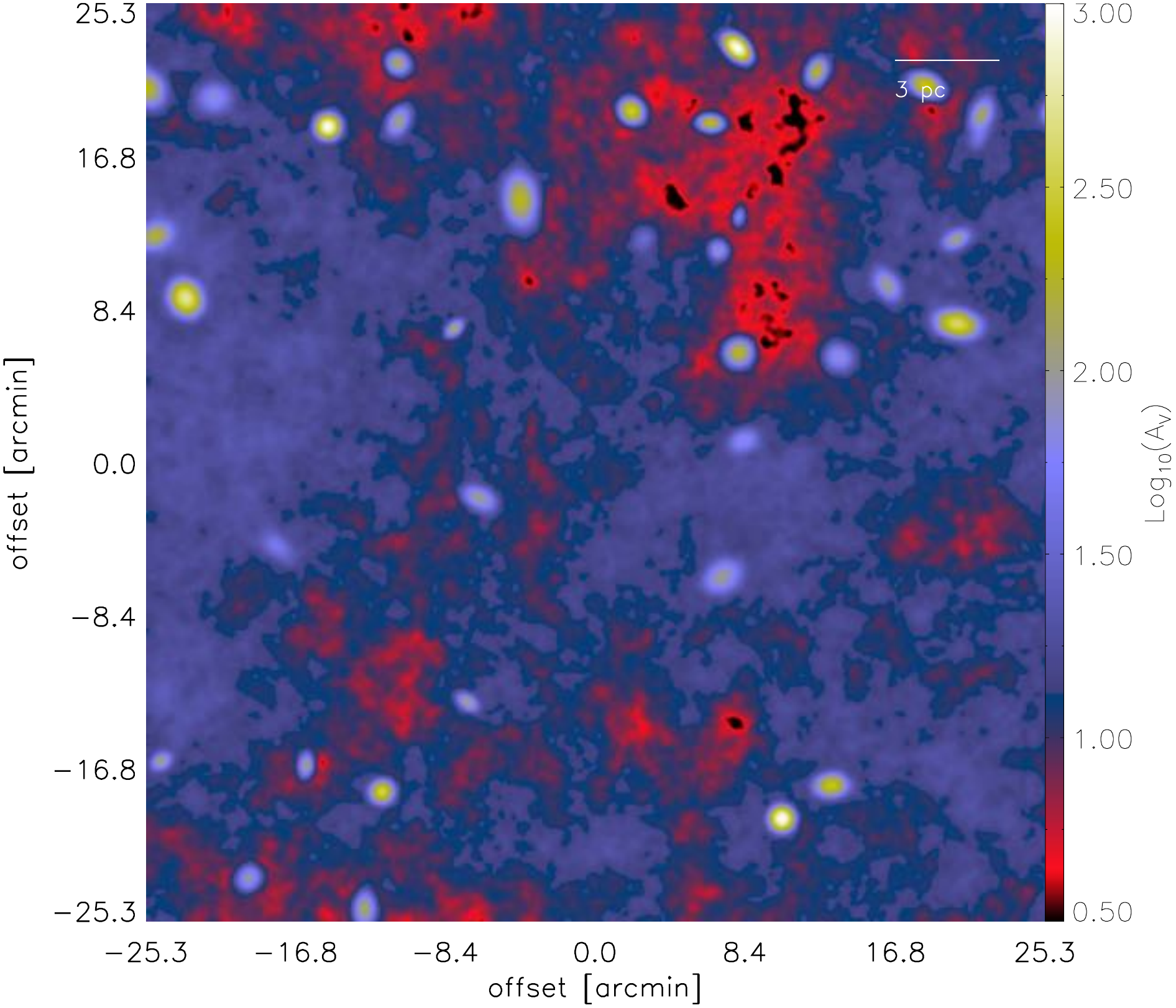}
\hspace{0.2cm}
\includegraphics[width=0.47\columnwidth]{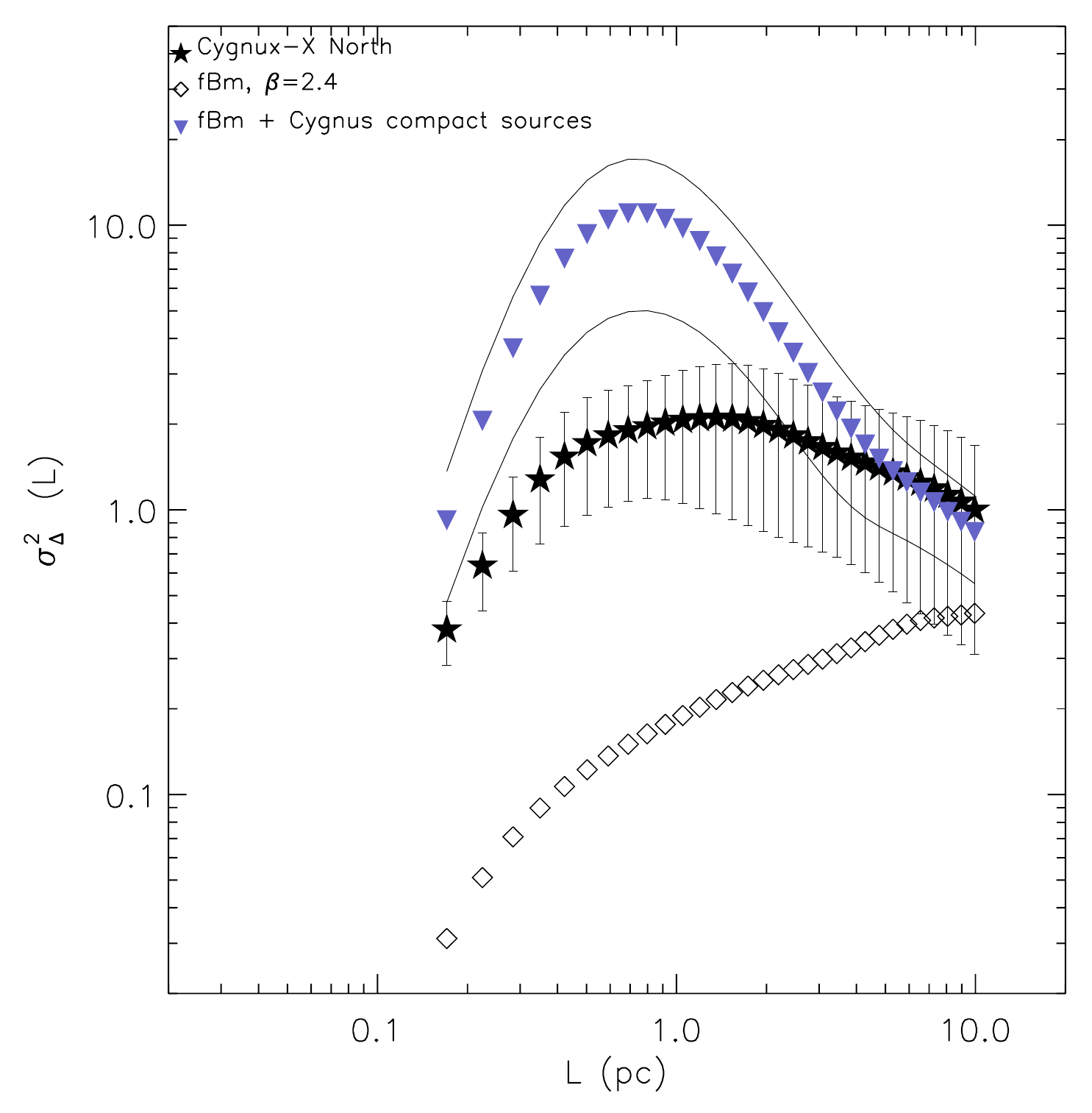}
\vspace{0.3cm}
\caption{Left: Synthetic column density map generated by reinjecting the compact source extracted from the Cygnus-X North map on top of an fBm image (convolved with a beam with FWHM=$18.2\arcsec$) with $\beta=2.4$. Right: $\Delta$-variance spectrum of the synthetic models using this approach. The $\Delta$-variance spectrum of the models is the average over 50 realizations, and the full lines represent the $1 \sigma$ dispersion around the mean. The synthetic $\Delta$-variance spectra are compared to that of the Cygnus-X North region and to the case of an fBm image with $\beta=2.4$.}
\label{fig20}
\end{figure}

\subsubsection{Contribution of compact sources to the $\Delta$-variance plateau in Cygnus-X}

In this section, we examine the contribution of compact sources to the observed plateau in the $\Delta$-variance spectrum of Cygnus-X North. In the previous section, we generated populations of structures whose properties were sampled from parameterized distribution functions and overlaid these structures on an fBm image. Here, we follow a different approach and extract the compact sources from the map of Cygnus-X North before reinjecting them onto the fBm image. To this purpose, we used a newly developed clump finding algorithm. The details of the code will be presented in a forthcoming paper (Bontemps et al., in preparation). Here we simply summarize its basic concepts. This code uses second-order spatial derivatives in order to recognize high curvature peaks where 2D Gaussian fits are applied, after subtracting a local background. It uses an improved determination of the background emission, thanks to a recently developed minimization, to interpolate an empty space (the footprint of a detected source) in a 2D map. Applying this code to the map of Cygnus-X North, we were able to detect a total of 1242 compact sources. The mean values of $\sigma_{1}$ and aspect ratio ($f$) for this sample of compact sources are $0.23$ pc and $1.32$, respectively. This implies a mean effective size for the compact sources of $D_{eff}=4 \sqrt{\sigma_{1}\sigma_{2}}\approx 0.82$ pc, which corresponds, roughly, to where the $\Delta$-variance of Cygnus has its peak. Because the synthetic maps we are using have $1000\times1000$ pixels and are thus $\approx 33$ times smaller than the map of Cygnus-X North, we injected a total of $(1242/33) \approx 38$ compact sources onto each synthetic map. Unlike synthetic maps generated earlier, only the underlying fBm is convolved with the Gaussian beam since the compact sources extracted from the Cygnus-X North map are already affected by beam smearing. 

We generated 50 synthetic maps such that each core is statistically selected at least once and, for each map, the 38 structures that are injected are randomly sampled from the list of structures that are extracted from the observational map and are assigned random positions and orientations on the map. Figure \ref{fig20} (left-hand panel) displays one of the realizations of the synthetic maps using this approach, and the mean $\Delta$-variance spectrum calculated from the 50 realizations is shown in the right-hand panel of Fig.~\ref{fig20}. The exponent of the fBm in this case is also taken to be $\beta=2.4$. We also generated other models with different values of the fBm exponent, in the range $\beta=[2,3]$ (figures not shown for redundancy). The $\Delta$-variance spectrum of the models (i.e., Fig.~\ref{fig20}) exhibits a peak at $\approx 0.6-0.8$ pc, which is at the lower end of the plateau found in the observations. However, there is no agreement between the models and the observations, neither in terms of the width of the $\Delta$-variance spectrum nor its amplitude. What Fig.~\ref{fig20} reveals is that the compact sources taken alone, despite having an important contribution to the signal at $\approx 0.6-0.8$ pc, cannot explain the full extent of the plateau that is observed in the $\Delta$-variance spectrum of Cygnus-X North; it further reveals that there is a need to consider a distribution of structures that includes both larger and more elongated objects in order to explain the observations. A broader distribution of sizes is required in order to reproduce the broad $\Delta$-variance spectrum in Cygnus-X North as well as to adjust the amplitude of the spectrum to the observed values since more extended structures can provide the intermediate column densities between the compact sources and the underlying fBm structure. 

\section{Discussion}\label{discussion}

The present study relies on the comparison of observations with synthetic maps that are generated using 2D fBm images with superimposed structures. As stated above, the structures are injected at random positions and are not necessarily associated with existing higher column density regions in the fBm images. Additionally, the random position and orientation that are assigned to each structure imply that the structures are not spatially correlated, and this may affect the signal on spatial scales on the order of the structures' effective separation. Furthermore, fBm images, albeit a good proxy for self-similar structures, are known to differ from real molecular clouds in terms of their multifractal nature, or, more specifically, their lack thereof (e.g., see details in Elia et al. 2018). There is, however, no reason to believe that any of these assumptions or simplifications are critical to the analysis. Our results demonstrate that the $\Delta$-variance spectrum of a complex region such as Cygnus-X North can be reproduced reasonably well using realistic distribution functions of the characteristics of these structures (size, contrast, aspect ratio). Nonetheless, it is important to stress that a more physical model is  still needed in order to tie the existence of these structures to the physical conditions that prevail in the gas and to the initial conditions of the gas when the molecular cloud has started to assemble. While some refinements can be made to the empirical models presented in this work, it is probably safe to state that numerical models that incorporate most or all of the necessary physics - and that preferably simulate galactic scales larger than the clouds themselves while resolving the internal structure of the clouds - constitute the next step for comparing models to the observations. Complex features, such as striations that are observed in molecular clouds (e.g., Heyer et al. 2016; Tritsis et al. 2018), can naturally emerge self-consistently in numerical models and are harder to implement in empirical models.

On the observational side, we recall that the Herschel maps presented in this work have been resampled to a higher resolution by a factor of two in order to match the resolution of the 250$\mu$m maps. This was done using the method detailed in Palmeirim et al. (2013). This approach has the advantage of increasing the dynamical range of the maps, but it may have introduced additional signal on small scales; this, in turn, may have contributed to worsening the agreement between the synthetic models and the observations of Cygnus-X on these scales. While revisiting this correction is well beyond the scope of this paper, this effect is possibly what is causing the $\Delta$-variance spectrum in Cygnus-X North to fall less sharply at smaller spatial scales than what is expected from the effects of beam smearing (i.e., the slope of the $\Delta$-variance spectrum in the first bin is shallower than the slope in the second bin of the spectrum). Another possible issue relates to the existence of an underlying self-similar regime in Cygnus-X. The simple experiments presented in Figs.~\ref{fig6}-\ref{fig9} when a single (or a few similar) structure(s) is (are) superimposed onto the fBm image show that the self-similarity in the $\Delta$-variance spectrum is preserved on scales that are either smaller or larger than the effective diameter of the structure(s). In the case of multiple structures with different sizes, contrasts, and elongations, the underlying self-similarity is perturbed on a larger range of spatial scales. Thus, the identification of a self-similar regime in Cygnus-X North, if it exists, would in principle require higher resolution observations in order to probe the shape of the $\Delta$-variance at smaller spatial scales and/or a larger map, possibly connecting to the \ion{H}{I} gas at the outer edges of the cloud in order to probe the shape of the spectrum at larger spatial scales.

\section {Conclusions}\label{conclusions}

The internal structure of molecular clouds holds important clues regarding the physical processes that lead to their formation and their subsequent dynamical evolution. While the overall morphology of a molecular cloud can be linked to its star formation activity (Dib \& Henning 2019) and thus provide hints about the cloud's assembly mechanism, the internal structure of the cloud also holds important information about the fragmentation process and the competition between different physical processes that redistribute matter within the cloud. Using the $\Delta$-variance spectrum, we have characterized the structure of the Cygnus-X North and Polaris Flare molecular clouds. These two clouds represent two extremes in terms of their star formation activity in the Milky Way. In Polaris, the structure of the cloud as revealed by the $\Delta$-variance is self-similar over more than one order of magnitude in spatial scales. In contrast, the $\Delta$-variance spectrum of Cygnus-X North exhibits an excess (compared to Polaris) and a plateau in the range of physical scales of $\approx 0.5-1.2$ pc. The departure from self-similarity in a region such as Cygnus-X North is due to the existence of over-dense structures, including compact sources (i.e., hubs and ridges), and more elongated clumps and filaments. In such a region, these structures may arise as a result of large-scale compressions (i.e., converging flows) before being dominated by their own self-gravity. They are also likely to be affected by the mechanical and radiative feedback from massive stars that form in the cloud.  

In order to explain the observations of Cygnus-X North, we built synthetic maps in which we overlaid a population of discrete structures (i.e., 2D Gaussians) on top of an fBm image. The properties of these structures, such as their major axis sizes, aspect ratios, and column density contrasts, are randomly drawn from parameterized probability distribution functions of these quantities. We show that the inclusion of discrete structures "on top" of a self-similar image increases the $\Delta-$variance, and this increment has its maximum on spatial scales that are equal to the effective size of the injected structures (or to an effective mean size of the structures if they have a spectrum of sizes and elongations). Using this forward modeling approach, we are able to show that, under very plausible assumptions, it is possible to reproduce a $\Delta$-variance spectrum that resembles that of the Cygnus-X North region. We also used a "reverse engineering" approach in which we extracted the compact structures in the Cygnus-X North cloud and reinjected them onto an fBm map. The calculated $\Delta$-variance spectrum using this approach deviates from the observations and is an indication that the range of characteristic scales ($\approx 0.5-1.2$ pc) observed in Cygnus-X North is not only due to the existence of compact sources, but is a signature of the whole population of structures that exist in the cloud, including more extended and more elongated structures such as ridges and hubs. Such structures are required in order to broaden the peak of the $\Delta$-variance spectrum and also because they provide the required intermediate column densities that reduce the contrast between the compact sources and the potentially underlying fBm, bringing the amplitude of the $\Delta$-variance in line with the observations. At present, it is relatively difficult to ascertain which physical process leads to the formation of structures with scales in the range $0.5-1.2$ pc. While gravity is the suspected culprit because its effect can precede from an evolutionary point of view over those of stellar feedback, an analysis of the pillars and globules in the Cygnus OB2 association has shown that these structures have typical sizes of $\approx 0.6$ pc (Schneider et al. 2016a). This indicates that feedback may be responsible, at least partially, for generating the peak in the $\Delta$-variance spectrum that is observed in the entire Cygnus-X North region. Further work should shed more light on the possible correlation between the shape of the $\Delta$-variance spectrum and star formation activity, such as the surface density of star formation and the intensity of the radiation field in different parts of the cloud. Independently, the application of the $\Delta$-variance method to numerical simulations of self-gravitation clouds with and without feedback effects will also help explain the dominant physical processes that can generate a structure similar to the one observed in  the Cygnus-X North molecular cloud. 

\begin{acknowledgements}
 
We thank the Referee for a careful reading of the paper and for useful suggestions. We would also like to thank Shantanu Basu, Bruce Elmegreen, Jan Palou\v{s}, Jo\~{a}o Alves, Nanda Kumar, and Alessio Traficante for their feedback and useful discussions on an early version of the paper. S. D., S. B., and N. S. acknowledge support from the french ANR and the german DFG through the project "GENESIS" (ANR-16-CE92-0035-01/DFG1591/2-1). Some of the calculations were performed on Copenhagen's University DCSC cluster which is supported by a research grant (VKR023406) from the Villum Foundation. This research has made use of data from the Herschel Gould Belt survey project (http://gouldbelt-herschel.cea.fr). The HGBS is a Herschel Key Project jointly carried out by SPIRE Specialist Astronomy Group 3 (SAG3), scientists of several institutes in the PACS Consortium (CEA Saclay, INAF-IAPS Rome and INAF-Arcetri, KU Leuven, MPIA Heidelberg), and scientists of the Herschel Science Center (HSC).

\end{acknowledgements}

%
%

\end{document}